\documentclass[12pt]{article}
\usepackage{amsmath,amssymb,amsthm,mathrsfs,amsfonts,mathtools,calculator,dsfont,bm} 
\usepackage[dvipsnames,svgnames, x11names]{xcolor}
\usepackage{natbib,bibentry,bibunits,lmodern}
\usepackage{grffile} 
\usepackage[bottom,hang,flushmargin]{footmisc} 
\usepackage{enumitem} 
\newlist{problems}{enumerate}{1}
\setlist[problems]{label={\arabic*.}, ref={\arabic{part}.\thechapter.\arabic*}}
\usepackage{pdfpages,selectp} 
\usepackage[hyphens]{url}
\usepackage[labelsep=period]{caption} 
\usepackage{subcaption}
\usepackage{fancyvrb} 
\usepackage{environ,trimspaces,xspace}
\usepackage[titletoc]{appendix}
\usepackage{fancyhdr,setspace,titlesec,titling} 
\usepackage{siunitx,booktabs,longtable,threeparttable,multirow,makecell,float,rotating} 
\usepackage{tikz,pgfplots} 
\pgfplotsset{compat=1.11}
\usetikzlibrary{arrows,intersections,plotmarks}
\usetikzlibrary{backgrounds,positioning,fit,petri}
\usetikzlibrary{mindmap,trees,calendar,external,shapes}
\usetikzlibrary{decorations.fractals,decorations.pathmorphing,shadows,shadings,patterns}
\usetikzlibrary{spy,calc} 
\usepgfplotslibrary{groupplots}
\usepackage[bookmarksnumbered,bookmarksopen,psdextra,unicode,hyperindex,breaklinks,hypertexnames=false]{hyperref}
\usepackage[open,openlevel=2,numbered,atend]{bookmark} 
\usepackage{tabularx}
\usepackage{cleveref}
\renewcommand{\thetable}{\Roman{table}}
\usepackage{pdflscape}
\usepackage{afterpage}

\definecolor{Ocean}{rgb}{0,0,0.75}
\makeatletter
\hypersetup{colorlinks, linkcolor = {NavyBlue}, citecolor = {NavyBlue}, urlcolor ={NavyBlue}}
\newtheorem{thm}{Theorem}[section]
\newtheorem{cor}{Corollary}[section]
\newtheorem{lem}{Lemma}[section]

\newtheorem{ass}{Assumption}[section]
\theoremstyle{definition}

\newtheorem{rem}{Remark}[section]

\theoremstyle{definition}
\newtheorem{ex}{Example}[section]
\newtheorem*{ex*}{Example}

\newtheoremstyle{exctd}
{\topsep} {\topsep}%
{\upshape}
{}
{\bfseries\scshape}
{.}
{1em}
{\thmname{#1} \thmnumber{ #2}\thmnote{#3} (cont.)}
\theoremstyle{exctd}

\newcommand{\amin}{\operatornamewithlimits{arg\,min}}

\pgfmathdeclarefunction{npdf}{3}{%
\pgfmathparse{1/(#3*sqrt(2*pi))*exp(-((#1-#2)^2)/(2*#3^2))}%
}
\tikzset{
    declare function={
        ncdf(\x,\m,\s)=1/(1 + exp(-0.07056*((\x-\m)/\s)^3 - 1.5976*(\x-\m)/\s));
    }
}


\defaultbibliographystyle{C:/Dropbox/Research_Related/BibMaker/ecta}
\allowdisplaybreaks 

\newcommand{\blind}{1}
\begin{document}

\begin{bibunit}
\def\spacingset#1{\renewcommand{\baselinestretch}%
{#1}\small\normalsize} \spacingset{1.8}

\pdfbookmark[1]{Title}{title}

\if1\blind
{
  \title{\bf A Unified Framework for Estimation of High-dimensional Conditional Factor Models}
  \author{Qihui Chen\\ School of Management and Economics\\ The Chinese University of Hong Kong, Shenzhen\\ qihuichen@cuhk.edu.cn}
  \date{}
  \maketitle
} \fi

\if0\blind
{
  \bigskip
  \bigskip
  \bigskip
\begin{center}
    {\LARGE\bf A Unified Framework for Estimation of High-dimensional Conditional Factor Models}
\end{center}
  \medskip
} \fi

\begin{abstract}
This paper presents a general framework for estimating high-dimensional conditional latent factor models via constrained nuclear norm regularization. We establish large sample properties of the estimators and provide efficient algorithms for their computation. To improve practical applicability, we propose a cross-validation procedure for selecting the regularization parameter. Our framework unifies the estimation of various conditional factor models, enabling the derivation of new asymptotic results while addressing limitations of existing methods, which are often model-specific or restrictive. Empirical analyses of the cross section of individual US stock returns suggest that imposing homogeneity improves the model's out-of-sample predictability, with our new method outperforming existing alternatives.
\end{abstract}


\noindent%
\vfill

\newpage
\section{Introduction}\label{Sec1}
\setlength{\abovedisplayskip}{4pt}
\setlength{\belowdisplayskip}{4pt}
\setlength{\abovedisplayshortskip}{4pt}
\setlength{\belowdisplayshortskip}{4pt}

In empirical asset pricing, a central question revolves around understanding why different assets yield varying average returns. Conditional factor models offer a comprehensive framework for integrating conditional information to address this inquiry \citep{Gagliardinietal_Timevarying_2016, Gagliardinietal_EstimationConditionalFactor_2019}. This paper delves into the investigation of a high-dimensional conditional factor model defined as follows:
\begin{align}\label{Eqn: Model}
\hspace{-0.2cm}y_{it} = \alpha_{it} + \beta_{it}^{\prime}f_{t} + \varepsilon_{it} \hspace{-0.05cm}\text{ with }  \hspace{-0.05cm}\alpha_{it} = a_i^{\prime}x_{it} \hspace{-0.05cm}\text{ and } \hspace{-0.05cm}\beta_{it} = B_i^{\prime} x_{it}, i\hspace{-0.05cm}=\hspace{-0.05cm}1,\ldots,N,t\hspace{-0.05cm}=\hspace{-0.05cm}1,\ldots,T.
\end{align}
Here, $y_{it}$ denotes the excess return of asset $i$ in time period $t$, $f_{t}$ represents a $K\times 1$ vector of \textit{unobserved latent} factors, $\alpha_{it}$ characterizes a pricing error, $\beta_{it}$ denotes a $K\times 1$ vector of risk exposures, $\varepsilon_{it}$ stands for an error term, $x_{it}$ is a $p\times 1$ vector of pre-specified explanatory variables known at the beginning of time period $t$ (such as constants, sieve transformations of asset characteristics, sieve transformations of macro state variables, and their interactions), and $a_i$ and $B_i$ are $p\times1$ vector and $p\times K$ matrix of unknown coefficients, respectively. This model captures time-variation in the risk exposures (i.e., $B_i^{\prime} x_{it}$) and the pricing error (i.e., $a_i^{\prime} x_{it}$) through their associations with $x_{it}$, while also allowing for the distinction between ``risk'' and ``mispricing'' explanations regarding the role of $x_{it}$ in predicting returns, thereby contributing to resolving the ongoing ``characteristics versus covariance'' debate \citep{DanielTitman_Characteristics_1997}. Moreover, given that $K$ can be significantly smaller than $p$, the model facilitates the condensation of information from a large dimension of $x_{it}$ into a smaller number of factors, thereby mitigating the so-called ``factor zoo'' that proliferate in the literature \citep{Cochrane_Presidential_2011}. However, the estimation of the model encounters at least two challenges: i) $\{f_t\}_{t\leq T}$ are unknown and unobservable; ii) the dimension of the unknown parameters $\{a_i\}_{i\leq N}$, $\{B_i\}_{i\leq N}$, and $\{f_t\}_{t\leq T}$ is high.

The model nests various factor models in the literature. Unlike homogeneous versions of conditional factor models \citep{Parketal_FactorDynamics_2009,Kellyetal_Characteristics_2019,Chenetal_SeimiparametricFactor_2021}, our model allows for heterogeneity of $a_i$ and $B_i$ across assets. Consequently, our model nests classical factor models \citep{Ross_APT_1976,ChamberlainRothschild_FactorStuctures_1982} where $x_{it} =1$ and $a_i=0$; semiparametric factor models \citep{Connoretal_EfficientFFFactor_2012,Fanetal_ProjectedPCA_2016,Kimetal_Arbitrage_2019} where $x_{it}$ comprises a constant and sieve transformations of asset's time-invariant characteristics, with homogeneity of $a_i$ and $B_i$ across assets for non-constant explanatory variables; and state-varying factor models \citep{PelgerXiong_State-varying_2019} where $x_{it}$ encompasses a constant and sieve transformations of macro state variables, with $a_i=0$. Unlike \citet{Gagliardinietal_Timevarying_2016}, our model does not necessitate observable $f_t$ and accommodates the presence of arbitrage and large $p$, referred to as the unconstrained conditional factor model.

We provide a general framework for the estimation of high-dimensional conditional factor models. Specifically, we develop a nuclear norm regularized estimation of the model in \eqref{Eqn: Model} with \textit{constraints} on $\{a_{i},B_{i}\}_{i\leq N}$. The estimation procedure comprises two steps: first, estimating an $Np\times T$ reduced rank matrix composed of block matrices $\{a_i + B_{i}f_t\}_{i\leq N,t\leq T}$ using nuclear norm regularization under the constraints; then, extracting estimators of $K$, $\{a_i\}_{i\leq N}$, $\{B_i\}_{i\leq N}$, and $\{f_t\}_{t\leq T}$ from the estimated matrix using eigenvalue decomposition. We establish asymptotic properties of the estimators under a restricted strong convexity condition. Our framework allows for both $p\to\infty$ and $K\to\infty$ and may accommodate the presence of missing values, which are prevalent in stock return datasets.

The general framework enables the estimation of the aforementioned nested models in a unified manner, overcoming limitations of existing methods that are often model-specific or restrictive.\footnote{For example, \citet{ConnorKorajczyk_Performance_1986}, \citet{StockWatson_PCA_2002}, and \cite{BaiNg_NumberofFactors_2002} estimate the classical factor model by principal component analysis (PCA), while \citet{Fanetal_largecovariance_2013} use a principal orthogonal complement thresholding method. \citet{Fanetal_ProjectedPCA_2016} propose a projected-PCA for the semiparametric factor model based on linear sieves, while \citet{Fanetal_StructuralDeep_2022} employ neural networks. \citet{PelgerXiong_State-varying_2019} estimate the state-varying factor by a local version of PCA based on kernel smoothing. \citet{Chenetal_SeimiparametricFactor_2021} develop a regressed-PCA for the homogenous conditional factor model; \citet{Parketal_FactorDynamics_2009} propose a Newton-Raphson algorithm; \citet{Kellyetal_Characteristics_2019} propose
an alternating least squares procedure; \citet{Guetal_Autoencoder_2021} propose an autoencoder method. \citet{Gagliardinietal_Timevarying_2016} require observable factors for estimating a conditional factor model with no arbitrage.\label{foot1}} By tailoring the general theory for each model and providing simple primitive conditions, we make several contributions. First, we offer a novel estimation approach for the homogeneous conditional factor model, allowing $p$ to grow as fast as $N$. Second, we accommodate time-varying characteristics, nonzero pricing errors, and non-noisy intercepts in both pricing errors and risk exposures for the semiparametric factor model. Third, our estimator is capable of consistently estimating the factor space in the state-varying factor model. Fourth, to the best of our knowledge, our paper is the first to provide an estimation method for the unconstrained conditional factor model.


To enhance the practical applicability of our estimation procedure, we offer two contributions. Firstly, we present an efficient computing algorithm for finding the constrained nuclear norm regularized estimator of the reduced rank matrix in each model. This contribution is particularly valuable as constrained nuclear norm regularization involves high-dimensional constrained nonsmooth convex minimization, where computational efficiency is crucial for practical implementation. Secondly, we propose a cross-validation (CV) procedure to determine the optimal regularization parameter and validate its effectiveness through a series of Monte Carlo simulations. This contribution is essential because the choice of regularization parameter significantly impacts the estimates, and a systematic method for its selection is necessary to ensure robustness and reliability of the results. Our simulation studies demonstrate that the finite sample performance of our estimators, using the CV-chosen regularization parameter, is satisfactory and encouraging. Our simulations also demonstrate the superiority of our estimators compared to existing ones. We apply our unified framework to analyze the cross section of individual stock returns in the US market. Our analysis reveals that imposing homogeneity of $a_i$ and $B_i$ across assets enhances the model's out-of-sample predictability, with our method outperforming existing approaches.


Nuclear norm regularization has been extensively employed for estimating reduced-rank matrices in the statistical literature, primarily focusing on estimating the reduced-rank matrix itself. For instance, \citet{NegahbanWainwright_Scaling_2011} investigate \textit{unconstrained} nuclear norm regularized estimation of trace linear regression models under a restricted strong convexity condition; \citet{RohdeTsybakov_Lowrank_2011} examine the same problem under a restricted isometry condition; \citet{Fanetal_Generalized_2019} study generalized trace regression models. Our work diverges from these studies in several key aspects. Firstly, we require constrained nuclear norm regularization, which entails extending  existing methodologies to accommodate constraints. Secondly, our parameters of interest are $K$, $\{a_i\}_{i\leq N}$, $\{B_i\}_{i\leq N}$, and $\{f_t\}_{t\leq T}$, rather than the reduced-rank matrix. This distinction introduces an additional step in the estimation procedure to estimate these parameters from the reduced-rank matrix.

There have been several recent studies in the econometric literature that utilize unconstrained nuclear norm regularization. \citet{BaiNg_RankRegularized_2019} use it to enhance estimation of the classical factor model. \citet{MoonWeidner_Nuclear_2018} leverage it to improve estimation of panel data models with interactive fixed effects. \citet{Chernozhukovetal_LowRank_2018} employ it to estimate panel data models with heterogeneous coefficients. \citet{Atheyetal_CompletionCausal_2021} adopt this approach in treatment effect estimation. For more examples, see \citet{MoonWeidner_Nuclear_2018}. To the best of our knowledge, the use of constrained nuclear norm regularization in estimating conditional factor models has not been studied previously.

The literature on the cross section of asset returns is extensive; here we focus on conditional factor models. While our paper emphasizes models with latent factors, a substantial portion of empirical asset pricing research relies on pre-specified observable factors. These factors are often constructed using portfolio-sorting approaches, such as those outlined in \citet{FamaFrench_Commonrisk_1993}, based on asset characteristics.\footnote{Notable works in this area include studies by \citet{Shanken_Intertemporal_1990}, \citet{FersonHarvey_Variation_1991,FersonHarvey_Conditioning_1999}, \citet{LettauLudvigson_Consumption_2001}, and \citet{Gagliardinietal_Timevarying_2016}, among others. For a comprehensive review, see \citet{Gagliardinietal_EstimationConditionalFactor_2019}.} This approach encounters challenges related to the ``characteristics versus covariances'' debate and the ``factor zoo'' problem. We contribute to the literature by presenting a unified method for estimating conditional factor models without the need for pre-specified factors, which are well-suited for addressing the debate and problem \citep{Kellyetal_Characteristics_2019,Chenetal_SeimiparametricFactor_2021}.

The structure of the paper is outlined as follows. Section \ref{Sec2} presents several nested models. Section \ref{Sec3} outlines the general estimation framework. Section \ref{Sec4} establishes the asymptotic properties of the estimators. Section \ref{Sec5} tailors the general theory for each model. Section \ref{Sec6} presents simulation studies. Section \ref{Sec7} analyzes the cross section of individual US stock returns. Finally, Section \ref{Sec8} provides a brief conclusion. The Supplementary Appendix presents proofs of main results, computing algorithms, additional discussions, and additional simulations.

\section{Nested Models}\label{Sec2}
Our model in \eqref{Eqn: Model} nests many factor models in the literature.

\begin{ex}[Classical Factor Models]\label{Ex: Classical}
The arbitrage pricing theory by \citet{Ross_APT_1976} and \citet{ChamberlainRothschild_FactorStuctures_1982} gives rise to the following model:
\begin{align}\label{Eqn: Classical}
y_{it} = \lambda_i^{\prime}f_t + e_{it},
\end{align}
where $\lambda_i$ represents an unknown vector of risk exposures and $e_{it}$ is the idiosyncratic component. Our model encompasses \eqref{Eqn: Classical} where $x_{it} =1$, $a_{i} = 0$, $B_{i} = \lambda_{i}^{\prime}$, and $\varepsilon_{it} = e_{it}$.
\end{ex}

\begin{ex}[Semiparametric Factor Models]\label{Ex: SemiTin}
The model examined by \citet{Connoretal_EfficientFFFactor_2012}, \citet{Fanetal_ProjectedPCA_2016}, and \citet{Kimetal_Arbitrage_2019} is as follows:\footnote{\citet{Fanetal_ProjectedPCA_2016} assume that $\phi(\cdot)=0$ and $\mu_i=0$, \citet{Connoretal_EfficientFFFactor_2012} additionally assume that $\Phi(\cdot)$ are univariate functions and $\lambda_{i}=0$, and \citet{Kimetal_Arbitrage_2019}  assume that $\mu_i=0$ and $\lambda_i=0$. }
\begin{align}\label{Eqn: SemiTin}
y_{it} = \phi(z_i)+\mu_i + (\Phi(z_i) + \lambda_{i})^{\prime}f_t + e_{it},
\end{align}
where $z_i$ represents a vector of asset's time-invariant characteristics, $\phi(\cdot)$ and $\Phi(\cdot)$ are unknown functions, $\mu_i$ and $\lambda_{i}$ are unknown scalar and vector (intercepts in the pricing errors and risk exposures, which are usually interpreted as the components that are not explained by the characteristics), and $e_{it}$ is the idiosyncratic component. Using sieve methods, $\phi(z_i) = \phi^{\prime}h(z_i) +\delta(z_i)$ and $\Phi(z_i) = \Phi^{\prime}h(z_i) +\Delta(z_i)$, where $h(z_i)$ denotes a vector of basis functions of $z_i$ (excluding constants), $\phi$ and $\Phi$ represent unknown vector and matrix of coefficients, and $\delta(z_i)$ and $\Delta(z_i)$ are negligible sieve approximation errors. Our model nests \eqref{Eqn: SemiTin} where $x_{it} = (1,h(z_i)^{\prime})^{\prime}$, $a_i = (\mu_i,\phi^{\prime})^{\prime}$, $B_i = (\lambda_{i},\Phi^{\prime})^{\prime}$, and $\varepsilon_{it} = e_{it} + \delta(z_i) + \Delta(z_{i})^{\prime}f_t$. Thus, the rows of $a_i$ and $B_i$ corresponding to $h(z_i)$ are homogenous across $i$, meaning that the coefficients for non-constant explanatory variables are homogenous across assets.
\end{ex}

\begin{ex}[State-varying Factor Models]\label{Ex: Statevarying}
\citet{PelgerXiong_State-varying_2019} examine the following model: 
\begin{align}\label{Eqn: Statevarying}
y_{it} = \Phi_i(z_{t})^{\prime}f_{t} + e_{it},
\end{align}
where $z_{t}$ represents a vector of constant and macro state variables known at the beginning of time period $t$, $\Phi_i(\cdot)$ is a vector of unknown functions, and $e_{it}$ is the idiosyncratic component. Employing sieve methods, $\Phi_i(z_t) = \Phi_i^{\prime}h(z_t) +\Delta_{i}(z_t)$, where $h(z_t)$ denotes a vector of basis functions of $z_t$ (which may include a constant), $\Phi_i$ is an unknown matrix of coefficients, and $\Delta_i(z_t)$ is a vector of negligible sieve approximation errors. Our model encompasses \eqref{Eqn: Statevarying} where $x_{it} = h(z_t)$, $a_i = 0$, $B_i = \Phi_i$, and $\varepsilon_{it} = e_{it} + \Delta_i(z_{t})^{\prime}f_t$.
\end{ex}

\begin{ex}[Homogeneous Conditional Factor Models]\label{Ex: HomoConditional}
\citet{Parketal_FactorDynamics_2009}, \citet{Kellyetal_Characteristics_2019}, and \citet{Chenetal_SeimiparametricFactor_2021} propose the following model:\footnote{\citet{Kellyetal_Characteristics_2019} assume that $\phi(\cdot)$ and $\Phi(\cdot)$ are linear functions.}
\begin{align}\label{Eqn: HomoConditional}
y_{it} = \phi_0(z_{it}) + \Phi_0(z_{it})^{\prime}f_{t} + e_{it},
\end{align}
where $z_{it}$ represents a vector of constant and asset characteristics known at the beginning of time period $t$, $\phi_0(\cdot)$ and $\Phi_0(\cdot)$ are unknown functions, and $e_{it}$ is the idiosyncratic component. Employing sieve methods, $\phi_0(z_{it}) = \phi_0^{\prime}h(z_{it}) +\delta(z_{it})$ and $\Phi_0(z_{it}) = \Phi_0^{\prime}h(z_{it}) +\Delta(z_{it})$, where $h(z_{it})$ denotes a vector of basis functions of $z_{it}$ (which may include a constant), $\phi_0$ and $\Phi_0$ are unknown vector and matrix of coefficients, and $\delta(z_{it})$ and $\Delta(z_{it})$ are negligible sieve approximation errors. Our model nests \eqref{Eqn: HomoConditional} where $x_{it} = h(z_{it})$, $a_i = \phi_0$, $B_i = \Phi_0$, and $\varepsilon_{it} = e_{it} + \delta(z_{it}) + \Delta(z_{it})^{\prime}f_t$. Thus, $a_i$ and $B_i$ are homogenous across $i$, meaning that the coefficients of explanatory variables are homogenous across assets.
\end{ex}

\begin{ex}[Unconstrained Conditional Factor Models]\label{Ex: ConitionalObs}
In the absence of arbitrage opportunities, \citet{Gagliardinietal_Timevarying_2016} propose the following model:
\begin{align}\label{Eqn: ConitionalObs}
y_{it} =z_{t}^{\prime}\Psi_i z_{t} + z_{it}^{\prime} \Upsilon_i z_{t} + z_{t}^{\prime}\Lambda_i f_t + z_{it}^{\prime}\Xi_i f_t + e_{it},
\end{align}
where $z_{t}$ represents a vector of constant and macro state variables known at the beginning of time period $t$, $z_{it}$ is a vector of asset characteristics known at the beginning of time period $t$, $\Psi_i$, $\Upsilon_i$, $\Lambda_i$, and $\Xi_i$ are unknown matrices of coefficients satisfying certain no arbitrage constraints, and $e_{it}$ is the idiosyncratic component. Our model encompasses \eqref{Eqn: ConitionalObs} without the no arbitrage constraints where $x_{it}$ consists of quadratic transformations of $z_t$ and $z_{it}$, $a_i$ and $B_i$ are functions of $\Phi_i$, $\Psi_i$, $\Upsilon_i$, and $\Lambda_i$, and $\varepsilon_{it} = e_{it}$. In contrast to their estimation method, which relies on observable $f_t$, our estimation procedure treats $f_t$ as latent factors and allows for the presence of arbitrage and large $p$.
\end{ex}

\section{Estimation Strategy}\label{Sec3}
For the convenience of the reader, we gather standard pieces of notation here, which will be utilized throughout the paper. We denote a $k\times k$ identity matrix as $I_k$. The Euclidean norm of a column vector $x$ is represented by $\|x\|$. For a symmetric matrix $A$, we denote its trace as $\mathrm{tr}(A)$, its smallest and largest eigenvalues as $\lambda_{\min}(A)$ and $\lambda_{\max}(A)$. The operator norm of a matrix $A$ is denoted by $\|A\|_2$, its Frobenius norm by  $\|A\|_{F}$, and its vectorization by $\mathrm{vec}(A)$. The Kronecker product of matrices $C$ and $D$ is denoted as $C \otimes D$. Unless specified, asymptotic statements in the paper shall be understood to hold as $N\to\infty$ with fixed $T$ or as $(N,T)\to\infty$, whenever appropriate.

We begin by reformulating the model in \eqref{Eqn: Model} using vectors/matrices. Let $a\equiv (a_1^{\prime},a_2^{\prime},\ldots, $ $a_{N}^{\prime})^{\prime}$ which is an $Np\times 1$ vector of unknown coefficients, $B \equiv (B_1^{\prime},B_2^{\prime},\ldots, B_{N}^{\prime})^{\prime}$ which is an $Np\times K$ matrix of unknown coefficients, and $F \equiv(f_1,f_2,\ldots, f_{T})^{\prime}$ which is a $T\times K$ matrix of latent factors. Let $\Pi$ be an $Np\times T$ unknown parameter matrix that collects the product of $(a_i, B_i)$ and $(1,f_t^{\prime})^{\prime}$, defined as
\begin{align}\label{Eqn: BigMatrixParameter}
\Pi\equiv \left(
            \begin{array}{c}
              (a_1, B_1) \\
              (a_2, B_2) \\
              \vdots \\
              (a_N, B_N) \\
            \end{array}
          \right) \left(
                    \begin{array}{cccc}
                      \left(
                        \begin{array}{c}
                          1 \\
                          f_1 \\
                        \end{array}
                      \right),
                       & \left(
                        \begin{array}{c}
                          1 \\
                          f_2 \\
                        \end{array}
                      \right), & \cdots, & \left(
                        \begin{array}{c}
                          1 \\
                          f_T \\
                        \end{array}
                      \right) \\
                    \end{array}
                  \right) \equiv a 1^{\prime}_{T} + BF^{\prime},
\end{align}
where $1_{T}$ is a $T\times 1$ vector of ones. Let $X_{it}\equiv (e_{N,i}\otimes x_{it}) e_{T,t}^{\prime}$ be an $Np\times T$ observed data matrix of $x_{it}$, where $e_{N,i}$ is the $i$th column of $I_N$ and $e_{T,t}$ is the $t$th column of $I_T$. Then $x_{it}^{\prime}a_i + x_{it}^{\prime}B_{i}f_{t} = \mathrm{tr}(X_{it}^{\prime}\Pi)$, so \eqref{Eqn: Model} can be succinctly expressed as
\begin{align}\label{Eqn: ModelRef}
y_{it} = \mathrm{tr}(X_{it}^{\prime}\Pi) + \varepsilon_{it}.
\end{align}
Since $\Pi$ has at most rank $K+1$, \eqref{Eqn: ModelRef} can be viewed as a trace linear regression model with reduced rank coefficient matrix $\Pi$ \citep{NegahbanWainwright_Scaling_2011, RohdeTsybakov_Lowrank_2011}. Thus, we first estimate $\Pi$ by using the nuclear norm regularization \citep{Fazel_MatrixRank_2002}, which employs the nuclear norm penalty as a surrogate function to enforce the reduced rank constraint. Our estimator of $\Pi$ is given by
\begin{align}\label{Eqn: NuclearNEst}
\hat{\Pi} = \amin_{\Gamma\in\mathcal{S}\subset\mathbf{R}^{Np\times T}}\frac{1}{2}\sum_{i=1}^{N}\sum_{t=1}^{T}(y_{it}-\mathrm{tr}(X_{it}^{\prime}\Gamma))^{2} +\lambda_{NT} \|\Gamma\|_{\ast},
\end{align}
where $\mathcal{S}\subset\mathbf{R}^{Np\times T}$ is convex, $\|\Gamma\|_{\ast}$ is the nuclear norm of $\Gamma$, and $\lambda_{NT}>0$ is a regularization parameter.\footnote{The nuclear norm of $\Gamma$ is $\|\Gamma\|_{\ast}=\sum_{j=1}^{\min\{Np,T\}}\sigma_{j}(\Gamma)$, corresponding to the sum of its singular values, where $\sigma_{j}(\Gamma)$'s are the singular values of $\Gamma$. The nuclear norm of $\Gamma$ is the convex envelope of the rank of $\Gamma$ over the set of matrices with spectral norm no greater than one; see, for example, \citet{Rechtetal_Guaranteed_2010}. } \textit{In particular, by introducing $\mathcal{S}$, which can be strictly smaller than $\mathbf{R}^{Np\times T}$, we can enforce the constraints of $\Pi$ induced by those of $a$ and $B$ \textemdash a critical aspect that has not been explored in the existing literature\textemdash enabling the estimation of various models within a unified framework.}  We set $\mathcal{S}=\mathbf{R}^{Np\times T}$ in Examples \ref{Ex: Classical}, \ref{Ex: Statevarying}, and \ref{Ex: ConitionalObs}, $\mathcal{S}=\mathcal{D}_{M}$ for $0<M<\infty$ (where $\mathcal{D}_{M}$ is given in \eqref{Eqn: Ex2: 1}) in Example \ref{Ex: SemiTin},  and $\mathcal{S} = \{1_{N}\otimes \Gamma: \Gamma\in \mathbf{R}^{p\times T}\}$ in Example \ref{Ex: HomoConditional}; see Section \ref{Sec5} for details. In the latter two cases,  $\mathcal{S}$ is strictly smaller than $\mathbf{R}^{Np\times T}$. Since \eqref{Eqn: NuclearNEst} involves constrained nonsmooth convex minimization, $\hat{\Pi}$ generally does not have an analytical closed form. Although several algorithms are available for solving convex minimization problems with a nuclear norm \citep{VandenbergheBoyd_Semidefinite_1996, Bertsekas_NonlinearProgramming_1999, LiuVandenberghe_Interior_2010, Maetal_Fixedpoint_2011}, they are not suitable for the high-dimensional settings with constraints in our context. In Appendix \ref{App: E}, we provide an efficient computing algorithm for each setting in Examples \ref{Ex: Classical}-\ref{Ex: ConitionalObs}.

We next proceed to derive estimators for $K$, $a$, $B$, and  $F$ from the nuclear norm regularized estimator $\hat{\Pi}$. Let $\hat{K}$, $\hat{a}$, $\hat{B}$, and $\hat{F}$ denote these estimators. Define $M_{T}\equiv I_{T} - 1_{T}1_{T}^{\prime}/T$. Since $\Pi M_{T} = BF^{\prime}M_{T}$, we can obtain $\hat{K}$ and $\hat{B}$ from the eigenvalues and eigenvectors of $\hat{\Pi} M_{T}\hat{\Pi}^{\prime}$. Specifically, $\hat{K}$ is given by
\begin{align}\label{Eqn: KEst}
\hat{K} = \sum_{j=1}^{Np}1\{\lambda_{j}(\hat{\Pi} M_{T}\hat{\Pi}^{\prime})\geq \delta_{NT}\},
\end{align}
where $\lambda_{j}(A)$ denotes the $j$th largest eigenvalue of $A$ and $\delta_{NT}>0$ is a threshold value. If $\hat{K} = 0$, $\hat{a} = {\hat{\Pi}1_{T}}/{T}$, $\hat{B} =0 $, and $\hat{F}=0$; otherwise we proceed as follows. To estimate $B$, we use the following normalization: $B^{\prime}B/N=I_K$ and $F^{\prime} M_{T} F/T$ being diagonal with diagonal entries in descending order. Then the columns of $\hat{B}/\sqrt{N}$ are given by the eigenvectors of $\hat{\Pi} M_{T}\hat{\Pi}^{\prime}$ corresponding to its largest $\hat{K}$ eigenvalues. To estimate $a$ and $F$, we impose the following condition: $a^{\prime} B = 0$. Since $a = (I_{Np}-B(B^{\prime}B)^{-1}B^{\prime})\Pi1_{T}/T$ and $F = \Pi^{\prime}B(B^{\prime}B)^{-1}$, we thus obtain
\begin{align}\label{Eqn: aFEst}
\hat{a} = \left(I_{Np}-\frac{\hat{B}\hat{B}^{\prime}}{N}\right)\frac{\hat{\Pi}1_{T}}{T} \text{ and } \hat{F} =  \frac{\hat{\Pi}^{\prime}\hat{B}}{N}.
\end{align}
It it noteworthy that in Examples \ref{Ex: SemiTin} and \ref{Ex: HomoConditional} there is no need to enforce the homogeneity restriction of $a$ and $B$ in extracting $\hat{a}$ and $\hat{B}$ from $\hat{\Pi}$ again to ensure the same homogeneity structure of $\hat{a}$ and $\hat{B}$; see Sections \ref{Sec52} and \ref{Sec53} for details.

Our estimation procedure is adaptable to accommodate the presence of missing values. In this case, the double summations in \eqref{Eqn: NuclearNEst} must be replaced with summations over non-missing data. This amounts to redefining the observations as $y_{it}m_{it}$ and $x_{it}m_{it}$, and the error term as $\varepsilon_{it}m_{it}$, where $m_{it} = 0$ when $y_{it}$ or $x_{it}$ are missing, and $1$ otherwise. Since $\hat{\Pi}$ accommodates missing values, we can employ a CV approach to choose the regularization parameter $\lambda_{NT}$ in \eqref{Eqn: NuclearNEst}. Specifically, we first randomly divide the observations into $L$ folds with observations indexed by $\{\mathcal{I}_{\ell}\}_{\ell\leq L}$, where $\mathcal{I}_{\ell}$ comprises observation indices in the $\ell$th fold, $\{\mathcal{I}_{\ell}\}_{\ell\leq L}$ are mutually exclusive, and $\cup_{\ell\leq L}\mathcal{I}_{\ell}=\mathcal{I}\equiv \{1,2,\cdots,N\}\times \{1,2,\cdots, T\}$. Rolling $\ell$ from $1$ to $L$, we then leave observations $\{(y_{it},x_{it}): (i,t)\in\mathcal{I}_{\ell}\}$ out, use observations $\{(y_{it},x_{it}): (i,t)\in\mathcal{I}/\mathcal{I}_{\ell}\}$ for training, and calculate the out-of-sample mean square error $\mathrm{MSE}_\ell$ for observations $\{(y_{it},x_{it}): (i,t)\in\mathcal{I}_{\ell}\}$. Finally, we choose $\lambda_{NT}$ by minimizing the average out-of-sample mean square error $\sum_{\ell=1}^{L}\mathrm{MSE}_\ell/L$.

\begin{rem}\label{Rem: ALS}
Enforcing the rank constraint directly is perhaps the most intuitive approach to incorporate the reduced-rank structure. This leads to the following problem:
\begin{align}\label{Eqn: LeastS}
\min_{{c_i}\in\mathbf{R}^{p},{D_i}\in\mathbf{R}^{p\times K},{g}_t\in\mathbf{R}^{K}}\frac{1}{2}\sum_{i=1}^{N}\sum_{t=1}^{T}(y_{it} - x_{it}^{\prime}c_i - x_{it}^{\prime}D_ig_t)^{2}.
\end{align}
However, solving \eqref{Eqn: LeastS} poses at least two challenges.\footnote{Enforcing the constraints of $a$ and $B$ in Examples \ref{Ex: SemiTin} and \ref{Ex: HomoConditional} does not resolve the challenges.} Firstly, it requires knowledge of $K$, which must be estimated prior to solving the problem. Secondly, \eqref{Eqn: LeastS} is nonconvex and its solution lacks an analytical closed form. These challenges not only complicate the design of computational algorithms to find the solution but also hinder derivation of its asymptotic properties. One potential approach to address the second challenge is alternating least squares; however, it may suffer from non-convergence issues due to the nonconvexity of \eqref{Eqn: LeastS} \citep{GolubVanLoan_MatrixComputation_2013, Chietal_Nonconvex_2019}. In contrast, the problem in \eqref{Eqn: NuclearNEst} is convex, and our estimators can be numerically solved efficiently without requiring prior knowledge of $K$, complemented by the asymptotic properties derived in Sections \ref{Sec4} and \ref{Sec5}.
\end{rem}

\section{Asymptotic Analysis}\label{Sec4}
In this section, we conduct an asymptotic analysis for our estimators in a general setup. Specifically, we establish consistency of $\hat{K}$ and a rate of convergence of $\hat{\Pi}$, $\hat{a}, \hat{B}$, and $\hat{F}$.

We begin by introducing the so-called ``restricted strong convexity'' condition \citep{Negahbanetal_UnifiedFramework_2012}. This condition ensures that the quadratic loss function in \eqref{Eqn: NuclearNEst} is strictly convex over a restricted set of ``low-rank'' matrices. To describe the set, we define some notation. Let $\Pi = U \Sigma V^{\prime}$ be a singular value decomposition of $\Pi$, where $U$ and $V$ are $Np\times Np$ and $T\times T$ orthonormal matrices, and $\Sigma$ is a diagonal matrix with singular values of $\Pi$ in the diagonal in descending order. Write $U = (U_1,U_2)$ and $V = (V_1,V_2)$, where the columns of $U_2$ and $V_2$ are singular vectors corresponding to the zero singular values of $\Pi$. For any $Np \times T$ matrix $\Delta$, let $\mathcal{P}(\Delta)\equiv U_2U_2^{\prime}\Delta  V_2V_2^{\prime}$ and $\mathcal{M}(\Delta)\equiv \Delta - \mathcal{P}(\Delta)$. Heuristically, $\mathcal{M}(\Delta)$ can be thought of as the projection of $\Delta$ onto the ``low-rank'' space of $\Pi$, and  $\mathcal{P}(\Delta)$ is the projection of $\Delta$ onto its orthogonal space. The restricted set of ``low-rank'' matrices is
\begin{align}\label{Eqn: RestrictedSet}
\mathcal{C}\equiv \{\Delta\in\mathcal{S}\ominus \mathcal{S}: \|\mathcal{P}(\Delta)\|_{\ast}\leq 3\|\mathcal{M}(\Delta)\|_{\ast}\},
\end{align} 
where $\mathcal{S}\ominus \mathcal{S}$ is the Minkowski difference between $\mathcal{S}$ and $\mathcal{S}$, that is, $\mathcal{S}\ominus \mathcal{S} = \{\Gamma_1-\Gamma_2: \Gamma_1,\Gamma_2\in\mathcal{S}\}$. We impose the restricted strong convexity condition as follows.

\begin{ass}\label{Ass: NuclearN}
(i) Assume that $\Pi\in\mathcal{S}\subset \mathbf{R}^{Np\times T}$. For any $\Delta\in\mathcal{S}\ominus \mathcal{S}$, the following decomposition holds:
\[\sum_{i=1}^{N}\sum_{t=1}^{T}|\mathrm{tr}(X_{it}^{\prime}\Delta)|^2 = \mathcal{Q}_{NT}(\Delta) + \mathcal{L}_{NT}(\Delta)\]
such that for some constant $0<\kappa<\infty$,
\[\mathcal{Q}_{NT}(\Delta) \geq \kappa \|\Delta\|_{F}^{2} \text{ for all } \Delta\in\mathcal{C},\]
and for some $r_{NT}>0$,
\[|\mathcal{L}_{NT}(\Delta)|\leq r_{NT} \|\Delta\|_{\ast}  \text{ for all } \Delta\in\mathcal{S}\ominus \mathcal{S}.\]
(ii) The following condition holds:
\[\left|\sum_{i=1}^{N}\sum_{t=1}^{T}\mathrm{tr}(\varepsilon_{it} X_{it}^{\prime}\Delta)\right|\leq \frac{1}{2}r_{NT} \|\Delta\|_{\ast} \text{ for all } \Delta\in\mathcal{S}\ominus \mathcal{S}.\]
\end{ass}

Assumption \ref{Ass: NuclearN} is weaker than the conditions of Corollary 1 in \citet{NegahbanWainwright_Scaling_2011}, which require $\mathcal{S}=\mathbf{R}^{Np \times T}$ and $\mathcal{L}_{NT}(\cdot)=0$, and are too restrictive in Examples \ref{Ex: SemiTin} and \ref{Ex: HomoConditional}. We refer to the condition: ``$\mathcal{Q}_{NT}(\Delta) \geq \kappa \|\Delta\|_{F}^{2}$ for all $\Delta\in\mathcal{C}$'' as the restricted strong convexity condition. Allowing $\mathcal{L}_{NT}(\cdot)\neq 0$ facilitates providing easy-to-verify primitive conditions for the restricted strong convexity condition in Example \ref{Ex: SemiTin}. The rate $r_{NT}$ plays an important role in determining the convergence rate of $\hat{\Pi}$, thus determining how fast $p$ and $K$ can grow.

\begin{ass}\label{Ass: NuclearKaBF}
There exist some constants $0<d_{\min}\leq d_{\max}<\infty$ such that: (i) $d_{\min}<\lambda_{\min}(B'B/N)\leq \lambda_{\max}(B'B/N)<d_{\max}$ for large $N$; (ii) $\max_{t\leq T}\|f_{t}\|< d_{\max}$; (iii) $\lambda_{\min}(F'M_{T}F/T)> d_{\min}$; (iv) $a^{\prime}a/N<d_{\max}$; (v) $a^{\prime}B = 0$.
\end{ass}

For the sake of clarity in presentation, we assume that $\{a_{i},B_{i}\}_{i\leq N}$ and $\{f_t\}_{t\leq T}$ are non-random. In other words, all stochastic statements are implicitly conditional on their realization. Assumption \ref{Ass: NuclearKaBF}(i) resembles the pervasive condition in \citet{StockWatson_PCA_2002} and \citet{BaiNg_NumberofFactors_2002}, which necessitates that $\{f_t\}_{t\leq T}$ are strong factors. Assumptions \ref{Ass: NuclearKaBF}(iv) and (v) are identification conditions for $a$, see Appendix \ref{App: F: 2} for discussion; similar assumptions are also used in \citet{Chenetal_SeimiparametricFactor_2021}. While Assumptions \ref{Ass: NuclearN} and \ref{Ass: NuclearKaBF} consist of high-level conditions for the general setup, in Section \ref{Sec5} we provide primitive conditions for each setting in Examples \ref{Ex: Classical}-\ref{Ex: ConitionalObs}.

\begin{thm}\label{Thm: NuclearNRate}
Suppose Assumption \ref{Ass: NuclearN} holds. Let $\hat{\Pi}$, $\hat{K}$, $\hat{a}, \hat{B}$, and $\hat{F}$ be given in \eqref{Eqn: NuclearNEst}-\eqref{Eqn: aFEst}. Assume that $0<K<\min\{Np,T\}-1$ and $\lambda_{NT}\geq 2r_{NT}$. (i) Then
\begin{align*}
\|\hat{\Pi}  - \Pi\|_{F}\leq \frac{3\sqrt{2(K+1)}\lambda_{NT}}{\kappa}.
\end{align*}
(ii) Suppose Assumption \ref{Ass: NuclearKaBF} also holds. Assume that $\delta_{NT}/(NT)\to 0$ and $\delta_{NT}/(K\lambda^{2}_{NT})\to \infty $. Let $H \equiv (F^{\prime}M_T\hat{F})(\hat{F}^{\prime}M_T\hat{F})^{-1}$. Then
\begin{align*}
P(\hat{K}= K) &\to 1,\\
\|\hat{a} - a\|&=O_{p}\left(\frac{\sqrt{K}\lambda_{NT}}{\sqrt{T}}\right),\notag\\
\|\hat{B} - B H\|_{F}&=O_{p}\left(\frac{\sqrt{K}\lambda_{NT}}{\sqrt{T}}\right),\notag\\
\|\hat{F}-F(H^{\prime})^{-1}\|_{F}&=O_{p}\left(\frac{\sqrt{K}\lambda_{NT}}{\sqrt{N}}\right).
\end{align*}
\end{thm}

Theorem \ref{Thm: NuclearNRate}(i) gives a deterministic statement about the estimation error of $\hat{\Pi}$, extending Corollary 1 of \citet{NegahbanWainwright_Scaling_2011} by allowing $\mathcal{L}_{NT}(\cdot)\neq 0$ and constraints on $\Pi$ (i.e., $\mathcal{S}\neq \mathbf{R}^{Np \times T}$) in addition to the reduced-rank constraint. While Assumption \ref{Ass: NuclearN} and $\lambda_{NT}\geq 2r_{NT}$ may not hold deterministically, they often hold with probability approaching one, as verified in Section \ref{Sec5}. In such cases, the result of Theorem \ref{Thm: NuclearNRate}(i) holds with probability approaching one, and the results of Theorem \ref{Thm: NuclearNRate}(ii) persist. Due to identification issues, $B$ and $F$ can only be consistently estimated up to a rotational transformation, as commonly encountered in high-dimensional factor analyses. The asymptotic results hold as $N\to\infty$ with fixed $T$ or as $(N,T)\to\infty$, as appropriate. Theorem \ref{Thm: NuclearNRate} is a theory for the general setup under high-level assumptions, which is applicable for each setting in Examples \ref{Ex: Classical}-\ref{Ex: ConitionalObs}. In Section \ref{Sec5}, we tailor Theorem \ref{Thm: NuclearNRate} for each model by providing low-level sufficient assumptions that are easier to verify. In all cases, $p$ and $K$ are permitted to grow with $N$ or $(N,T)$ for the consistency of the estimators, and the presence of missing values is allowed.

\section{Revisiting Nested Models}\label{Sec5}
For simplicity of notation, we continue to use $x_{it}$ representing the vector of explanatory variables in all models, rather than each model's specific notation in Section \ref{Sec2}.

\subsection{Examples \ref{Ex: Classical}, \ref{Ex: Statevarying}, and \ref{Ex: ConitionalObs}}\label{Sec51}

Our objective is to estimate $a$, $B$, $F$, and $K$.\footnote{For simplicity of presentation, we continue to use $a$ and $B$ representing the coefficients of interest in all three examples, rather than each example's specific notation in Section \ref{Sec2}, and ignore the sieve approximation error in Example \ref{Ex: Statevarying} (so $\Delta_{i}(\cdot) = 0$). This allows us to unify results in one theorem. One may account for the sieve approximation error as similar to Corollaries \ref{Cor: Ex2: 1} and \ref{Cor: Ex4: 1}.} No constraints are imposed on $a$ and $B$ and we set $\mathcal{S} = \mathbf{R}^{Np\times T}$ in \eqref{Eqn: NuclearNEst}. In the scenario  when $x_{it}=1$, we can obtain an analytical closed form for $\hat{\Pi}$. Let $Y$ be an $N\times T$ matrix with the $it$th entry $y_{it}$. Consider the singular value decomposition $Y = U\Sigma V^{\prime}$, where $U$ and $V$ are $N\times N$ and $T\times T$ orthonormal matrices and $\Sigma$ is an $N\times T$ diagonal matrix with singular values $\sigma_{j}(Y)$'s in the diagonal in descending order. For $x>0$, define $\Sigma_{x}$ be an $N\times T$ diagonal matrix with $\max\{0,\sigma_j(Y)-x\}$ in descending order. Consequently, $\hat{\Pi} = U\Sigma_{\lambda_{NT}/2} V^{\prime}$, as described in \citet{Caietal_SVD_2010} and \citet{Maetal_Fixedpoint_2011}. However, an analytical closed form is not available for general cases. An efficient algorithm for finding $\hat{\Pi}$ is provided in Appendix \ref{App: E}.

To provide primitive conditions, we impose the following assumptions.

\begin{ass}\label{Ass: Ex135: 1}
(i) There exists some constant $0<\kappa<\infty$ such that
\[\sum_{i=1}^{N}\sum_{t=1}^{T}|\mathrm{tr}(X_{it}^{\prime}\Delta)|^2\geq \kappa \|\Delta\|_{F}^{2} \text{ for all } \Delta\in\mathcal{D},\]
where $\mathcal{D}\equiv\{\Delta\in\mathbf{R}^{Np\times T}: \|\mathcal{P}(\Delta)\|_{\ast}\leq 3\|\mathcal{M}(\Delta)\|_{\ast}\}$. (ii) $\{(x_{1t}^{\prime}e_{1t},x_{2t}^{\prime}e_{2t},\ldots,x_{Nt}^{\prime}e_{Nt})'\}_{t\leq T}$ is a sequence of independent sub-Gaussian vectors.\footnote{Independence is not necessary here and also in Assumptions \ref{Ass: Eqn: Ex2: 1}(iv), (v) and \ref{Ass: Eqn: Ex4: 1}(iii). We may allow for weak dependence over $t$; see Lemma \ref{Lem: UseD1}. }
\end{ass}

A condition similar to Assumption \ref{Ass: Ex135: 1} has been imposed in \citet{MoonWeidner_Nuclear_2018} and \citet{Chernozhukovetal_LowRank_2018}. We apply Theorem \ref{Thm: NuclearNRate} to obtain the following corollary.

\begin{cor}\label{Cor: Ex135: 1}
Suppose Assumption \ref{Ass: Ex135: 1}(ii) holds. Let $\hat{\Pi}$, $\hat{K}$, $\hat{a}, \hat{B}$, and $\hat{F}$ be given in \eqref{Eqn: NuclearNEst}-\eqref{Eqn: aFEst} with $\mathcal{S} = \mathbf{R}^{Np\times T}$ and $\lambda_{NT}=\sqrt{(Np+T)\log N}$. Assume that $0<K<\min\{Np,T\}-1$. (i) If $x_{it} =1$ or Assumption \ref{Ass: Ex135: 1}(i) holds, then as $(N,T)\to\infty$,
\begin{align*}
\frac{1}{\sqrt{NT}}\|\hat{\Pi}  - \Pi\|_{F}= O_{p}\left(\sqrt{\frac{K(Np+T)\log N}{NT}}\right).
\end{align*}
(ii) Suppose Assumptions \ref{Ass: NuclearKaBF}(i)-(iii) additionally hold. Assume that as $(N,T)\to\infty$, \hspace{-0.05cm}$\delta_{NT}/(NT)\hspace{-0.05cm}\to \hspace{-0.05cm}0$ \hspace{-0.05cm}and\hspace{-0.05cm} $\delta_{NT}/[K(Np+T)\log N]\to \hspace{-0.05cm}\infty\hspace{-0.05cm} $. \hspace{-0.1cm}Let $H \hspace{-0.1cm}\equiv \hspace{-0.1cm}(F^{\prime}M_T\hat{F})(\hat{F}^{\prime}M_T\hat{F})^{-1}$. If $a=0$ or Assumptions \ref{Ass: NuclearKaBF}(iv)-(v) hold, then as $(N,T)\to\infty$,
\begin{align*}
P(\hat{K}= K) &\to 1,\\
\frac{1}{\sqrt{N}}\|\hat{a} - a\|&=O_{p}\left(\sqrt{\frac{K(Np+T)\log N}{NT}}\right),\notag\\
\frac{1}{\sqrt{N}}\|\hat{B} - B H\|_{F}&=O_{p}\left(\sqrt{\frac{K(Np+T)\log N}{NT}}\right),\notag\\
\frac{1}{\sqrt{T}}\|\hat{F}-F(H^{\prime})^{-1}\|_{F}&=O_{p}\left(\sqrt{\frac{K(Np+T)\log N}{NT}}\right).
\end{align*}
\end{cor}

Corollary \ref{Cor: Ex135: 1} requires large $N$ and large $T$. In particular, $K(Np+T)\log N=o(NT)$ is required for the consistency of the estimators. This implies that $p$ is allowed to grow as $(N,T)\to\infty$.
While the result for Example \ref{Ex: Classical} is well-documented in the literature, the results for Examples \ref{Ex: Statevarying} and \ref{Ex: ConitionalObs} are novel. Distinct from \citet{PelgerXiong_State-varying_2019}, we offer an estimator capable of consistently estimate $F$ up to a common rotational transformation, which is not state-specific. In other words, we can consistently estimate the factor space. Moreover, our method allows for large $p$. In contrast to \citet{Gagliardinietal_Timevarying_2016}, we provide an estimation approach that does not necessitate observable $f_t$ and permits the presence of arbitrage and large $p$. Notably, there is no available method for estimating the unconstrained conditional latent factor model in the literature.


\subsection{Example \ref{Ex: SemiTin}}\label{Sec52}
Our objective is to estimate $\mu\equiv (\mu_1,\mu_2,\ldots,\mu_N)^{\prime}$, $\Lambda\equiv (\lambda_{1}, \lambda_{2},\ldots,\lambda_{N})^{\prime}$, $\phi$, $\Phi$, $F$, and $K$. Since $a = (\mu_1, \phi^{\prime}, \mu_2, \phi^{\prime},\ldots, \mu_N, \phi^{\prime})^{\prime}$ and $B = (\lambda_1, \Phi^{\prime}, \lambda_2, \Phi^{\prime}, \ldots, \lambda_N, \Phi^{\prime})^\prime$, we have $\Pi =a 1^{\prime}_{T} + BF^{\prime}= ((\pi_1,\Pi^{\ast\prime}), (\pi_2,\Pi^{\ast\prime}),\ldots, (\pi_N,\Pi^{\ast\prime}))^{\prime}$, where $\pi_i\equiv\mu_i1_{T}+ F\lambda_i$ and $\Pi^{\ast} \equiv \phi1_{T}^{\prime} + \Phi F^{\prime}$, which are $ T\times 1$ vector and $(p-1)\times T$ matrix, respectively. Then we set
\begin{align}\label{Eqn: Ex2: 1}
\hspace{-0.25cm}\mathcal{S} = \mathcal{D}_M \equiv \left\{\left(
                    \begin{array}{c}
                      \gamma_1^{\prime} \\
                      \Gamma^{\ast} \\
                      \gamma_2^{\prime} \\
                      \Gamma^{\ast} \\
                      \vdots \\
                      \gamma_N^{\prime} \\
                      \Gamma^{\ast} \\
                    \end{array}
                  \right)
: \left(
    \begin{array}{c}
    \gamma_1^{\prime} \\
    \gamma_2^{\prime} \\
      \vdots \\
    \gamma_N^{\prime} \\
    \end{array}
  \right)
\in\mathbf{R}^{N\times T}, \Gamma^{\ast}\in \mathbf{R}^{(p-1)\times T} \text{ and } \|\Gamma^{\ast}\|_{\max}\leq M\right\}
\end{align}
for $0<M<\infty$ in \eqref{Eqn: NuclearNEst}, where $\|\Gamma^{\ast}\|_{\max}$ denotes the largest absolute value of the entries of $\Gamma^{\ast}$.\footnote{\label{foot7}Imposing $\|\Gamma^{\ast}\|_{\infty}\leq M$ facilitates providing easy-to-verify primitive conditions for Assumption \ref{Ass: NuclearN}(i).} Since $\hat{\Pi}\in\mathcal{D}_{M}$, we can write $\hat{\Pi} = ((\hat{\pi}_1,\hat{\Pi}^{\ast\prime}), (\hat{\pi}_2,\hat{\Pi}^{\ast\prime}),\ldots, (\hat{\pi}_N,\hat{\Pi}^{\ast\prime}))^{\prime}$, where $\hat{\pi}_i$ is an estimator of $\pi_i$ and $\hat{\Pi}^{\ast}$ is an estimator of $\Pi^{\ast}$. Let ${\Pi}^{\diamond}\equiv({\pi}_{1},{\pi}_{2}, \ldots, {\pi}_{N})^{\prime}$ and $\hat{\Pi}^{\diamond}\equiv(\hat{\pi}_{1},\hat{\pi}_{2}, \ldots, \hat{\pi}_{N})^{\prime}$. An efficient algorithm for finding $\hat{\Pi}^{\diamond}$ and $\hat{\Pi}^{\ast}$ is provided in Appendix \ref{App: E}.

By Lemma \ref{Lem: TechD2} (iv) and simple algebra, we can write
\begin{align}\label{Eqn: Ex2: Estimators}
\hat{a} = ((\hat{\mu}_{1},\hat{\phi}^{\prime}), (\hat{\mu}_{2},\hat{\phi}^{\prime}),\ldots, (\hat{\mu}_{N},\hat{\phi}^{\prime}))^{\prime} \text{ and } \hat{B} = ((\hat{\lambda}_{1},\hat{\Phi}^{\prime}),(\hat{\lambda}_{2},\hat{\Phi}^{\prime}),\ldots,(\hat{\lambda}_{N},\hat{\Phi}^{\prime}))^{\prime},
\end{align}
where $\hat{\mu}_i$ is a scalar, $\hat{\phi}$ is a $(p-1)\times 1$ vector, $\hat{\lambda}_i$ is a $\hat{K} \times 1$ vector, and $\hat{\Phi}$ is a $(p-1)\times \hat{K}$ matrix. Thus, $\hat{a}$ and $\hat{B}$ share the same homogeneity structure with $a$ and $B$, respectively. It is not necessary to enforce the homogeneity restriction of $a$ and $B$ in extracting $\hat{a}$ and $\hat{B}$ from $\hat{\Pi}$ to ensure the same homogeneity structure, as the homogeneity structure of $\hat{\Pi}$ inherited from $a$ and $B$ automatically passes to $\hat{a}$ and $\hat{B}$. We define the estimators of  ${\mu}$, $\Lambda$, $\phi$, and $\Phi$ as  $\hat{\mu}\equiv (\hat{\mu}_1, \hat{\mu}_2, \ldots, \hat{\mu}_{N})^{\prime}$, $\hat{\Lambda} \equiv (\hat{\lambda}_{1},\hat{\lambda}_{2},\ldots, \hat{\lambda}_{N})^{\prime}$, $\hat{\phi}$, and $\hat{\Phi}$, respectively.

A convergence rate for $\hat{\Pi}^{\diamond}$, $\hat{\Pi}^{\ast}$, $\hat{\mu}$, $\hat{\Lambda}$, $\hat{\phi}$, and $\hat{\Phi}$ follows immediately from Theorem \ref{Thm: NuclearNRate}, as we have $\|\hat{\Pi}  - \Pi\|^{2}_{F} = \|\hat{\Pi}^{\diamond}-{\Pi}^{\diamond}\|^{2}_{F} + N\|\hat{\Pi}^{\ast}  - \Pi^{\ast}\|^{2}_{F}$, $\|\hat{a} - a\|^{2} = \|\hat{\mu} - \mu\|^2 + N\|\hat{\phi} - \phi\|^2$, and $\|\hat{B} - BH\|_{F}^{2} = \|\hat{\Lambda} - \Lambda H\|_{F}^{2} + N\|\hat{\Phi} - \Phi H\|^2_{F}$. To provide primitive conditions, we impose the following assumptions.

\begin{ass}\label{Ass: Eqn: Ex2: 1}
(i) Write $x_{it} = (1,x_{it}^{\ast\prime})^{\prime}$.\footnote{We allow for time-varying characteristics, so we write $x_{it}$ rather than $x_i$.} There are positive constants $c_{\min}$ and $c_{\max}$ such that: with probability approaching one as $(N,T)\to\infty$,
\[c_{\min} \leq \min_{t\leq T}\lambda_{\min}\left(\frac{1}{N}\sum_{i=1}^{N}x^{\ast}_{it}x_{it}^{\ast\prime}\right)\leq \max_{t\leq T}\lambda_{\max}\left(\frac{1}{N}\sum_{i=1}^{N}x^{\ast}_{it}x_{it}^{\ast\prime}\right)\leq c_{\max} .\]
(ii) $\max_{t\leq T}\hspace{-0.1cm}\|\phi + \Phi f_{t}\|_{\infty}$ is bounded. \hspace{-0.1cm}(iii) $\max_{t\leq T}\hspace{-0.1cm}E[\|\sum_{i=1}^{N}x^{\ast}_{it}e_{it}/\hspace{-0.1cm}\sqrt{Np}\|^{2}]$ is bounded. (iv) $\{(x_{1t}^{\ast\prime},x_{2t}^{\ast\prime},\ldots,x_{Nt}^{\ast\prime})'\}_{t\leq T}$ is a sequence of independent sub-Gaussian vectors. (v) $\{(e_{1t},e_{2t},$ $\ldots,e_{Nt})'\}_{t\leq T}$ is a sequence of independent sub-Gaussian vectors. (vi) $\sup_{z}|\delta(z)| = O(p^{-s})$ and $\sup_{z}\|\Delta(z)\| = O(p^{-s})$ for some constant $s>0$.
\end{ass}

\begin{ass}\label{Ass: Eqn: Ex2: 2}
There are constants $0<d_{\min}\leq d_{\max}<\infty$ such that: (i) $\lambda_{\min}(\Phi'\Phi+\Lambda^{\prime}\Lambda/N)>d_{\min}$; (ii) $\lambda_{\max}(\Phi'\Phi)\hspace{-0.05cm}<\hspace{-0.05cm}d_{\max}/2$; (iii) $\lambda_{\max}(\Lambda^{\prime}\Lambda/N)\hspace{-0.05cm}<\hspace{-0.05cm}d_{\max}/2$; (iv) $\max_{t\leq T}\|f_{t}\|$ $< d_{\max}$; (v) $\lambda_{\min}(F'M_{T}F/T)> d_{\min}$; (vi) $\|\phi\|^{2}<d_{\max}/2$; (vii) $\|\mu\|^{2}/N<d_{\max}/2$; (viii) $\phi^{\prime}\Phi = 0$; (ix) $\mu^{\prime}\Lambda = 0$.
\end{ass}

Assumption \ref{Ass: Eqn: Ex2: 1} involves no multicolinearity, finite moments, weak dependence, and small sieve approximation errors, all of which are standard in the literature. Conditions similar to Assumption \ref{Ass: Eqn: Ex2: 1}(i), (iii), and (vi) have been imposed in \citet{Fanetal_ProjectedPCA_2016}. We apply Theorem \ref{Thm: NuclearNRate} to obtain the following corollary.

\begin{cor}\label{Cor: Ex2: 1}
Suppose Assumption \ref{Ass: Eqn: Ex2: 1} holds. Let $\hat{\Pi}$ be given in \eqref{Eqn: NuclearNEst} with $\mathcal{S} = \mathcal{D}_M$ and $\lambda_{NT}= [M\sqrt{(Np^2+ Tp)} + \sqrt{NT} p^{-s}]\sqrt{\log N} $. Let $\hat{\Pi}^{\diamond}$ and $\hat{\Pi}^{\ast}$ be given below \eqref{Eqn: Ex2: 1}.  Let  $\hat{K}$, $\hat{F}$, $\hat{\mu}$, $\hat{\Lambda}$, $\hat{\phi}$, and $\hat{\Phi}$ be given in \eqref{Eqn: KEst}, \eqref{Eqn: aFEst}, and \eqref{Eqn: Ex2: Estimators}. Assume that $0<K<\min\{N+p-1,T\}-1$. (i) Then as $(N,T)\to\infty$,
\begin{align*}
\frac{1}{\sqrt{NT}}\|\hat{\Pi}^{\diamond} - \Pi^{\diamond}\|_{F}&= O_{p}\left(M\sqrt{\frac{K(Np^{2}+Tp)\log N}{NT}} + \frac{\sqrt{K \log N}}{p^{s}}\right),\notag\\
\frac{1}{\sqrt{T}}\|\hat{\Pi}^{\ast}  - \Pi^{\ast}\|_{F}&= O_{p}\left(M\sqrt{\frac{K(Np^{2}+Tp)\log N}{NT}}+ \frac{\sqrt{K \log N}}{p^{s}}\right).
\end{align*}
(ii) Suppose Assumption \ref{Ass: Eqn: Ex2: 2} additionally holds. \hspace{-0.1cm}Assume that as $(N,T)\hspace{-0.05cm}\to\infty$, $\delta_{NT}/(NT)$ $\to 0$ and $\delta_{NT}/\{K[M^{2}(Np^{2}+Tp)+NT p^{-2s}]\log N \}\to \infty $. Let $H \equiv (F^{\prime}M_T\hat{F})(\hat{F}^{\prime}M_T\hat{F})^{-1}$. Then as $(N,T)\to\infty$,
\begin{align*}
P(\hat{K}= K) &\to 1,\\
\frac{1}{\sqrt{N}}\|\hat{\mu} - \mu\|&=O_{p}\left(M\sqrt{\frac{K(Np^{2}+Tp)\log N}{NT}}+ \frac{\sqrt{K \log N}}{p^{s}}\right),\notag\\
\frac{1}{\sqrt{N}}\|\hat{\Lambda} - \Lambda H\|_{F}&=O_{p}\left(M\sqrt{\frac{K(Np^{2}+Tp)\log N}{NT}}+ \frac{\sqrt{K \log N}}{p^{s}}\right),\notag\\
\|\hat{\phi} - \phi\|&=O_{p}\left(M\sqrt{\frac{K(Np^{2}+Tp)\log N}{NT}}+ \frac{\sqrt{K \log N}}{p^{s}}\right),\notag\\
\|\hat{\Phi} - \Phi H\|_{F}&=O_{p}\left(M\sqrt{\frac{K(Np^{2}+Tp)\log N}{NT}}+ \frac{\sqrt{K \log N}}{p^{s}}\right),\notag\\
\frac{1}{\sqrt{T}}\|\hat{F}-F(H^{\prime})^{-1}\|_{F}&=O_{p}\left(M\sqrt{\frac{K(Np^{2}+Tp)\log N}{NT}}+ \frac{\sqrt{K \log N}}{p^{s}}\right).
\end{align*}
\end{cor}

The slower rate in Corollary \ref{Cor: Ex2: 1} compared to Corollary \ref{Cor: Ex135: 1} is attributed to the reliance on a set of easier-to-verify conditions, namely Assumption \ref{Ass: Eqn: Ex2: 1}, rather than a version of Assumption \ref{Ass: Ex135: 1}. However, it is noteworthy that the rate can be improved to $O_{p}(\sqrt{K(N+p+T)\log N/(NT)} + \sqrt{K \log N}/p^{s})$ under Assumption \ref{Ass: Ex135: 1}.  The second term $\sqrt{K \log N}/p^{s}$ arises from sieve approximation errors. Our results differ from \citet{Fanetal_ProjectedPCA_2016} in several aspects. First, we allow for $\mu_i\neq 0 $ and $\phi\neq 0$, which are crucial to capture pricing errors in asset pricing. Second, we permit $x_{it}$ to vary over $t$, a critical feature in asset pricing as many stock characteristics (e.g., book to market ratio and momentum) change from month to month. Our simulations in Appendix \ref{App: G: sub4} show that \citet{Fanetal_ProjectedPCA_2016}'s projected-PCA fails in the presence of time-varying $x_{it}$. Third, we do not require that $\lambda_i$ has zero mean and weak cross-sectional dependence (in such cases $\lambda_i$ can be interpreted as a vector of noises), which is barely justified in practice. We allow for non-noisy intercepts $\mu_i$ and $\lambda_i$ in pricing errors and risk exposures. Fourth, we allow $K\to\infty$. In addition, our results extend  \citet{Chenetal_SeimiparametricFactor_2021} by allowing for the heterogeneity of $\mu_i$ and $\lambda_i$ across $i$.

\subsection{Example \ref{Ex: HomoConditional}}\label{Sec53}
Our objective is to estimate $\phi_0$, $\Phi_0$, $F$, and $K$. Since $a = 1_{N}\otimes \phi_0$ and $B = 1_{N}\otimes \Phi_0$, we have $\Pi =a 1^{\prime}_{T} + BF^{\prime}= 1_{N} \otimes \Pi_0$, where $\Pi_0\equiv \phi_0 1^{\prime}_{T} + \Phi_0 F^{\prime}$, which is a $p\times T$ matrix. Then we set $\mathcal{S} = \{1_{N}\otimes \Gamma: \Gamma\in \mathbf{R}^{p\times T}\}$ in \eqref{Eqn: NuclearNEst}. Since $\hat{\Pi}\in\mathcal{S}$, we can write $\hat\Pi = 1_{N} \otimes \hat{\Pi}_0$, where $\hat{\Pi}_0$ is an estimator of ${\Pi}_0$. An efficient algorithm for finding $\hat{\Pi}_0$ is provided in Appendix \ref{App: E}.

By Lemma \ref{Lem: TechD1}(iv) and simple algebra, we can write
\begin{align}\label{Eqn: Ex4: Estimators}
\hat{a} = 1_{N} \otimes \hat{\phi}_0 \text{ and } \hat{B} = 1_{N} \otimes\hat{\Phi}_0,
\end{align}
where $\hat{\phi}_0$ is a $p\times 1$ vector and $\hat{\Phi}_0$ is a $p\times \hat{K}$ matrix. For the same reason as in Example \ref{Ex: SemiTin}, there is no need to enforce the homogeneity restriction of $a$ and $B$ in extracting $\hat{a}$ and $\hat{B}$ from $\hat{\Pi}$. We define the estimators of $\phi_0$ and $\Phi_0$ as $\hat{\phi}_0$ and $\hat{\Phi}_0$, respectively.

A convergence rate for $\hat{\Pi}_0$, $\hat{\phi}_0$, and $\hat{\Phi}_0$ follows immediately from Theorem \ref{Thm: NuclearNRate}, as we have $\|\hat{\Pi}  - \Pi\|_{F} = \sqrt{N}\|\hat{\Pi}_0  - \Pi_0\|_{F}$, $\|\hat{a} - a\| = \sqrt{N}\|\hat{\phi}_0 - \phi_0\| $, and $\|\hat{B} - BH\|_{F} = \sqrt{N}\|\hat{\Phi}_0 - \Phi_0H\|_{F}$. To provide primitive conditions, we impose the following assumptions.

\begin{ass}\label{Ass: Eqn: Ex4: 1}
(i) There are positive constants $c_{\min}$ and $c_{\max}$ such that: with probability approaching one as $N\to\infty$ with fixed $T$ or as $(N,T)\to\infty$,
\[c_{\min} \leq \min_{t\leq T}\lambda_{\min}\left(\frac{1}{N}\sum_{i=1}^{N}x_{it}x_{it}^{\prime}\right)\leq \max_{t\leq T}\lambda_{\max}\left(\frac{1}{N}\sum_{i=1}^{N}x_{it}x_{it}^{\prime}\right)\leq c_{\max} .\]
(ii) $E[\|\sum_{i=1}^{N}x_{it}e_{it}/\sqrt{Np}\|^2]$ is bounded for each $t\leq T$. (iii) $\{\sum_{i=1}^{N}x_{it}e_{it}/\sqrt{N}\}_{t\leq T}$ is a sequence of independent sub-Gaussian vectors. (iv) $\sup_{z}|\delta(z)| = O(p^{-s})$ and $\sup_{z}\|\Delta(z)\| = O(p^{-s})$ for some constant $s>0$.
\end{ass}

\begin{ass}\label{Ass: Eqn: Ex4: 2}
There are constants $0<d_{\min}\leq d_{\max}<\infty$ such that: (i) $d_{\min}<\lambda_{\min}(\Phi_0^{\prime}\Phi_0)\leq \lambda_{\max}(\Phi_0^{\prime}\Phi_0)<d_{\max}$; (ii) $\max_{t\leq T}\|f_{t}\|< d_{\max}$; (iii) $\lambda_{\min}(F'M_{T}F/T)> d_{\min}$; (iv) $\|\phi_0\|^2<d_{\max}$; (v) $\phi_0^{\prime}\Phi_0 = 0$.
\end{ass}

Assumption \ref{Ass: Eqn: Ex4: 1} involves no multicolinearity, finite moments, weak dependence, and small sieve approximation errors, all of which are standard in the literature. Assumptions \ref{Ass: Eqn: Ex4: 1}(i), (ii), (iv), and \ref{Ass: Eqn: Ex4: 2} have been imposed in \citet{Chenetal_SeimiparametricFactor_2021}. We apply Theorem \ref{Thm: NuclearNRate} to obtain the following corollary.

\begin{cor}\label{Cor: Ex4: 1}
Suppose Assumptions \ref{Ass: Eqn: Ex4: 1}(i), (ii), and (iv) hold. Let $\hat\Pi$ be given in \eqref{Eqn: NuclearNEst} with $\mathcal{S} = \{1_{N}\otimes \Gamma: \Gamma\in \mathbf{R}^{p\times T}\}$ and $\lambda_{NT}=(\sqrt{p+T} +\sqrt{NT}p^{-s})\sqrt{\log N}$. Let $\hat{\Pi}_0 $ be given above \eqref{Eqn: Ex4: Estimators}. Let $\hat{K}$, $\hat{F}$, $\hat{\phi}_0$, and $\hat{\Phi}_0$ be given in \eqref{Eqn: KEst}, \eqref{Eqn: aFEst}, and \eqref{Eqn: Ex4: Estimators}. Assume $0<K<\min\{p,T\}-1$. (i) Then as $N\to\infty$ with fixed $T$,
\begin{align*}
\frac{1}{\sqrt{T}}\|\hat{\Pi}_0  - \Pi_0\|_{F}= O_{p}\left(\sqrt{\frac{K(p+T)\log N}{NT}} + \frac{\sqrt{K \log N}}{p^{s}}\right).
\end{align*}
(ii) Suppose Assumption \ref{Ass: Eqn: Ex4: 2} additionally holds. Assume that as $N\to\infty$ with fixed $T$, $\delta_{NT}/(NT)\hspace{-0.05cm}\to \hspace{-0.05cm}0$ and $\delta_{NT}/[K(p+T+ NTp^{-2s})\log N]\hspace{-0.05cm}\to \hspace{-0.05cm}\infty$. \hspace{-0.1cm}Let $H \hspace{-0.1cm}\equiv \hspace{-0.1cm}(F^{\prime}M_T\hat{F})(\hat{F}^{\prime}M_T\hat{F})^{-1}$. Then as $N\to\infty$ with fixed $T$,
\begin{align*}
P(\hat{K}= K) &\to 1,\\
\|\hat{\phi}_0 - \phi_0\|&=O_{p}\left(\sqrt{\frac{K(p+T)\log N}{NT}} + \frac{\sqrt{K \log N}}{p^{s}}\right),\notag\\
\|\hat{\Phi}_0 - \Phi_0 H\|_{F}&=O_{p}\left(\sqrt{\frac{K(p+T)\log N}{NT}} + \frac{\sqrt{K \log N}}{p^{s}}\right),\notag\\
\frac{1}{\sqrt{T}}\|\hat{F}-F(H^{\prime})^{-1}\|_{F}&=O_{p}\left(\sqrt{\frac{K(p+T)\log N}{NT}} + \frac{\sqrt{K \log N}}{p^{s}}\right).
\end{align*}
(iii) If Assumption \ref{Ass: Eqn: Ex4: 1}(ii) is replaced with Assumption \ref{Ass: Eqn: Ex4: 1}(iii), then (i) and (ii) continue to hold by replacing ``as $N\to\infty$ with fixed $T$'' with ``as $(N,T)\to\infty$'' in all places.
\end{cor}

Corollary \ref{Cor: Ex4: 1} establishes a convergence rate of $\hat{\Pi}_0$, $\hat{K}$, $\hat{\phi}_0, \hat{\Phi}_0$, and $\hat{F}$ either under large $N$ with fixed $T$ or scenarios with both large $N$ and large $T$. In particular, $K(p+T)\log N = o(NT)$ is required for the consistency. This implies that $p$ can be as large as $N$ up to $\log N$. Such a result represents a significant improvement from similar results in \citet{Chenetal_SeimiparametricFactor_2021}, which require that $p$ grows at a rate slower than $N^{1/3}$. Our simulations in Appendix \ref{App: G: sub4} show that \citet{Chenetal_SeimiparametricFactor_2021}'s regressed-PCA exhibits poor performance when $p$ is close to $N$. The rate $\sqrt{K \log N}/p^{s}$ arises from sieve approximation errors. In addition, our framework accommodates the scenario where $K$ tends to infinity and allows for weak cross-sectional dependence of $x_{it}$.


\section{Simulation Studies}\label{Sec6}
In this section, we conduct Monte Carlo simulations to investigate the finite sample performance of our estimators. We consider settings with $p=37$, $N = 500, 1000, 2000$, and $T = 250, 500$, which are comparable with those in the empirical analysis in Section \ref{Sec7}.

We consider three different data generating processes (DGPs), which correspond to the settings in Examples \ref{Ex: SemiTin}, \ref{Ex: HomoConditional},  and \ref{Ex: ConitionalObs}. In all three DGPs, we let
\begin{align}\label{Eqn: Covariates}
x_{it,1}=1, x_{it,2}=\sigma_{t}u_{it,1}, x_{it,3} = 0.3x_{i(t-1),3} + u_{it,2}, x_{it,4}=u_{it,3}, \ldots, x_{it,37}=u_{it,36},
\end{align}
where $u_{it} = (u_{it,1},u_{it,2},\ldots, u_{it,36})^{\prime}$ are i.i.d. $N(0,I_{36})$ across both $i$ and $t$, $\sigma_{t}$'s are i.i.d. $U(1,2)$ over $t$, and $x_{i0,3}$'s are i.i.d. $N(0,1)$ across $i$. Let $x_{it} = (x_{it,1},x_{it,2},\ldots,x_{it,37})^{\prime}$, hence $p=37$. Let $f_{t} = 0.3 f_{t-1} + \eta_t$, where $\eta_t$'s are i.i.d. $N(1_2,I_2)$ over $t$ and $f_0\sim N(1_2/0.7,I_2/0.91)$, resulting in $K=2$. The errors $\varepsilon_{it}$'s be i.i.d. $N(0,4)$ across both $i$ and $t$. In the first DGP (DGP1), we assume
\begin{align}\label{Eqn: alphabeta1}
a_i &= \left(
         \begin{array}{ccccccccccc}
             0 & \theta_{1i} & 0 & 0 & \theta_{2i} & 0 & 0 &\cdots & \theta_{12i} & 0 & 0 \\
         \end{array}
       \right)^{\prime} \text{ and }
\notag\\
B_i &=\left(
       \begin{array}{ccccccccccc}
         0 & 0 & \varrho_{1i} & 0 & 0 & \varrho_{2i} & 0 & \cdots & 0 & \varrho_{12i} & 0\\
         \varrho_{13i} & 0 & 0 &  \varrho_{14i} & 0 & 0 & \varrho_{15i} & \cdots & 0 & 0 & \varrho_{25i}\\
       \end{array}
     \right)^{\prime},
\end{align}
where $\theta_{ji}$'s are i.i.d. $N(0,1/4)$ across both $i$ and $j=1,2,\ldots, 12$ and $\varrho_{ji}$'s are i.i.d. $U(1/3,1)$ across both $i$ and $j=1,2,\ldots, 25$. In DGP1, $a_i$ and $B_i$ are heterogenous across $i$, which is the setting in Example \ref{Ex: ConitionalObs}. We are interested in $a$, $B$, $F$, and $K$. In the second DGP (DGP2), we assume
\begin{align}\label{Eqn: alphabeta2}
a_i &= \left(
         \begin{array}{cc}
            \mu_i & \phi^{\prime} \\
         \end{array}
       \right)^{\prime} = \left(
         \begin{array}{ccccccccccc}
            0 & 1/2 & 0 & 0 & 1/2 & 0 & 0 & \cdots & 1/2 & 0 & 0 \\
         \end{array}
       \right)^{\prime} \text{ and }
\notag\\
B_i &=\left(
         \begin{array}{cc}
            \lambda_i & \Phi^{\prime} \\
         \end{array}
       \right)^{\prime} = \left(
       \begin{array}{ccccccccccc}
         0 & 0 & 2/3 & 0 & 0 & 2/3 & 0 & \cdots & 0 & 2/3 & 0\\
         \vartheta_{i} &  0 & 0 & 2/3 & 0 & 0 & 2/3 & \cdots &  0 & 0 & 2/3 \\
       \end{array}
     \right)^{\prime},
\end{align}
where $\vartheta_i$'s are i.i.d. $U(1,3)$ across $i$. In DGP2, the rows of $a_i$ and $B_i$ corresponding to the nonconstant part of $x_{it}$ are homogenous across $i$, which is the setting in Example \ref{Ex: SemiTin}.  We are interested in $\mu$, $\phi$, $\Lambda$, $\Phi$, $F$, and $K$.
In the third DGP (DGP3), we assume
\begin{align}\label{Eqn: alphabeta3}
a_i &=\phi_0= \left(
         \begin{array}{ccccccccccc}
          0 & 1/2 & 0 & 0 & 1/2 & 0 & 0 & \cdots & 1/2 & 0& 0 \\
         \end{array}
       \right)^{\prime} \text{ and }
\notag\\
B_i &=\Phi_0 = \left(
       \begin{array}{ccccccccccc}
         0 & 0 & 2/3 & 0 & 0 & 2/3 & 0 & \cdots & 0 & 2/3 & 0\\
         2/3 & 0 & 0 &  2/3 & 0 & 0 & 2/3 & \cdots & 0 & 0 & 2/3\\
       \end{array}
     \right)^{\prime}.
\end{align}
In DGP3, $a_i$ and $B_i$ are homogenous across $i$, which is the setting in Example \ref{Ex: HomoConditional}. We are interested in $\phi_0$, $\Phi_0$, $F$, and $K$. Here, $u_{it}$'s, $\sigma_{t}$'s, $x_{i0,3}$'s, $\eta_t$'s, $f_0$, $\theta_{ji}$'s, $\varrho_{ji}$'s, $\vartheta_{i}$'s,  and $\varepsilon_{it}$'s are mutually independent. We generate $y_{it}$ according to the model in \eqref{Eqn: Model}.

For DGP1, we implement the estimators in \eqref{Eqn: NuclearNEst}-\eqref{Eqn: aFEst} with $\mathcal{S} = \mathbf{R}^{Np\times T}$. We assess the performance of $\hat{\Pi}$, $\hat{a}, \hat{B}$, $\hat{F}$, and $\hat{K}$. By Corollary \ref{Cor: Ex135: 1}, we set $\lambda_{NT} = c\sqrt{(Np+T)\log N}$ and  $\delta_{NT} = 2(Np+T)\log N$ for some $c>0$. For DGP2, we implement the estimators in \eqref{Eqn: NuclearNEst}-\eqref{Eqn: aFEst} and \eqref{Eqn: Ex2: Estimators} as well as below \eqref{Eqn: Ex2: 1} with $\mathcal{S} = \mathcal{D}_\infty$. We evaluate the performance of $\hat{\Pi}^{\diamond}$, $\hat{\Pi}^{\ast}$, $\hat{\mu}$, $\hat\Lambda$, $\hat{\phi}, \hat{\Phi}$, $\hat{F}$, and $\hat{K}$. By the discussion after Corollary \ref{Cor: Ex2: 1}, we set $\lambda_{NT} = c\sqrt{(N+p+T)\log N}$ and  $\delta_{NT} = 2(N+p+T)\log N$ for some $c>0$. For DGP3, we implement the estimators in \eqref{Eqn: NuclearNEst}-\eqref{Eqn: aFEst} and \eqref{Eqn: Ex4: Estimators} with $\mathcal{S} = \{1_{N}\otimes \Gamma: \Gamma\in \mathbf{R}^{p\times T}\}$. We evaluate the performance of $\hat{\Pi}_0$, $\hat{\phi}_0, \hat{\Phi}_0$, $\hat{F}$, and $\hat{K}$. By Corollary \ref{Cor: Ex4: 1}, we set $\lambda_{NT} = c\sqrt{(p+T)\log N}$ and $\delta_{NT} = 2\sqrt{N}(p+T)\log(N)$ for some $c>0$.

To determine the optimal value of $c$, we employ the $5$-fold CV approach, as outlined in Section \ref{Sec3} with $L=5$. The mean square errors of the regularized estimators ($\hat{\Pi}$, $(\hat{\Pi}^{\diamond\prime}, \sqrt{N}\hat{\Pi}^{\ast\prime})$, and $\hat{\Pi}_0$) are assessed both with fixed values of $c$ and using the CV method, with $c$ confined to $[0,2]$.\footnote{Specifically, we consider the grid set $\{0, 0.05, 0.1, 0.2, \ldots, 0.9,1,1.5,2\}$.} All simulation results are based on $200$ simulation replications. The main findings are as follows.
\begin{itemize}
  \item Nuclear norm regularization significantly enhances the performance of the estimators. In DGP1 and DPG2 (see Figures \ref{Fig: DGP1} and \ref{Fig: DGP2}), the mean square error of the unregularized estimator (i.e., $c=0$) remains relatively constant as both $N$ and $T$ increase (the value stays constantly around 40 in DGP1 and above 10 in DGP2), indicating potential inconsistency. Conversely, applying appropriate nuclear norm regularization (e.g., $c=1$) not only reduces the mean square error for each combination of $(N,T)$ but also drive the error towards zero as both $N$ and $T$ increase (e.g., the value for $c=1$ is getting closer to zero as $N$ and $T$ increase). This suggests that regularized estimators with a well-chosen $c$ value are consistent, aligning with Corollaries \ref{Cor: Ex135: 1} and \ref{Cor: Ex2: 1}. In DGP3 (see Figure \ref{Fig: DGP3}), although the mean square error of the unregularized estimator decreases with increasing $N$ (note that the scale of the vertical axis changes across columns of graphs), a properly chosen $c$ value (e.g., $c=0.6$ or $0.7$) leads to smaller errors. Thus, the simulations underscore the crucial role of nuclear norm regularization.
  \item The regularized estimators exhibit high sensitivity to the choice of $c$. For instance, selecting $c>2$ in DGP3 can result in a larger mean square error than that of the unregularized estimator across all $(N,T)$ combinations (as seen in Figure \ref{Fig: DGP3}). Therefore, careful consideration is essential when choosing $c$ in practice.
  \item The CV approach proves effective in selecting $c$ to minimize mean square error. Across all the three DGPs, the mean square error of the regularized estimator using the CV-selected $c$ value closely approximates the smallest error obtained with fixed $c$ values (as evidenced by the blue line closely tracking the lowest point of the dash-dotted line in Figures \ref{Fig: DGP1}-\ref{Fig: DGP3}), irrespective of $(N,T)$ combinations.
\end{itemize}
Overall, these findings highlight the importance of nuclear norm regularization, the sensitivity of estimators to $c$, and the efficacy of the CV approach in selecting an optimal $c$ value for minimizing mean square error.

We proceed to assess the performance of estimators other than $\hat{\Pi}$, $\hat{\Pi}^{\diamond}$, $\hat{\Pi}^{\ast}$, and $\hat{\Pi}_0$ by utilizing the CV-selected value of $c$. Tables \ref{Tab: MSEDGP1}-\ref{Tab: MSEDGP3} present their mean square errors or correct rates. The main findings are summarized as follows.
First, the number factor estimators (i.e., $\hat{K}$) consistently perform well across all cases, only one correct rate falling below $100\%$. This indicates their reliability in estimating the number of factors. Second, in DPG1 and DPG2, all mean square errors decrease as both $N$ and $T$ increase, consistent with Corollaries \ref{Cor: Ex135: 1} and \ref{Cor: Ex2: 1}. Similarly, in DGP3, mean square errors decrease as $N$ increases, indicating consistency as $N\to\infty$, aligning with Corollary \ref{Cor: Ex4: 1}. Third, increasing $N$ consistently reduces the mean square errors of the factor estimators (i.e., $\hat{F}$) across all cases, while increasing $T$ may not have a similar effect. In addition, increasing either $N$ or $T$ tends to reduce the mean square errors of $\hat{\Pi}^{\ast}$, $\hat{\phi}$, $\hat{\Phi}$, $\hat{\phi}_0$, and $\hat{\Phi}_0$. While this phenomenon is not explicitly explained by Corollaries \ref{Cor: Ex2: 1} and \ref{Cor: Ex4: 1}, it may be attributed to the homogeneity of the estimands (${\Pi}^{\ast}$, ${\phi}$, ${\Phi}$, ${\phi}_0$, and ${\Phi}_0$). In conclusion, our estimators exhibit promising performance in finite sample settings. The same findings are observed in settings with sparse $a$ and $B$, as well as in scenarios with small $p$, $N$, and $T$; see Appendix \ref{App: G} for additional simulation results. Furthermore, Appendix \ref{App: G} demonstrates the superiority of our estimators compared to existing ones.

\section{Empirical Analysis}\label{Sec7}
In this section, we analyze the cross section of individual stock returns in the US market using the same dataset as in \citet{Chenetal_SeimiparametricFactor_2021}, originally derived from \citet{Freybergeretal_Dissecting_2017}. The dataset comprises monthly returns and 36 characteristics of 12,813 individual US stocks spanning from September 1968 to May 2014. Due to a significant proportion of missing values in many stocks, we opt to exclude stocks with a sample length less than 200 to ensure that the proportion of missing values remains manageable. This results in an unbalanced panel with $N = 2,121$ and $T = 549$. Each time period includes at least 580 stocks with observations on both returns and the 36 characteristics, while each stock has observations in at least 200 time periods. Additionally, we transform the values of each characteristic to relative ranking values within the range $[-0.5, 0.5]$ in each time period.

In our analysis, we consider six different model specifications. The first three specifications, denoted S1, S2, and S3, include $x_{it}$ comprising a constant and the 36 characteristics. The remaining three specifications, denoted S4, S5, and S6, involve $x_{it}$ consisting of a constant and linear B-splines of 18 characteristics with one internal knot, as studied in \citet{Chenetal_SeimiparametricFactor_2021}. Refer to their paper for the 18 characteristics. In S1 and S4, we explore an unconstrained conditional factor model (corresponding to the setup in Example \ref{Ex: ConitionalObs}), where $a_{i}$ and $B_i$ can vary heterogeneously across $i$. For S2 and S5, we investigate a semiparametric conditional factor model (corresponding to the setup in Example \ref{Ex: SemiTin}), where the rows of $a_i$ and $B_i$ corresponding to the nonconstant explanatory variables in $x_{it}$ are constrained to be homogeneous. Lastly, S3 and S6 examine a homogeneous conditional factor model (corresponding to the setup in Example \ref{Ex: HomoConditional}), where $a_i$ and $B_i$ are constrained to be homogeneous. We estimate the models for $K=1,2,\ldots, 10$ by using our new method and select the regularization parameter using the 5-fold CV approach as outlined in Section \ref{Sec3}. Specifically, we set $\lambda_{NT} = c\sqrt{(Np+T)\log N}$ for S1 and S4, $\lambda_{NT} = c\sqrt{(N+p+T)\log N}$ for S2 and S5, $\lambda_{NT} = c\sqrt{(p+T)\log N}$ for S3 and S6, and choose $c$ from the set $\{0, 0.01, 0.02, 0.05, 0.1, 0.2, 0.5,1,2,5\}/100$. For a comparison, we also evaluate \citet{Chenetal_SeimiparametricFactor_2021}'s regressed-PCA method, denoted R1 and R2, alongside the homogeneous conditional factor models (S3 and S6).

To assess the performance of the models, we adopt various goodness-of-fit measures. First, we consider different types of in-sample $R^2$ measures:
\begin{align}
R^2 & = 1-\frac{\sum_{i, t}(y_{it}- x_{it}^{\prime}\hat{a}_{i} - x_{it}^{\prime}\hat{B}_{i}\hat{f}_t)^2}{\sum_{i, t} y_{it}^{2}}, \label{Eqn: R21}\\
R^2_{T,N} & = 1 - \frac{1}{N} \sum_{i} \frac{\sum_{t}(y_{it}- x_{it}^{\prime}\hat{a}_{i} - x_{it}^{\prime}\hat{B}_{i}\hat{f}_t)^2}{\sum_{t}y_{it}^{2}},\label{Eqn: R22}\\
R^2_{N,T} & = 1 - \frac{1}{T} \sum_{t} \frac{\sum_{i}(y_{it}- x_{it}^{\prime}\hat{a}_{i} - x_{it}^{\prime}\hat{B}_{i}\hat{f}_t)^2}{\sum_{i}y_{it}^{2}},\label{Eqn: R23}
\end{align}
where $\hat{a}\equiv (\hat{a}_1^{\prime},\hat{a}_2^{\prime},\ldots, \hat{a}_{N}^{\prime})^{\prime}$, $\hat{B} \equiv (\hat{B}_1^{\prime},\hat{B}_2^{\prime},\ldots, \hat{B}_{N}^{\prime})^{\prime}$, and $\hat{F} \equiv(\hat{f}_1,\hat{f}_2,\ldots, \hat{f}_{T})^{\prime}$. Here, the first one is total $R^2$, measuring the overall explanatory power of the models. The second one measures the cross-sectional average of time series $R^2$ across all stocks, reflecting the ability of the models to capture common variation in asset returns. The third one measures the time series average of cross-sectional $R^2$, which is the one of interest for evaluating the models' ability to explain the cross-section of average returns.
Second, we assess out-of-sample prediction. For $t\geq 300$, we utilize the data up to $t-1$ for estimation and obtain estimators, say $\hat{a}_{it}$, $\hat{B}_{it}$, and $\hat{F}_{t}\equiv (\hat{f}^{(t)}_{1},\hat{f}^{(t)}_{2},\ldots, \hat{f}^{(t)}_{t-1})^{\prime}$. The out-of-sample prediction of $y_{it}$ is then computed as $x_{it}^{\prime}\hat{a}_{it} - x_{it}^{\prime}\hat{B}_{it}\hat{\lambda}_t$, where $\hat{\lambda}_{t} = \sum_{s\leq t-1}\hat{f}^{(t)}_s/(t-1)$. Analogously, we define  three types of out-of-sample predictive $R^2$'s by replacing $\hat{a}_i$, $\hat{B}_i$ and $\hat{f}_t$ with $\hat{a}_{it}$, $\hat{B}_{it}$ and $\hat{\lambda}_t$:
\begin{align}
R^{2}_{O} & = 1-\frac{\sum_{i, t\geq 300}(y_{it}- x_{it}^{\prime}\hat{a}_{it} - x_{it}^{\prime}\hat{B}_{it}\hat{\lambda}_t)^2}{\sum_{i, t\geq 300} y_{it}^{2}}, \label{Eqn: R21predictive}\\
R^2_{T,N,O} & = 1 - \frac{1}{N} \sum_{i} \frac{\sum_{t\geq 300}(y_{it}- x_{it}^{\prime}\hat{a}_{it} - x_{it}^{\prime}\hat{B}_{it}\hat{\lambda}_t)^2}{\sum_{t\geq 300}y_{it}^{2}},\label{Eqn: R22predictive}\\
R^2_{N,T,O} & = 1 - \frac{1}{T-299} \sum_{t\geq 300} \frac{\sum_{i}(y_{it}- x_{it}^{\prime}\hat{a}_{it} - x_{it}^{\prime}\hat{B}_{it}\hat{\lambda}_t)^2}{\sum_{i}y_{it}^{2}}.\label{Eqn: R23predictive}
\end{align}

The results depicted in Figure \ref{Fig: Emprical1} yield several key observations. Firstly, the in-sample $R^2$ values of our methods (S1, S2, S3, S4, S5, and S6) exhibit an increasing trend as the number of factors $K$ rises, while the out-of-sample $R^2$ metrics remain unaffected by changes in $K$. This constancy arises from the fact that $\hat{\lambda} = \sum_{t\leq T}\hat{f}_t/T = \hat{F}^{\prime}1_{T}/T = \hat{B}^{\prime}\hat{\Pi}1_{T}/(NT)$, $\hat{a} + \hat{B}\hat{\lambda} = \hat{\Pi}1_{T}/T$, rendering the out-of-sample predictions of $y_{it}$ independent of $K$. Secondly, among the linear models (S1, S2, and S3), S1 consistently outperforms others in terms of in-sample $R^2$ values across all tested values of $K$. Conversely, S3 emerges as the top performer in out-of-sample $R^2$ metrics for all configurations. This suggests that enforcing  homogeneity of $a_i$ and $B_i$ across $i$ may improve the model's out-of-sample predictability, despite potentially compromising the in-sample fit. Similarly, for the spline models (S4, S5, and S6), enforcing homgogeneity of $a_i$ and $B_i$ across $i$ yields improvements in out-of-sample predictability. Thirdly, S5 and S6 demonstrate superior out-of-sample performance compared to S2 and S3, respectively. This underscores the potential benefits of incorporating spline transformations of characteristics, emphasizing the significance of capturing nonlinear relationships. Lastly, the importance of nonlinearity is also observed for the regressed-PCA method; R2 has larger out-of-sample $R^2$ values than R1. However, S3 and S6 exhibit better both in-sample and out-of-sample performance than R1 and R2, respectively. This implies that our method outperforms the regressed-PCA method. In conclusion, while S1 exhibits the most favorable in-sample performance, S6 stands out for its superior out-of-sample predictive capabilities.

\section{Concluding Remarks}\label{Sec8}
In this paper, we introduced a nuclear norm regularized estimation approach for high-dimensional conditional factor models and established large sample properties of the estimators. Our method provides a unified framework for estimating various conditional factor models, facilitating the derivation of new asymptotic results while addressing the limitations of existing methods, which are often model-specific or restrictive. We applied this method to analyze the cross section of individual US stock returns, uncovering potential improvements in out-of-sample performance by enforcing homogeneity of $a_i$ and $B_i$ across $i$. Our results also show that the proposed method outperforms existing alternatives.

In asset pricing, addressing key inference problems\textemdash such as testing for zero pricing errors and conducting specification tests for risk exposure functions\textemdash is crucial for evaluating and comparing factor models. Previous studies, including \citet{XiaYuan_InfMatrix_2019}, \citet{Chenetal_Uncertainty_2019}, and  \citet{Chernozhukovetal_InferenceLowRank_2021}, have investigated debiasing techniques in trace linear regression models with $p=1$ and $a_i=0$, with applications to matrix completion, PCA with missing data, and heterogeneous treatment effects. However, these methods are not applicable to our framework, which accommodates large $p$ and $a_i\neq 0$, as is often the case in asset pricing. Developing a general inferential method within this framework presents an intriguing avenue for future research.



\spacingset{1.75}

\addcontentsline{toc}{section}{References}
\putbib
\end{bibunit}

\pgfplotstableread{
c	d1	d2	d3	d4	d5	d6	d10	d20	d30	d40	d50	d60
0	43.14426637	43.33627923	43.50645384	41.59508512	41.76457064	41.97837082	4.169800373	3.996109432	3.850109822	1.821119546	1.686205982	1.594316174
0.05	32.08304342	32.03894122	32.06602237	29.30623291	28.95795319	29.00921376	4.169800373	3.996109432	3.850109822	1.821119546	1.686205982	1.594316174
0.1	23.54430716	23.36835109	23.36198416	20.10658389	19.39373486	19.32940007	4.169800373	3.996109432	3.850109822	1.821119546	1.686205982	1.594316174
0.2	12.23444796	11.98405134	11.9600371	8.540870773	7.7832915	7.672400098	4.169800373	3.996109432	3.850109822	1.821119546	1.686205982	1.594316174
0.3	6.595638549	6.415623745	6.394547331	3.49027754	3.068663151	2.992567961	4.169800373	3.996109432	3.850109822	1.821119546	1.686205982	1.594316174
0.4	4.395456642	4.232468312	4.179969769	1.928527141	1.73600281	1.659238317	4.169800373	3.996109432	3.850109822	1.821119546	1.686205982	1.594316174
0.5	4.059342044	3.845888286	3.73065423	1.80548605	1.680349044	1.564735503	4.169800373	3.996109432	3.850109822	1.821119546	1.686205982	1.594316174
0.6	4.639974264	4.49352774	4.390555511	2.427700319	2.35902577	2.207595137	4.169800373	3.996109432	3.850109822	1.821119546	1.686205982	1.594316174
0.7	5.390295758	5.250412363	5.115788629	2.804322671	2.608810821	2.519225758	4.169800373	3.996109432	3.850109822	1.821119546	1.686205982	1.594316174
0.8	5.778735496	5.546078172	5.365294694	2.653498884	2.361544595	2.238620608	4.169800373	3.996109432	3.850109822	1.821119546	1.686205982	1.594316174
0.9	5.729392572	5.442801024	5.289179373	2.366988798	2.176505816	2.187699483	4.169800373	3.996109432	3.850109822	1.821119546	1.686205982	1.594316174
1	5.565479792	5.356919464	5.407878306	2.306518885	2.395930596	2.601045541	4.169800373	3.996109432	3.850109822	1.821119546	1.686205982	1.594316174
1.5	7.71740683	8.151418736	8.582979736	3.929517516	4.16301821	4.405527265	4.169800373	3.996109432	3.850109822	1.821119546	1.686205982	1.594316174
2	10.73069694	11.37161622	12.00254787	5.752482952	6.111023504	6.498363794	4.169800373	3.996109432	3.850109822	1.821119546	1.686205982	1.594316174
}\dgpa

\begin{figure}[htbp]
\centering
\begin{subfigure}[b]{0.32\textwidth}
\centering
\resizebox{\linewidth}{3cm}{
\begin{tikzpicture}
\begin{axis}[
    legend style={draw=none},
    grid = minor,
    xmax=2,xmin=0,
    ymax=40,ymin=0,
    xtick={0,1,2},
    ytick={20,40},
    title={$N=500,T=250$},
    tick label style={/pgf/number format/fixed},
legend style={at={(0.2,0.9)},anchor=north,
    row sep = 3pt}]
\addplot[smooth,tension=0.5,color=black, line width=0.75pt,dashdotted] table[x = c,y=d1] from \dgpa;
\addplot[smooth,tension=0.5,no markers, color=blue, line width=0.75pt] table[x = c,y=d10] from \dgpa;
\legend{\footnotesize Fixed $c$, \footnotesize CV}
\end{axis}
\end{tikzpicture}}
\end{subfigure}
\begin{subfigure}[b]{0.32\textwidth}
\centering
\resizebox{\linewidth}{3cm}{
\begin{tikzpicture}
\begin{axis}[
    legend style={draw=none},
    grid = minor,
    xmax=2,xmin=0,
    ymax=40,ymin=0,
    xtick={0,1,2},
    ytick={20,40},
    title={$N=1000,T=250$},
    tick label style={/pgf/number format/fixed},
legend style={at={(0.2,0.9)},anchor=north,
    row sep = 3pt}]
\addplot[smooth,tension=0.5,color=black, line width=0.75pt,dashdotted] table[x = c,y=d2] from \dgpa;
\addplot[smooth,tension=0.5,no markers, color=blue, line width=0.75pt] table[x = c,y=d20] from \dgpa;
\legend{\footnotesize Fixed $c$, \footnotesize CV}
\end{axis}
\end{tikzpicture}}
\end{subfigure}
\begin{subfigure}[b]{0.32\textwidth}
\centering
\resizebox{\linewidth}{3cm}{
\begin{tikzpicture}
\begin{axis}[
    legend style={draw=none},
    grid = minor,
    xmax=2,xmin=0,
    ymax=40,ymin=0,
    xtick={0,1,2},
    ytick={20,40},
    title={$N=2000,T=250$},
    tick label style={/pgf/number format/fixed},
legend style={at={(0.2,0.9)},anchor=north,
    row sep = 3pt}]
\addplot[smooth,tension=0.5,color=black, line width=0.75pt,dashdotted] table[x = c,y=d3] from \dgpa;
\addplot[smooth,tension=0.5,no markers, color=blue, line width=0.75pt] table[x = c,y=d30] from \dgpa;
\legend{\footnotesize Fixed $c$, \footnotesize CV}
\end{axis}
\end{tikzpicture}}
\end{subfigure}

\begin{subfigure}[b]{0.32\textwidth}
\centering
\resizebox{\linewidth}{3cm}{
\begin{tikzpicture}
\begin{axis}[
    legend style={draw=none},
    grid = minor,
    xmax=2,xmin=0,
    ymax=40,ymin=0,
    xtick={0,1,2},
    ytick={20,40},
    title={$N=500,T=500$},
    tick label style={/pgf/number format/fixed},
legend style={at={(0.2,0.9)},anchor=north,
    row sep = 3pt}]
\addplot[smooth,tension=0.5,color=black, line width=0.75pt,dashdotted] table[x = c,y=d4] from \dgpa;
\addplot[smooth,tension=0.5,no markers, color=blue, line width=0.75pt] table[x = c,y=d40] from \dgpa;
\legend{\footnotesize Fixed $c$, \footnotesize CV}
\end{axis}
\end{tikzpicture}}
\end{subfigure}
\begin{subfigure}[b]{0.32\textwidth}
\centering
\resizebox{\linewidth}{3cm}{
\begin{tikzpicture}
\begin{axis}[
    legend style={draw=none},
    grid = minor,
    xmax=2,xmin=0,
    ymax=40,ymin=0,
    xtick={0,1,2},
    ytick={20,40},
    title={$N=1000,T=500$},
    tick label style={/pgf/number format/fixed},
legend style={at={(0.2,0.9)},anchor=north,
    row sep = 3pt}]
\addplot[smooth,tension=0.5,color=black, line width=0.75pt,dashdotted] table[x = c,y=d5] from \dgpa;
\addplot[smooth,tension=0.5,no markers, color=blue, line width=0.75pt] table[x = c,y=d50] from \dgpa;
\legend{\footnotesize Fixed $c$, \footnotesize CV}
\end{axis}
\end{tikzpicture}}
\end{subfigure}
\begin{subfigure}[b]{0.32\textwidth}
\centering
\resizebox{\linewidth}{3cm}{
\begin{tikzpicture}
\begin{axis}[
    legend style={draw=none},
    grid = minor,
    xmax=2,xmin=0,
    ymax=40,ymin=0,
    xtick={0,1,2},
    ytick={20,40},
    title={$N=2000,T=500$},
    tick label style={/pgf/number format/fixed},
legend style={at={(0.2,0.9)},anchor=north,
    row sep = 3pt}]
\addplot[smooth,tension=0.5,color=black, line width=0.75pt,dashdotted] table[x = c,y=d6] from \dgpa;
\addplot[smooth,tension=0.5,no markers, color=blue, line width=0.75pt] table[x = c,y=d60] from \dgpa;
\legend{\footnotesize Fixed $c$, \footnotesize CV}
\end{axis}
\end{tikzpicture}}
\end{subfigure}
\caption{Mean square errors of $\hat{\Pi}$ when using fixed $c$ and CV: DGP1}\label{Fig: DGP1}
\end{figure}

\pgfplotstableread{
c	d1	d2	d3	d4	d5	d6  d10	d20	d30	d40	d50	d60
0	24.55839437	24.16194186	23.96283721	23.83049028	23.47120882	23.28934132	0.204140592	0.150309638	0.150082636	0.133793707	0.109463196	0.07745866
0.05	3.571337437	3.578098296	3.576812316	3.502824797	3.478312025	3.482357528	0.204140592	0.150309638	0.150082636	0.133793707	0.109463196	0.07745866
0.1	2.970929389	3.000093571	3.010431768	2.942819685	2.922942773	2.935466205	0.204140592	0.150309638	0.150082636	0.133793707	0.109463196	0.07745866
0.2	2.142551935	2.128320948	2.11247134	2.097298787	2.037780035	2.022078369	0.204140592	0.150309638	0.150082636	0.133793707	0.109463196	0.07745866
0.3	1.489770499	1.434977475	1.394806768	1.446744207	1.361641069	1.310895211	0.204140592	0.150309638	0.150082636	0.133793707	0.109463196	0.07745866
0.4	1.001448585	0.915966195	0.850249751	0.962615026	0.866175217	0.789574242	0.204140592	0.150309638	0.150082636	0.133793707	0.109463196	0.07745866
0.5	0.653469242	0.556268875	0.475153265	0.61705589	0.52233315	0.439336609	0.204140592	0.150309638	0.150082636	0.133793707	0.109463196	0.07745866
0.6	0.420501601	0.328901216	0.254188009	0.38371882	0.300425231	0.227704386	0.204140592	0.150309638	0.150082636	0.133793707	0.109463196	0.07745866
0.7	0.27854637	0.20363607	0.150144267	0.23857852	0.171805881	0.118650583	0.204140592	0.150309638	0.150082636	0.133793707	0.109463196	0.07745866
0.8	0.204489441	0.15025438	0.120881442	0.159256028	0.109669903	0.077293761	0.204140592	0.150309638	0.150082636	0.133793707	0.109463196	0.07745866
0.9	0.177889305	0.141664271	0.128880179	0.125945557	0.090375619	0.073115778	0.204140592	0.150309638	0.150082636	0.133793707	0.109463196	0.07745866
1	0.181336614	0.155644406	0.147303773	0.1217567	0.094184922	0.08247404	0.204140592	0.150309638	0.150082636	0.133793707	0.109463196	0.07745866
1.5	0.315932678	0.281357944	0.270269968	0.209020799	0.169102144	0.150959432	0.204140592	0.150309638	0.150082636	0.133793707	0.109463196	0.07745866
2	0.509283795	0.457452371	0.442318719	0.337662003	0.274968678	0.247084728	0.204140592	0.150309638	0.150082636	0.133793707	0.109463196	0.07745866
}\dgpb

\begin{figure}[htbp]
\centering
\begin{subfigure}[b]{0.32\textwidth}
\centering
\resizebox{\linewidth}{3cm}{
\begin{tikzpicture}
\begin{axis}[
    grid = minor,
    xmax=2,xmin=0,
    ymax=10,ymin=0,
    xtick={0,1,2},
    ytick={5,10},
    title={$N=500,T=250$},
    tick label style={/pgf/number format/fixed},
    legend style={draw=none, at={(0.2,0.9)},anchor=north,
    row sep = 3pt}]
\addplot[smooth,tension=0.1,color=black, line width=0.75pt,dashdotted] table[x = c,y=d1] from \dgpb;
\addplot[smooth,tension=0.1,no markers, color=blue, line width=0.75pt] table[x = c,y=d10] from \dgpb;
\legend{\footnotesize Fixed $c$, \footnotesize CV}
\end{axis}
\end{tikzpicture}}
\end{subfigure}
\begin{subfigure}[b]{0.32\textwidth}
\centering
\resizebox{\linewidth}{3cm}{
\begin{tikzpicture}
\begin{axis}[
    grid = minor,
    xmax=2,xmin=0,
    ymax=10,ymin=0,
    xtick={0,1,2},
    ytick={5,10},
    title={$N=1000,T=250$},
    tick label style={/pgf/number format/fixed},
    legend style={draw=none, at={(0.2,0.9)},anchor=north,
    row sep = 3pt}]
\addplot[smooth,tension=0.1,color=black, line width=0.75pt,dashdotted] table[x = c,y=d2] from \dgpb;
\addplot[smooth,tension=0.1,no markers, color=blue, line width=0.75pt] table[x = c,y=d20] from \dgpb;
\legend{\footnotesize Fixed $c$, \footnotesize CV}
\end{axis}
\end{tikzpicture}}
\end{subfigure}
\begin{subfigure}[b]{0.32\textwidth}
\centering
\resizebox{\linewidth}{3cm}{
\begin{tikzpicture}
\begin{axis}[
    grid = minor,
    xmax=2,xmin=0,
    ymax=10,ymin=0,
    xtick={0,1,2},
    ytick={5,10},
    title={$N=2000,T=250$},
    tick label style={/pgf/number format/fixed},
    legend style={draw=none, at={(0.2,0.9)},anchor=north,
    row sep = 3pt}]
\addplot[smooth,tension=0.1,color=black, line width=0.75pt,dashdotted] table[x = c,y=d3] from \dgpb;
\addplot[smooth,tension=0.1,no markers, color=blue, line width=0.75pt] table[x = c,y=d30] from \dgpb;
\legend{\footnotesize Fixed $c$, \footnotesize CV}
\end{axis}
\end{tikzpicture}}
\end{subfigure}

\begin{subfigure}[b]{0.32\textwidth}
\centering
\resizebox{\linewidth}{3cm}{
\begin{tikzpicture}
\begin{axis}[
    grid = minor,
    xmax=2,xmin=0,
    ymax=10,ymin=0,
    xtick={0,1,2},
    ytick={5,10},
    title={$N=500,T=500$},
    tick label style={/pgf/number format/fixed},
    legend style={draw=none, at={(0.2,0.9)},anchor=north,
    row sep = 3pt}]
\addplot[smooth,tension=0.1,color=black, line width=0.75pt,dashdotted] table[x = c,y=d4] from \dgpb;
\addplot[smooth,tension=0.1,no markers, color=blue, line width=0.75pt] table[x = c,y=d40] from \dgpb;
\legend{\footnotesize Fixed $c$, \footnotesize CV}
\end{axis}
\end{tikzpicture}}
\end{subfigure}
\begin{subfigure}[b]{0.32\textwidth}
\centering
\resizebox{\linewidth}{3cm}{
\begin{tikzpicture}
\begin{axis}[
    grid = minor,
    xmax=2,xmin=0,
    ymax=10,ymin=0,
    xtick={0,1,2},
    ytick={5,10},
    title={$N=1000,T=500$},
    tick label style={/pgf/number format/fixed},
    legend style={draw=none, at={(0.2,0.9)},anchor=north,
    row sep = 3pt}]
\addplot[smooth,tension=0.1,color=black, line width=0.75pt,dashdotted] table[x = c,y=d5] from \dgpb;
\addplot[smooth,tension=0.1,no markers, color=blue, line width=0.75pt] table[x = c,y=d50] from \dgpb;
\legend{\footnotesize Fixed $c$, \footnotesize CV}
\end{axis}
\end{tikzpicture}}
\end{subfigure}
\begin{subfigure}[b]{0.32\textwidth}
\centering
\resizebox{\linewidth}{3cm}{
\begin{tikzpicture}
\begin{axis}[
    grid = minor,
    xmax=2,xmin=0,
    ymax=10,ymin=0,
    xtick={0,1,2},
    ytick={5,10},
    title={$N=2000,T=500$},
    tick label style={/pgf/number format/fixed},
    legend style={draw=none, at={(0.2,0.9)},anchor=north,
    row sep = 3pt}]
\addplot[smooth,tension=0.1,color=black, line width=0.75pt,dashdotted] table[x = c,y=d6] from \dgpb;
\addplot[smooth,tension=0.1,no markers, color=blue, line width=0.75pt] table[x = c,y=d60] from \dgpb;
\legend{\footnotesize Fixed $c$, \footnotesize CV}
\end{axis}
\end{tikzpicture}}
\end{subfigure}
\caption{Mean square errors of $(\hat{\Pi}^{\diamond\prime}, \sqrt{N}\hat{\Pi}^{\ast\prime})$ when using fixed $c$ and CV: DGP2}\label{Fig: DGP2}
\end{figure}

\pgfplotstableread{
c	d1	d2	d3	d4	d5	d6	d10	d20	d30	d40	d50	d60
0	0.315186412	0.151644977	0.074242933	0.315283062	0.151342761	0.074203126	0.053402803	0.02746442	0.01344317	0.048648191	0.025929072	0.012648708
0.05	0.274626226	0.132247697	0.06456282	0.275495043	0.132342158	0.064718108	0.053402803	0.02746442	0.01344317	0.048648191	0.025929072	0.012648708
0.1	0.238040197	0.114630754	0.055761851	0.239448271	0.115010093	0.056056517	0.053402803	0.02746442	0.01344317	0.048648191	0.025929072	0.012648708
0.2	0.176131736	0.084579088	0.040755953	0.17797218	0.085208616	0.0411667	0.053402803	0.02746442	0.01344317	0.048648191	0.025929072	0.012648708
0.3	0.128212659	0.061191003	0.029141246	0.129704323	0.061663814	0.029457328	0.053402803	0.02746442	0.01344317	0.048648191	0.025929072	0.012648708
0.4	0.093138731	0.044170528	0.020841481	0.093585177	0.044103773	0.020858001	0.053402803	0.02746442	0.01344317	0.048648191	0.025929072	0.012648708
0.5	0.069865494	0.033221388	0.0157499	0.068656402	0.032279573	0.015298531	0.053402803	0.02746442	0.01344317	0.048648191	0.025929072	0.012648708
0.6	0.056914134	0.027646015	0.013456611	0.054046877	0.02592295	0.01265732	0.053402803	0.02746442	0.01344317	0.048648191	0.025929072	0.012648708
0.7	0.05207068	0.026234379	0.013270325	0.048375435	0.024216932	0.012382856	0.053402803	0.02746442	0.01344317	0.048648191	0.025929072	0.012648708
0.8	0.053097026	0.027734333	0.014472316	0.049233867	0.025705647	0.013595737	0.053402803	0.02746442	0.01344317	0.048648191	0.025929072	0.012648708
0.9	0.058026694	0.031014026	0.01644803	0.054154238	0.028934535	0.01550308	0.053402803	0.02746442	0.01344317	0.048648191	0.025929072	0.012648708
1	0.065243985	0.035206175	0.018792668	0.061072756	0.032884238	0.017706274	0.053402803	0.02746442	0.01344317	0.048648191	0.025929072	0.012648708
1.5	0.116005859	0.063515101	0.034433462	0.108498852	0.05938283	0.032383674	0.053402803	0.02746442	0.01344317	0.048648191	0.025929072	0.012648708
2	0.187507243	0.103322792	0.056400679	0.175321907	0.096647806	0.052983099	0.053402803	0.02746442	0.01344317	0.048648191	0.025929072	0.012648708
}\dgpc

\begin{figure}[htbp]
\centering
\begin{subfigure}[b]{0.32\textwidth}
\centering
\resizebox{\linewidth}{3cm}{
\begin{tikzpicture}
\begin{axis}[
    legend style={draw=none},
    grid = minor,
    xmax=2,xmin=0,
    ymax=0.4,ymin=0,
    xtick={0,1,2},
    ytick={0.2,0.4},
    title={$N=500,T=250$},
    tick label style={/pgf/number format/fixed},
legend style={at={(0.2,0.9)},anchor=north,
    row sep = 3pt}]
\addplot[smooth,tension=0.5,color=black, line width=0.75pt,dashdotted] table[x = c,y=d1] from \dgpc;
\addplot[smooth,tension=0.5,no markers, color=blue, line width=0.75pt] table[x = c,y=d10] from \dgpc;
\legend{\footnotesize Fixed $c$, \footnotesize CV}
\end{axis}
\end{tikzpicture}}
\end{subfigure}
\begin{subfigure}[b]{0.32\textwidth}
\centering
\resizebox{\linewidth}{3cm}{
\begin{tikzpicture}
\begin{axis}[
    legend style={draw=none},
    grid = minor,
    xmax=2,xmin=0,
    ymax=0.2,ymin=0,
    xtick={0,1,2},
    ytick={0.1,0.2},
    title={$N=1000,T=250$},
    tick label style={/pgf/number format/fixed},
legend style={at={(0.2,0.9)},anchor=north,
    row sep = 3pt}]
\addplot[smooth,tension=0.5,color=black, line width=0.75pt,dashdotted] table[x = c,y=d2] from \dgpc;
\addplot[smooth,tension=0.5,no markers, color=blue, line width=0.75pt] table[x = c,y=d20] from \dgpc;
\legend{\footnotesize Fixed $c$, \footnotesize CV}
\end{axis}
\end{tikzpicture}}
\end{subfigure}
\begin{subfigure}[b]{0.32\textwidth}
\centering
\resizebox{\linewidth}{3cm}{
\begin{tikzpicture}
\begin{axis}[
    legend style={draw=none},
    grid = minor,
    xmax=2,xmin=0,
    ymax=0.1,ymin=0,
    xtick={0,1,2},
    ytick={0.05,0.1},
    title={$N=2000,T=250$},
    tick label style={/pgf/number format/fixed},
legend style={at={(0.2,0.9)},anchor=north,
    row sep = 3pt}]
\addplot[smooth,tension=0.5,color=black, line width=0.75pt,dashdotted] table[x = c,y=d3] from \dgpc;
\addplot[smooth,tension=0.5,no markers, color=blue, line width=0.75pt] table[x = c,y=d30] from \dgpc;
\legend{\footnotesize Fixed $c$, \footnotesize CV}
\end{axis}
\end{tikzpicture}}
\end{subfigure}

\begin{subfigure}[b]{0.32\textwidth}
\centering
\resizebox{\linewidth}{3cm}{
\begin{tikzpicture}
\begin{axis}[
    legend style={draw=none},
    grid = minor,
    xmax=2,xmin=0,
    ymax=0.4,ymin=0,
    xtick={0,1,2},
    ytick={0.2,0.4},
    title={$N=500,T=500$},
    tick label style={/pgf/number format/fixed},
legend style={at={(0.2,0.9)},anchor=north,
    row sep = 3pt}]
\addplot[smooth,tension=0.5,color=black, line width=0.75pt,dashdotted] table[x = c,y=d4] from \dgpc;
\addplot[smooth,tension=0.5,no markers, color=blue, line width=0.75pt] table[x = c,y=d40] from \dgpc;
\legend{\footnotesize Fixed $c$, \footnotesize CV}
\end{axis}
\end{tikzpicture}}
\end{subfigure}
\begin{subfigure}[b]{0.32\textwidth}
\centering
\resizebox{\linewidth}{3cm}{
\begin{tikzpicture}
\begin{axis}[
    legend style={draw=none},
    grid = minor,
    xmax=2,xmin=0,
    ymax=0.2,ymin=0,
    xtick={0,1,2},
    ytick={0.1,0.2},
    title={$N=1000,T=500$},
    tick label style={/pgf/number format/fixed},
legend style={at={(0.2,0.9)},anchor=north,
    row sep = 3pt}]
\addplot[smooth,tension=0.5,color=black, line width=0.75pt,dashdotted] table[x = c,y=d5] from \dgpc;
\addplot[smooth,tension=0.5,no markers, color=blue, line width=0.75pt] table[x = c,y=d50] from \dgpc;
\legend{\footnotesize Fixed $c$, \footnotesize CV}
\end{axis}
\end{tikzpicture}}
\end{subfigure}
\begin{subfigure}[b]{0.32\textwidth}
\centering
\resizebox{\linewidth}{3cm}{
\begin{tikzpicture}
\begin{axis}[
    legend style={draw=none},
    grid = minor,
    xmax=2,xmin=0,
    ymax=0.1,ymin=0,
    xtick={0,1,2},
    ytick={0.05,0.1},
    title={$N=2000,T=500$},
    tick label style={/pgf/number format/fixed},
legend style={at={(0.2,0.9)},anchor=north,
    row sep = 3pt}]
\addplot[smooth,tension=0.5,color=black, line width=0.75pt,dashdotted] table[x = c,y=d6] from \dgpc;
\addplot[smooth,tension=0.5,no markers, color=blue, line width=0.75pt] table[x = c,y=d60] from \dgpc;
\legend{\footnotesize Fixed $c$, \footnotesize CV}
\end{axis}
\end{tikzpicture}}
\end{subfigure}
\caption{Mean square errors of $\hat{\Pi}_0$ when using fixed $c$ and CV: DGP3}\label{Fig: DGP3}
\end{figure}

\setlength{\tabcolsep}{18pt}
\begin{table}[htbp]
\centering
\resizebox{0.99\textwidth}{!}{
\begin{threeparttable}
\renewcommand{\arraystretch}{1.35}
\caption{Mean square errors of $\hat{\Pi}$, $\hat{a}$, $\hat{B}$, and $\hat{F}$, and correct rates of $\hat{K}$: DGP1\tnote{\dag}}\label{Tab: MSEDGP1}
\begin{tabular}{cccccccccc}
\hline\hline
&(N,T)&&$\hat{\Pi}$&$\hat{a}$&$\hat{B}$&$\hat{F}$&&$\hat{K}$&\\
\cline{2-9}
&$(500,250)$    &&4.170&2.295&0.853&0.183&&0.950&\\
&$(1000,250)$   &&3.996&2.233&0.800&0.171&&1.000&\\
&$(2000,250)$   &&3.850&2.188&0.759&0.154&&1.000&\\
\cline{2-9}
&$(500,500)$   &&1.821&1.641&0.243&0.088&&1.000&\\
&$(1000,500)$  &&1.686&1.595&0.222&0.066&&1.000&\\
&$(2000,500)$  &&1.584&1.543&0.210&0.053&&1.000&\\
\hline\hline
\end{tabular}
\begin{tablenotes}
      \small
      \item[\dag] The mean square errors of $\hat{\Pi}$, $\hat{a}$ , $\hat{B}$, and $\hat{F}$ are given by $\sum_{\ell=1}^{200}\|\hat{\Pi}^{(\ell)}-\Pi\|_{F}^2/200NT$, $\sum_{\ell=1}^{200}\|\hat{a}^{(\ell)}-a\|^2/200N$, $\sum_{\ell=1}^{200}\|\hat{B}^{(\ell)}-BH^{(\ell)}\|_{F}^2/200N$ and $\sum_{\ell=1}^{200}\|\hat{F}^{(\ell)}- F(H^{{(\ell)}\prime})^{-1}\|_{F}^2/{200T}$, where $\hat{\Pi}^{(\ell)}$, $\hat{a}^{(\ell)}$, $\hat{B}^{(\ell)}$, and $\hat{F}^{(\ell)}$ are estimates in the $\ell$th simulation replication, and $H^{(\ell)}\equiv (F^{\prime}M_T\hat{F}^{(\ell)})(\hat{F}^{{(\ell)}\prime}M_T\hat{F}^{(\ell)})^{-1}$ is a rotational transformation matrix. The value of $c$ is chosen from $\{0, 0.05, 0.1, 0.2, \ldots, 0.9,1,1.5,2\}$ by using the 5-fold CV method as outlined in Section \ref{Sec3}.
    \end{tablenotes}
\end{threeparttable}
}
\end{table}

\setlength{\tabcolsep}{12pt}
\begin{table}[htbp]
\centering
\resizebox{0.99\textwidth}{!}{
\begin{threeparttable}
\renewcommand{\arraystretch}{1.35}
\caption{Mean square errors of $\hat{\Pi}^{\diamond}$, $\hat{\Pi}^{\ast}$, $\hat{\mu}$, $\hat\Lambda$, $\hat{\phi}, \hat{\Phi}$, and $\hat{F}$, and correct rates of $\hat{K}$: DGP2\tnote{\dag}}\label{Tab: MSEDGP2}
\begin{tabular}{ccccccccccccc}
\hline\hline
&(N,T)&&$\hat{\Pi}^{\diamond}$&$\hat{\Pi}^{\ast}$&$\hat{\mu}$&$\hat\Lambda$&$\hat{\phi}$&$\hat{\Phi}$&$\hat{F}$&&$\hat{K}$&\\
\cline{2-12}
&$(500,250)$     &&0.108&0.096&0.061&0.005&0.157&0.009&0.038&&1.000&\\
&$(1000,250)$    &&0.077&0.073&0.062&0.005&0.133&0.008&0.028&&1.000&\\
&$(2000,250)$    &&0.095&0.055&0.065&0.005&0.104&0.006&0.020&&1.000&\\
\cline{2-12}
&$(500,500)$    &&0.060&0.074&0.031&0.003&0.109&0.006&0.032&&1.000&\\
&$(1000,500)$   &&0.061&0.048&0.033&0.002&0.076&0.004&0.020&&1.000&\\
&$(2000,500)$   &&0.040&0.038&0.033&0.002&0.065&0.004&0.014&&1.000&\\
\hline\hline
\end{tabular}
\begin{tablenotes}
      \small
      \item[\dag] The mean square errors of $\hat{\Pi}^{\diamond}$, $\hat{\Pi}^{\ast}$, $\hat{\mu}$, $\hat\Lambda$, $\hat{\phi}, \hat{\Phi}$, and $\hat{F}$ are given by $\sum_{\ell=1}^{200}\|\hat{\Pi}^{\diamond(\ell)}-\Pi^{\diamond}\|_{F}^2/200NT$, $\sum_{\ell=1}^{200}\|\hat{\Pi}^{\ast(\ell)}-\Pi^{\ast}\|_{F}^2/200T$,$\sum_{\ell=1}^{200}\|\hat{\mu}^{(\ell)}-\mu\|^2/200N$, $\sum_{\ell=1}^{200}\|\hat{\Lambda}^{(\ell)}-\Lambda H^{(\ell)}\|_{F}^2/200N$, $\sum_{\ell=1}^{200}\|\hat{\phi}^{(\ell)}-\phi\|^2/200$, $\sum_{\ell=1}^{200}\|\hat{\Phi}^{(\ell)}-\Phi H^{(\ell)}\|^2/200$ and $\sum_{\ell=1}^{200}\|\hat{F}^{(\ell)}- F(H^{{(\ell)}\prime})^{-1}\|_{F}^2/{200T}$, where $\hat{\Pi}^{\diamond(\ell)}$, $\hat{\Pi}^{\ast(\ell)}$, $\hat{\mu}^{(\ell)}$, $\hat{\Lambda}^{(\ell)}$, $\hat{\phi}^{(\ell)}$, $\hat{\Phi}^{(\ell)}$, and $\hat{F}^{(\ell)}$ are estimates in the $\ell$th simulation replication, and $H^{(\ell)}\equiv (F^{\prime}M_T\hat{F}^{(\ell)})(\hat{F}^{{(\ell)}\prime}M_T\hat{F}^{(\ell)})^{-1}$ is a rotational transformation matrix. The value of $c$ is chosen from $\{0, 0.05, 0.1, 0.2, \ldots, 0.9,1,1.5,2\}$ by using the 5-fold CV method as outlined in Section \ref{Sec3}.
    \end{tablenotes}
\end{threeparttable}
}
\end{table}

\setlength{\tabcolsep}{18pt}
\begin{table}[htbp]
\centering
\resizebox{0.99\textwidth}{!}{
\begin{threeparttable}
\renewcommand{\arraystretch}{1.35}
\caption{Mean square errors of $\hat{\Pi}_0$, $\hat{\phi}_0$, $\hat{\Phi}_0$, and $\hat{F}$ ($\times 10^{-2}$), and correct rates of $\hat{K}$: DGP3\tnote{\dag}}\label{Tab: MSEDGP3}
\begin{tabular}{cccccccccc}
\hline\hline
&(N,T)&&$\hat{\Pi}_0$&$\hat{\phi}_0$&$\hat{\Phi}_0$&$\hat{F}$&&$\hat{K}$&\\
\cline{2-9}
&$(500,250)$    &&5.340&4.007&0.271&2.224&&1.000&\\
&$(1000,250)$   &&2.746&1.785&0.121&1.124&&1.000&\\
&$(2000,250)$   &&1.344&0.974&0.065&0.580&&1.000&\\
\cline{2-9}
&$(500,500)$   &&4.865&3.482&0.234&2.187&&1.000&\\
&$(1000,500)$  &&2.594&1.477&0.099&1.064&&1.000&\\
&$(2000,500)$  &&1.265&0.810&0.054&0.559&&1.000&\\
\hline\hline
\end{tabular}
\begin{tablenotes}
      \small
      \item[\dag] The mean square errors of $\hat{\Pi}_0$, $\hat{\phi}_0$ , $\hat{\Phi}_0$, and $\hat{F}$ are given by $\sum_{\ell=1}^{200}\|\hat{\Pi}_0^{(\ell)}-\Pi_0\|_{F}^2/200T$, $\sum_{\ell=1}^{200}\|\hat{\phi}_0^{(\ell)}-\phi\|^2/200$, $\sum_{\ell=1}^{200}\|\hat{\Phi}_0^{(\ell)}-\Phi H^{(\ell)}\|_{F}^2/200$ and $\sum_{\ell=1}^{200}\|\hat{F}^{(\ell)}- F(H^{{(\ell)}\prime})^{-1}\|_{F}^2/{200T}$, where $\hat{\Pi}_0^{(\ell)}$, $\hat{\phi}_0^{(\ell)}$, $\hat{\Phi}_0^{(\ell)}$, and $\hat{F}^{(\ell)}$ are estimates in the $\ell$th simulation replication, and $H^{(\ell)}\equiv (F^{\prime}M_T\hat{F}^{(\ell)})(\hat{F}^{{(\ell)}\prime}M_T\hat{F}^{(\ell)})^{-1}$ is a rotational transformation matrix. The value of $c$ is chosen from $\{0, 0.05, 0.1, 0.2, \ldots, 0.9,1,1.5,2\}$ by using the 5-fold CV method as outlined in Section \ref{Sec3}.
    \end{tablenotes}
\end{threeparttable}
}
\end{table}

\pgfplotstableread{
k	x1	x2	x3	x4	x5	x6	x7	x8	x9	x10	x11	x12
1	19.78397619	16.0496283	22.03292447	19.73344125	15.9938032	21.98069301	0.425273488	0.311054379	0.573209396	0.050534939	0.055825105	0.05223146
2	23.0014627	17.73511348	24.76457163	22.9527652	17.67916671	24.71132158	0.425273488	0.311054379	0.573209396	0.0486975	0.055946769	0.053250046
3	25.0268463	19.28833767	26.56515449	24.97852174	19.23405106	26.51237451	0.425273488	0.311054379	0.573209396	0.048324558	0.054286606	0.052779978
4	26.60842464	20.53462093	27.95754052	26.56039318	20.48145688	27.90540099	0.425273488	0.311054379	0.573209396	0.048031457	0.053164058	0.052139525
5	27.82976075	20.96279622	28.53417785	27.78323551	20.91129621	28.48331217	0.425273488	0.311054379	0.573209396	0.046525246	0.051500006	0.050865675
6	28.88765005	21.47669566	29.37062689	28.84371213	21.42903585	29.32555334	0.425273488	0.311054379	0.573209396	0.043937918	0.047659804	0.045073547
7	29.7444564	22.12044479	30.12003611	29.71221227	22.0872535	30.08730266	0.425273488	0.311054379	0.573209396	0.032244139	0.033191291	0.032733452
8	30.59685879	22.65088575	30.85109124	30.56593078	22.62033581	30.8204353	0.425273488	0.311054379	0.573209396	0.030928004	0.030549947	0.030655941
9	31.39774395	23.13030608	31.48915112	31.36997237	23.10203218	31.46277561	0.425273488	0.311054379	0.573209396	0.027771583	0.028273903	0.026375514
10	32.21667299	23.51575945	31.98281204	32.18937347	23.48885673	31.95690381	0.425273488	0.311054379	0.573209396	0.02729952	0.026902723	0.025908226
}\empa

\pgfplotstableread{
k	x1	x2	x3	x4	x5	x6	x7	x8	x9
1	18.76569193	15.4511116	21.01375672	18.68086381	15.36174128	20.91939556	0.431590505	0.291440272	0.60915766
2	21.52130469	16.83284757	23.43484188	21.43867791	16.74233284	23.33699895	0.431590505	0.291440272	0.60915766
3	23.33659308	18.03321005	24.83143466	23.25387779	17.94206342	24.73359332	0.431590505	0.291440272	0.60915766
4	24.79577503	19.14549108	26.14152398	24.71289728	19.05490673	26.04440973	0.431590505	0.291440272	0.60915766
5	25.96589867	19.72514425	26.85217679	25.88448709	19.63622678	26.75547804	0.431590505	0.291440272	0.60915766
6	26.96752425	20.17916008	27.58643068	26.88682413	20.0923324	27.49223402	0.431590505	0.291440272	0.60915766
7	27.82991203	20.68437029	28.36175757	27.74897961	20.59804742	28.2672491	0.431590505	0.291440272	0.60915766
8	28.66900065	21.04614467	28.84771355	28.58994915	20.96091933	28.75562084	0.431590505	0.291440272	0.60915766
9	29.40747168	21.5512486	29.58116539	29.33651189	21.47932124	29.50298396	0.431590505	0.291440272	0.60915766
10  30.15945774	21.88639719	30.05057626	30.08881386	21.81445376	29.97273641	0.431590505	0.291440272	0.60915766
}\empb


\pgfplotstableread{
k	x1	x2	x3	x4	x5	x6	x7	x8	x9	x10	x11	x12
1	18.20072892	15.26663258	19.16387575	17.82355183	14.79504704	18.7366448	0.834675501	0.556749775	1.07	0.377177084	0.471585541	0.427230945
2	19.47372843	16.14815566	20.48679072	19.08796113	15.67154105	20.04761936	0.834675501	0.556749775	1.07	0.385767298	0.476614612	0.439171365
3	21.12188542	17.6823269	22.84629057	20.7800035	17.19922972	22.41672007	0.834675501	0.556749775	1.07	0.341881917	0.48309718	0.429570499
4	22.54645031	18.7879371	24.6562571	22.21722146	18.3402633	24.25647883	0.834675501	0.556749775	1.07	0.329228846	0.447673799	0.399778265
5	22.88209261	19.15667531	25.0520791	22.67599493	18.89002953	24.78176698	0.834675501	0.556749775	1.07	0.206097677	0.266645782	0.270312118
6	23.34220596	19.59630166	25.6214229	23.22021748	19.45253597	25.50152092	0.834675501	0.556749775	1.07	0.12198848	0.14376569	0.119901987
7	23.66748845	19.94946119	26.04318542	23.54643426	19.80776749	25.92288627	0.834675501	0.556749775	1.07	0.121054191	0.141693703	0.120299146
8	23.91039198	20.20174572	26.27570259	23.88275982	20.16815315	26.23980135	0.834675501	0.556749775	1.07	0.027632156	0.033592571	0.035901235
9	24.06037721	20.3581503	26.44275535	24.03348048	20.32607118	26.40716528	0.834675501	0.556749775	1.07	0.026896735	0.032079122	0.035590068
10	24.22196944	20.53850597	26.61357558	24.19339303	20.50467011	26.5743102	0.834675501	0.556749775	1.07	0.028576409	0.033835869	0.039265382
}\empc

\pgfplotstableread{
k	x1	x2	x3  x7	x8	x9
1	2.4151775	0.270744452	2.11800967	0.794638557	0.462437015	0.978870287
2	3.61323705	0.858619263	3.441763274	0.794638557	0.462437015	0.978870287
3	4.559943135	1.324080283	4.437781153	0.794638557	0.462437015	0.978870287
4	4.765972671	1.656090297	4.623180378	0.794638557	0.462437015	0.978870287
5	8.747233654	4.616366892	8.737732062	0.794638557	0.462437015	0.978870287
6	11.37130751	7.207234276	11.08379845	0.794638557	0.462437015	0.978870287
7	13.40334322	9.618156033	13.06671806	0.794638557	0.462437015	0.978870287
8	16.95898943	13.07758168	16.97741614	0.794638557	0.462437015	0.978870287
9	19.23457836	15.66545377	19.6535871	0.794638557	0.462437015	0.978870287
10  20.03926363	16.49919903	20.35534774	0.794638557	0.462437015	0.978870287
}\empregressedpca

\pgfplotstableread{
k	x1	x2	x3	x4	x5	x6	x7	x8	x9	x10	x11	x12
1	18.76673286	15.30625399	20.95104112	18.73792622	15.2719071	20.91618163	0.404136198	0.304530481	0.577825976	0.028806637	0.034346886	0.034859486
2	21.6242784	16.72190856	23.50696865	21.59610295	16.68812602	23.47142938	0.404136198	0.304530481	0.577825976	0.028175446	0.033782536	0.035539267
3	23.34308699	17.90151103	24.95455185	23.31523509	17.8699816	24.91962685	0.404136198	0.304530481	0.577825976	0.0278519	0.031529431	0.034924996
4	24.59322027	18.79766041	26.07427425	24.56672625	18.76828182	26.04197985	0.404136198	0.304530481	0.577825976	0.02649402	0.02937859	0.032294399
5	25.63147032	19.16174473	26.5829391	25.60611835	19.13341529	26.55128872	0.404136198	0.304530481	0.577825976	0.025351971	0.028329446	0.031650379
6	26.47450801	19.50841809	27.24773126	26.45065155	19.48133058	27.2192703	0.404136198	0.304530481	0.577825976	0.023856461	0.027087509	0.028460957
7	27.21814799	19.96035984	27.88220827	27.19451124	19.93430927	27.85354374	0.404136198	0.304530481	0.577825976	0.023636755	0.026050576	0.02866453
8	27.88940635	20.31018598	28.44764624	27.87457921	20.29377989	28.43109794	0.404136198	0.304530481	0.577825976	0.014827139	0.016406092	0.0165483
9	28.48601845	20.60910367	28.89216548	28.47179514	20.59521179	28.87680388	0.404136198	0.304530481	0.577825976	0.014223309	0.013891884	0.015361596
10	29.0886451	20.88821919	29.35572407	29.07768941	20.87695932	29.34472081	0.404136198	0.304530481	0.577825976	0.010955691	0.011259878	0.011003255
}\empaspline

\pgfplotstableread{
k	x1	x2	x3	x4	x5	x6	x7	x8	x9
1	18.21376464	15.08225972	19.26981639	18.14018907	14.97738242	19.17183983	0.466663562	0.30740805	0.682441868
2	21.39172538	16.8779238	23.21987308	21.32001502	16.79694204	23.12958975	0.466663562	0.30740805	0.682441868
3	23.21036393	18.09507539	24.6618819	23.13834286	18.01638289	24.57154117	0.466663562	0.30740805	0.682441868
4	24.60064279	19.09498298	25.87211299	24.52912277	19.01716806	25.78375924	0.466663562	0.30740805	0.682441868
5	25.77610958	19.68258797	26.58531642	25.7068148	19.60723891	26.49836989	0.466663562	0.30740805	0.682441868
6	26.8378879	20.15756436	27.44367613	26.77045494	20.0850586	27.36156266	0.466663562	0.30740805	0.682441868
7	27.75956697	20.68667443	28.41417755	27.70003269	20.62507016	28.34869038	0.466663562	0.30740805	0.682441868
8	28.60887668	21.04634712	28.93592615	28.55213651	20.98544788	28.87342997	0.466663562	0.30740805	0.682441868
9	29.37462613	21.40000969	29.43111884	29.31848225	21.33905417	29.36910247	0.466663562	0.30740805	0.682441868
10  30.11243206	21.9021766	30.1198585	30.05793753	21.84589042	30.06123271	0.466663562	0.30740805	0.682441868
}\empbspline


\pgfplotstableread{
k	x1	x2	x3	x4	x5	x6	x7	x8	x9	x10	x11	x12
1	13.10776141	9.670879077	12.33359277	12.69000145	9.123067886	11.87889585	0.864456072	0.604413358	1.080307192	0.417759955	0.547811191	0.454696914
2	15.01725349	11.70886333	13.86303032	14.59759032	11.17794811	13.39901714	0.864456072	0.604413358	1.080307192	0.419663172	0.530915224	0.464013174
3	17.81232534	14.6544419	17.69184302	17.42751736	14.1291686	17.21712697	0.864456072	0.604413358	1.080307192	0.384807981	0.525273295	0.474716052
4	18.60071732	15.08747128	18.68605936	18.21071773	14.70728754	18.25149473	0.864456072	0.604413358	1.080307192	0.389999586	0.380183732	0.434564631
5	20.44039553	16.53566919	21.32885478	20.16753269	16.20164243	21.02411215	0.864456072	0.604413358	1.080307192	0.272862839	0.334026752	0.30474263
6	20.87068851	16.99828725	21.72060871	20.74406607	16.8352111	21.55560507	0.864456072	0.604413358	1.080307192	0.126622439	0.163076146	0.165003644
7	21.90544273	18.04922017	23.28705198	21.69665669	17.79859404	22.98446656	0.864456072	0.604413358	1.080307192	0.208786038	0.250626122	0.302585418
8	22.72418108	18.75859548	24.22033442	22.5712626	18.57500396	24.00162082	0.864456072	0.604413358	1.080307192	0.152918472	0.18359152	0.218713602
9	23.04429504	19.11544664	24.58025596	22.8526714	18.89543952	24.27495593	0.864456072	0.604413358	1.080307192	0.191623645	0.220007119	0.305300034
10	23.44119958	19.64001011	25.17929301	23.28802816	19.43136666	24.93887761	0.864456072	0.604413358	1.080307192	0.153171413	0.208643453	0.240415398
}\empcspline

\pgfplotstableread{
k	x1	x2	x3  x7	x8	x9
1	3.730955782	1.457065701	3.445342152	0.841677528	0.514120961	1.016947396
2	9.803144174	6.509265921	9.005506164	0.841677528	0.514120961	1.016947396
3	11.3797279	7.958427723	10.11507774	0.841677528	0.514120961	1.016947396
4	15.31450372	11.87442339	14.35138896	0.841677528	0.514120961	1.016947396
5	16.14869052	12.70308811	15.42907235	0.841677528	0.514120961	1.016947396
6	16.54821647	13.03980857	15.86031005	0.841677528	0.514120961	1.016947396
7	17.96624755	14.09265656	17.85728909	0.841677528	0.514120961	1.016947396
8	18.28952	14.42483044	18.17414921	0.841677528	0.514120961	1.016947396
9	19.18600499	15.32249453	19.28049809	0.841677528	0.514120961	1.016947396
10	19.50205769	15.62131374	19.54298715	0.841677528	0.514120961	1.016947396
}\empregressedpcaspline

\begin{figure}[H]
\centering
\begin{subfigure}[b]{0.32\textwidth}
\centering
\resizebox{\linewidth}{!}{
\pgfplotsset{title style={at={(0.5,0.91)}}}
\begin{tikzpicture}
\begin{axis}[
        ybar,
    height=13cm,
    width=13cm,
    enlarge y limits=false,
    axis lines*=left,
    xmax=10,xmin=1,
    ymax=40,ymin=0,
    xtick={1,2,3,4,5,6,7,8,9,10},
    ytick={0,20,40},
        title={$R^{2}$},
     legend style={at={(0.51,0.9)},
        anchor=north,legend columns=-1,
        /tikz/every even column/.append style={column sep=0.5cm}
        },
    ]
\addplot[smooth,tension=0.3, mark=star,color=blue, line width=1.2pt] table[x = k,y=x1] from \empa;
\addplot[smooth,tension=0.3, mark=square,color=red, line width=1.2pt] table[x = k,y=x1] from \empb;
\addplot[smooth,tension=0.3, mark=triangle,color=black, line width=1.2pt] table[x = k,y=x1] from \empc;
\addplot[smooth,tension=0.3, mark=star,color=blue, line width=1.2pt,dashdotted] table[x = k,y=x1] from \empaspline;
\addplot[smooth,tension=0.3, mark=square,color=red, line width=1.2pt,dashdotted] table[x = k,y=x1] from \empbspline;
\addplot[smooth,tension=0.3, mark=triangle,color=black, line width=1.2pt,dashdotted] table[x = k,y=x1] from \empcspline;
\addplot[smooth,tension=0.3, mark=o,color=green, line width=1.2pt] table[x = k,y=x1] from \empregressedpca;
\addplot[smooth,tension=0.3, mark=o,color=green, line width=1.2pt,dashdotted] table[x = k,y=x1] from \empregressedpcaspline;
\legend{S1,S2,S3,S4,S5,S6,R1,R2}
  \end{axis}
\end{tikzpicture}}
\end{subfigure}
\begin{subfigure}[b]{0.32\textwidth}
\centering
\resizebox{\linewidth}{!}{
\pgfplotsset{title style={at={(0.5,0.91)}}}
\begin{tikzpicture}
\begin{axis}[
        ybar,
    height=13cm,
    width=13cm,
    enlarge y limits=false,
    axis lines*=left,
    xmax=10,xmin=1,
    ymax=40,ymin=0,
    xtick={1,2,3,4,5,6,7,8,9,10},
    ytick={0,20,40},
        title={$R^{2}_{T,N}$},
     legend style={at={(0.51,0.9)},
        anchor=north,legend columns=-1,
        /tikz/every even column/.append style={column sep=0.5cm}
        },
    ]
\addplot[smooth,tension=0.3, mark=star,color=blue, line width=1.2pt] table[x = k,y=x3] from \empa;
\addplot[smooth,tension=0.3, mark=square,color=red, line width=1.2pt] table[x = k,y=x3] from \empb;
\addplot[smooth,tension=0.3, mark=triangle,color=black, line width=1.2pt] table[x = k,y=x3] from \empc;
\addplot[smooth,tension=0.3, mark=star,color=blue, line width=1.2pt,dashdotted] table[x = k,y=x3] from \empaspline;
\addplot[smooth,tension=0.3, mark=square,color=red, line width=1.2pt,dashdotted] table[x = k,y=x3] from \empbspline;
\addplot[smooth,tension=0.3, mark=triangle,color=black, line width=1.2pt,dashdotted] table[x = k,y=x3] from \empcspline;
\addplot[smooth,tension=0.3, mark=o,color=green, line width=1.2pt] table[x = k,y=x3] from \empregressedpca;
\addplot[smooth,tension=0.3, mark=o,color=green, line width=1.2pt,dashdotted] table[x = k,y=x3] from \empregressedpcaspline;
\legend{S1,S2,S3,S4,S5,S6,R1,R2}
  \end{axis}
\end{tikzpicture}}
\end{subfigure}
\begin{subfigure}[b]{0.32\textwidth}
\centering
\resizebox{\linewidth}{!}{
\pgfplotsset{title style={at={(0.5,0.91)}}}
\begin{tikzpicture}
\begin{axis}[
        ybar,
    height=13cm,
    width=13cm,
    enlarge y limits=false,
    axis lines*=left,
    xmax=10,xmin=1,
    ymax=40,ymin=0,
    xtick={1,2,3,4,5,6,7,8,9,10},
    ytick={0,20,40},
        title={$R^{2}_{N,T}$},
     legend style={at={(0.51,0.9)},
        anchor=north,legend columns=-1,
        /tikz/every even column/.append style={column sep=0.5cm}
        },
    ]
\addplot[smooth,tension=0.3, mark=star,color=blue, line width=1.2pt] table[x = k,y=x2] from \empa;
\addplot[smooth,tension=0.3, mark=square,color=red, line width=1.2pt] table[x = k,y=x2] from \empb;
\addplot[smooth,tension=0.3, mark=triangle,color=black, line width=1.2pt] table[x = k,y=x2] from \empc;
\addplot[smooth,tension=0.3, mark=star,color=blue, line width=1.2pt,dashdotted] table[x = k,y=x2] from \empaspline;
\addplot[smooth,tension=0.3, mark=square,color=red, line width=1.2pt,dashdotted] table[x = k,y=x2] from \empbspline;
\addplot[smooth,tension=0.3, mark=triangle,color=black, line width=1.2pt,dashdotted] table[x = k,y=x2] from \empcspline;
\addplot[smooth,tension=0.3, mark=o,color=green, line width=1.2pt] table[x = k,y=x2] from \empregressedpca;
\addplot[smooth,tension=0.3, mark=o,color=green, line width=1.2pt,dashdotted] table[x = k,y=x2] from \empregressedpcaspline;
\legend{S1,S2,S3,S4,S5,S6,R1,R2}
  \end{axis}
\end{tikzpicture}}
\end{subfigure}

\begin{subfigure}[b]{0.32\textwidth}
\centering
\resizebox{\linewidth}{!}{
\pgfplotsset{title style={at={(0.5,0.91)}}}
\begin{tikzpicture}
\begin{axis}[
        ybar,
    height=13cm,
    width=13cm,
    enlarge y limits=false,
    axis lines*=left,
    xmax=10,xmin=1,
    ymax=1,ymin=0.2,
    xtick={1,2,3,4,5,6,7,8,9,10},
    ytick={0.2,1},
        title={$R^{2,O}$},
     legend style={at={(0.51,0.12)},
        anchor=north,legend columns=-1,
        /tikz/every even column/.append style={column sep=0.5cm}
        },
    ]
\addplot[smooth,tension=0.3, mark=star,color=blue, line width=1.2pt] table[x = k,y=x7] from \empa;
\addplot[smooth,tension=0.3, mark=square,color=red, line width=1.2pt] table[x = k,y=x7] from \empb;
\addplot[smooth,tension=0.3, mark=triangle,color=black, line width=1.2pt] table[x = k,y=x7] from \empc;
\addplot[smooth,tension=0.3, mark=star,color=blue, line width=1.2pt,dashdotted] table[x = k,y=x7] from \empaspline;
\addplot[smooth,tension=0.3, mark=square,color=red, line width=1.2pt,dashdotted] table[x = k,y=x7] from \empbspline;
\addplot[smooth,tension=0.3, mark=triangle,color=black, line width=1.2pt,dashdotted] table[x = k,y=x7] from \empcspline;
\addplot[smooth,tension=0.3, mark=o,color=green, line width=1.2pt] table[x = k,y=x7] from \empregressedpca;
\addplot[smooth,tension=0.3, mark=o,color=green, line width=1.2pt,dashdotted] table[x = k,y=x7] from \empregressedpcaspline;
\legend{S1,S2,S3,S4,S5,S6,R1,R2}
  \end{axis}
\end{tikzpicture}}
\end{subfigure}
\begin{subfigure}[b]{0.32\textwidth}
\centering
\resizebox{\linewidth}{!}{
\pgfplotsset{title style={at={(0.5,0.91)}}}
\begin{tikzpicture}
\begin{axis}[
        ybar,
    height=13cm,
    width=13cm,
    enlarge y limits=false,
    axis lines*=left,
    xmax=10,xmin=1,
    ymax=1.2,ymin=0.2,
    xtick={1,2,3,4,5,6,7,8,9,10},
    ytick={0.2,1.2},
        title={$R^{2}_{T,N,O}$},
     legend style={at={(0.51,0.12)},
        anchor=north,legend columns=-1,
        /tikz/every even column/.append style={column sep=0.5cm}
        },
    ]
\addplot[smooth,tension=0.3, mark=star,color=blue, line width=1.2pt] table[x = k,y=x9] from \empa;
\addplot[smooth,tension=0.3, mark=square,color=red, line width=1.2pt] table[x = k,y=x9] from \empb;
\addplot[smooth,tension=0.3, mark=triangle,color=black, line width=1.2pt] table[x = k,y=x9] from \empc;
\addplot[smooth,tension=0.3, mark=star,color=blue, line width=1.2pt,dashdotted] table[x = k,y=x9] from \empaspline;
\addplot[smooth,tension=0.3, mark=square,color=red, line width=1.2pt,dashdotted] table[x = k,y=x9] from \empbspline;
\addplot[smooth,tension=0.3, mark=triangle,color=black, line width=1.2pt,dashdotted] table[x = k,y=x9] from \empcspline;
\addplot[smooth,tension=0.3, mark=o,color=green, line width=1.2pt] table[x = k,y=x9] from \empregressedpca;
\addplot[smooth,tension=0.3, mark=o,color=green, line width=1.2pt,dashdotted] table[x = k,y=x9] from \empregressedpcaspline;
\legend{S1,S2,S3,S4,S5,S6,R1,R2}
  \end{axis}
\end{tikzpicture}}
\end{subfigure}
\begin{subfigure}[b]{0.32\textwidth}
\centering
\resizebox{\linewidth}{!}{
\pgfplotsset{title style={at={(0.5,0.91)}}}
\begin{tikzpicture}
\begin{axis}[
        ybar,
    height=13cm,
    width=13cm,
    enlarge y limits=false,
    axis lines*=left,
    xmax=10,xmin=1,
    ymax=0.7,ymin=0.2,
    xtick={1,2,3,4,5,6,7,8,9,10},
    ytick={0.2,0.7},
        title={$R^{2}_{N,T,O}$},
     legend style={at={(0.51,0.12)},
        anchor=north,legend columns=-1,
        /tikz/every even column/.append style={column sep=0.5cm}
        },
    ]
\addplot[smooth,tension=0.3, mark=star,color=blue, line width=1.2pt] table[x = k,y=x8] from \empa;
\addplot[smooth,tension=0.3, mark=square,color=red, line width=1.2pt] table[x = k,y=x8] from \empb;
\addplot[smooth,tension=0.3, mark=triangle,color=black, line width=1.2pt] table[x = k,y=x8] from \empc;
\addplot[smooth,tension=0.3, mark=star,color=blue, line width=1.2pt,dashdotted] table[x = k,y=x8] from \empaspline;
\addplot[smooth,tension=0.3, mark=square,color=red, line width=1.2pt,dashdotted] table[x = k,y=x8] from \empbspline;
\addplot[smooth,tension=0.3, mark=triangle,color=black, line width=1.2pt,dashdotted] table[x = k,y=x8] from \empcspline;
\addplot[smooth,tension=0.3, mark=o,color=green, line width=1.2pt] table[x = k,y=x8] from \empregressedpca;
\addplot[smooth,tension=0.3, mark=o,color=green, line width=1.2pt,dashdotted] table[x = k,y=x8] from \empregressedpcaspline;
\legend{S1,S2,S3,S4,S5,S6,R1,R2}
  \end{axis}
\end{tikzpicture}}
\end{subfigure}
\caption{In-sample and out-of-sample $R^2$'s} \label{Fig: Emprical1}
\end{figure}

\clearpage
\newpage

\begin{bibunit}
\def\spacingset#1{\renewcommand{\baselinestretch}%
{#1}\small\normalsize} \spacingset{1.8}
\setlength{\abovedisplayskip}{4pt}
\setlength{\belowdisplayskip}{4pt}
\setlength{\abovedisplayshortskip}{4pt}
\setlength{\belowdisplayshortskip}{4pt}
\begin{appendices} \sloppy
\bookmarksetup{open=false}
\allowdisplaybreaks 
\titleformat{\section}{\Large\center}{{\sc Appendix} \thesection}{0.25em}{- }
\setcounter{page}{1}
\setcounter{section}{0}
\setcounter{equation}{0}
\setcounter{table}{0}
\setcounter{figure}{0}
\numberwithin{equation}{section}
\numberwithin{table}{section}
\numberwithin{figure}{section}
\renewcommand{\thetable}{\thesection.\Roman{table}}
\renewcommand\thefigure{\thesection.\arabic{figure}}

\emptythanks
\phantomsection
\pdfbookmark[1]{Appendix Title}{title1}

\if1\blind
{
  \title{\bf Supplementary Appendix to ``A Unified Framework for Estimation of High-dimensional Conditional Factor Models''}
  \author{Qihui Chen\\ School of Management and Economics\\ The Chinese University of Hong Kong, Shenzhen\\ qihuichen@cuhk.edu.cn}
  \date{}
  \maketitle
} \fi

\if0\blind
{
  \bigskip
  \bigskip
  \bigskip
\begin{center}
    {\LARGE\bf Supplementary Appendix to ``A Unified Framework for Estimation of High-dimensional Conditional Factor Models''}
\end{center}
  \medskip
} \fi

This supplementary appendix is structured as follows. Appendices \ref{App: A} - \ref{App: D} collect proofs of main results, Appendix \ref{App: E} presents computing algorithms, Appendix \ref{App: F} provides additional discussions, and Appendix \ref{App: G} consists of additional simulation results.

\section{Proof of Theorem \ref{Thm: NuclearNRate}}\label{App: A}
\renewcommand{\theequation}{A.\arabic{equation}}
\setcounter{equation}{0}
\noindent{\sc Proof of Theorem \ref{Thm: NuclearNRate}:} (i) The proof closely follows the proof of Corollary 1 in \citet{NegahbanWainwright_Scaling_2011}. By the definition of $\hat{\Pi}$,
\begin{align}\label{Eqn: Thm: NuclearNRate: 1}
\hspace{-0.2cm}\frac{1}{2}\sum_{i=1}^{N}\sum_{t=1}^{T}(y_{it}-\mathrm{tr}(X_{it}^{\prime}\hat{\Pi}))^{2} \hspace{-0.05cm}+\hspace{-0.05cm}\lambda_{NT} \|\hat{\Pi}\|_{\ast}\hspace{-0.05cm}\leq\hspace{-0.05cm} \frac{1}{2}\sum_{i=1}^{N}\sum_{t=1}^{T}(y_{it}-\mathrm{tr}(X_{it}^{\prime}\Pi))^{2} \hspace{-0.05cm}+\hspace{-0.05cm}\lambda_{NT} \|\Pi\|_{\ast}.
\end{align}
Let $\Delta\equiv \hat{\Pi}-\Pi\in\mathcal{S}\ominus \mathcal{S}$. Noting that $\sum_{i=1}^{N}\sum_{t=1}^{T}|\mathrm{tr}(X_{it}^{\prime}\Delta)|^2 = \mathcal{Q}_{NT}(\Delta) + \mathcal{L}_{NT}(\Delta)$, we may rearrange \eqref{Eqn: Thm: NuclearNRate: 1} to obtain
\begin{align}\label{Eqn: Thm: NuclearNRate: 2}
\frac{1}{2} \mathcal{Q}_{NT}(\Delta) &\leq -\frac{1}{2} \mathcal{L}_{NT}(\Delta) +\sum_{i=1}^{N} \sum_{t=1}^{T}\mathrm{tr}(\varepsilon_{it} X_{it}^{\prime}\Delta) +\lambda_{NT} \|\Pi\|_{\ast} -\lambda_{NT} \|\Pi +\Delta\|_{\ast}\notag\\
&\leq r_{NT}\|\Delta\|_{\ast}+\lambda_{NT} \|\Pi\|_{\ast} -\lambda_{NT} \|\Pi +\Delta\|_{\ast}\notag\\
&\leq \lambda_{NT}\left(\frac{1}{2}\|\Delta\|_{\ast}+\|\Pi\|_{\ast} - \|\Pi +\Delta\|_{\ast}\right),
\end{align}
where the first inequality follows by Assumption \ref{Ass: NuclearN} and the second inequality follows since $\lambda_{NT}\geq 2r_{NT}$. Since $\Delta = \mathcal{P}(\Delta) +\mathcal{M}(\Delta)$, it follows that
\begin{align}\label{Eqn: Thm: NuclearNRate: 3}
\|\Pi\|_{\ast}-\|\Pi +\Delta\|_{\ast} &= \|\Pi\|_{\ast}-\|\Pi +\mathcal{P}(\Delta) +\mathcal{M}(\Delta)\|_{\ast}\notag\\
&\leq \|\Pi\|_{\ast} - \|\Pi +\mathcal{P}(\Delta)\|_{\ast} +\|\mathcal{M}(\Delta)\|_{\ast}\notag\\
& = \|\mathcal{M}(\Delta)\|_{\ast}- \|\mathcal{P}(\Delta)\|_{\ast}
\end{align}
where the inequality follows by the triangle inequality and the second equality follows by Lemma \ref{Lem: TechA1}(i). Since $\|\Delta\|_{\ast}\leq \|\mathcal{P}(\Delta)\|_{\ast} + \|\mathcal{M}(\Delta)\|_{\ast}$, combining \eqref{Eqn: Thm: NuclearNRate: 2} and \eqref{Eqn: Thm: NuclearNRate: 3} gives
\begin{align}\label{Eqn: Thm: NuclearNRate: 4}
0\leq \frac{1}{2} \mathcal{Q}_{NT}(\Delta)\leq  \lambda_{NT}\left(\frac{3}{2}\|\mathcal{M}(\Delta)\|_{\ast}-\frac{1}{2}\|\mathcal{P}(\Delta)\|_{\ast}\right).
\end{align}
Therefore, $\|\mathcal{P}(\Delta)\|_{\ast}\leq 3 \|\mathcal{M}(\Delta)\|_{\ast}$ and $\Delta\in\mathcal{C}$. This in turn together with \eqref{Eqn: Thm: NuclearNRate: 4} and Assumption \ref{Ass: NuclearN}(i) implies that
\begin{align}\label{Eqn: Thm: NuclearNRate: 5}
\frac{1}{2}\kappa \|\Delta\|_{F}^{2} &\leq \lambda_{NT}\left(\frac{3}{2}\|\mathcal{M}(\Delta)\|_{\ast}-\frac{1}{2}\|\mathcal{P}(\Delta)\|_{\ast}\right)\leq \frac{3}{2} \lambda_{NT} \|\mathcal{M}(\Delta)\|_{\ast}\notag\\
&\leq \frac{3}{2} \lambda_{NT}\sqrt{2(K+1)}\|\mathcal{M}(\Delta)\|_{F}\leq  \frac{3}{2} \lambda_{NT}\sqrt{2(K+1)}\|\Delta\|_{F},
\end{align}
where the second inequality follows since $\|\mathcal{P}(\Delta)\|_{\ast}\geq 0$, the third inequality follows by the Cauchy-Schwartz inequality (i.e., $\|A\|_{\ast}\leq \sqrt{\mathrm{rank}(A)}\|A\|_{F}$) and Lemma \ref{Lem: TechA1}(ii), and the last inequality follows by Lemma \ref{Lem: TechA1}(iii). Thus, the result follows by \eqref{Eqn: Thm: NuclearNRate: 5}.

(ii) Let $\sigma_{j}(A)$ denote the $j$th largest singular value of $A$, so $\lambda_{j}(\hat{\Pi}M_{T}\hat{\Pi}^{\prime}) = \sigma^{2}_{j}(\hat{\Pi}M_{T})$. If $\hat{K}\neq K$, then $\lambda_{K}(\hat{\Pi}M_{T}\hat{\Pi}^{\prime})<\delta_{NT}$ or $\lambda_{K+1}(\hat{\Pi}M_{T}\hat{\Pi}^{\prime})\geq \delta_{NT}$, equivalently, $\sigma_{K}(\hat{\Pi}M_{T})<\sqrt{\delta_{NT}}$ or $\sigma_{K+1}(\hat{\Pi}M_{T})\geq \sqrt{\delta_{NT}}$. Thus, we obtain
\begin{align}\label{Eqn: Thm: NuclearNRate: 6}
P(\hat{K}\neq K)\leq P(\sigma_{K}(\hat{\Pi}M_{T})<\sqrt{\delta_{NT}}) + P(\sigma_{K+1}(\hat{\Pi}M_{T})\geq \sqrt{\delta_{NT}}).
\end{align}
By the Weyl's inequality, we have
\begin{align}\label{Eqn: Thm: NuclearNRate: 7}
\sup_{j\leq \min\{Np,T\}}|\sigma_{j}(\hat{\Pi}M_{T}) -\sigma_{j}(\Pi M_{T})|\leq \|\hat{\Pi}M_{T} -\Pi M_{T}\|_{F}\leq \|\hat{\Pi}  - \Pi\|_{F},
\end{align}
where the second inequality follows since $\|CD\|_{F}\leq \|C\|_{F}\|D\|_{2}$ and $\|M_T\|_{2} = 1$. It then follows from \eqref{Eqn: Thm: NuclearNRate: 7} and Theorem \ref{Thm: NuclearNRate}(i) that with probability approaching one,
\begin{align}\label{Eqn: Thm: NuclearNRate: 8}
\sigma_{K}(\hat{\Pi}M_{T})\geq \sigma_{K}(\Pi M_{T}) - O_{p}(\sqrt{K}\lambda_{NT})\geq \sqrt{\delta_{NT}}
\end{align}
and
\begin{align}\label{Eqn: Thm: NuclearNRate: 9}
\sigma_{K+1}(\hat{\Pi}M_{T})\leq \sigma_{K+1}(\Pi M_{T}) + O_{p}(\sqrt{K}\lambda_{NT})<\sqrt{\delta_{NT}},
\end{align}
where the second equality in \eqref{Eqn: Thm: NuclearNRate: 8} follows since $\delta_{NT}/(K\lambda^{2}_{NT})\to\infty$, $\delta_{NT}/(NT)\to0$ and $\sigma^{2}_{K}(\Pi M_{T}/\sqrt{NT}) =\lambda_{\min}((B^{\prime}B/N)(F^{\prime} M_{T}F/T))$ $>d_{\min}^{2}$, and the second equality in \eqref{Eqn: Thm: NuclearNRate: 9} follows since $\sigma_{K+1}(\Pi M_{T})=0$ and $\delta_{NT}/(K\lambda^{2}_{NT})\to\infty$. Thus, the first result follows from \eqref{Eqn: Thm: NuclearNRate: 6}, \eqref{Eqn: Thm: NuclearNRate: 8} and \eqref{Eqn: Thm: NuclearNRate: 9}.

It is without loss of generality to assume that $\hat{K} = K$. Let $V$ be a $K\times K$ diagonal matrix of the first $K$ largest eigenvalues of $\hat{\Pi}M_T\hat{\Pi}^{\prime}/(NT)$. By the definitions of $\hat{B}$,
\begin{align}\label{Eqn: Thm: NuclearNRate: 10}
\hat{B} &= \frac{1}{NT}\hat{\Pi}M_{T}\hat{\Pi}^{\prime}\hat{B}V^{-1} = BH +  \frac{1}{NT}(\hat{\Pi}-\Pi)M_{T}\hat{\Pi}^{\prime}\hat{B}V^{-1},
\end{align}
where the second equality follows since $\hat{F}^{\prime}M_T\hat{F}/T =V$, $\Pi M_{T} = BF^{\prime}M_{T}$ and $\hat{F} = \hat{\Pi}^{\prime}\hat{B}$. By Assumptions \ref{Ass: NuclearKaBF}(i), (ii) and (iv), $\|\Pi/\sqrt{NT}\|_{F}$ is bounded. Since $\sqrt{K}\lambda_{NT}/\sqrt{NT} = o(1)$, $\|\hat{\Pi}/\sqrt{NT}\|_{F} = O_{p}(1)$ by Theorem \ref{Thm: NuclearNRate}(i). Thus, the thid result follows from \eqref{Eqn: Thm: NuclearNRate: 10}, Lemma \ref{Lem: TechA1}(i) and Theorem \ref{Thm: NuclearNRate}(i). By the definition of $\hat{a}$,
\begin{align}\label{Eqn: Thm: NuclearNRate: 11}
\hat{a} &= a-\frac{1}{N}\hat{B}(\hat{B}-BH)^{\prime}a-\left(I_{Np}-\frac{\hat{B}\hat{B}^{\prime}}{N}\right)(\hat{B}-BH)H^{-1}\frac{1}{T}F^{\prime}1_{T}\notag\\
&\hspace{0.5cm}+\left(I_{Np}-\frac{\hat{B}\hat{B}^{\prime}}{N}\right)\frac{1}{T}(\hat{\Pi} - \Pi)1_{T},
\end{align}
where we have used $a^{\prime}B=0$ and $\Pi = a 1_{T}^{\prime} + BF^{\prime}$. By Assumptions \ref{Ass: NuclearKaBF}(ii) and (iv), $\|F^{\prime}1_{T}/T\|$ and $\|a/\sqrt{N}\|$ are bounded. Thus, the second result follows from \eqref{Eqn: Thm: NuclearNRate: 11}, the second result, Lemma \ref{Lem: TechA1}(ii) and Theorem \ref{Thm: NuclearNRate}(i). By the definition of $\hat{F}$,
\begin{align}\label{Eqn: Thm: NuclearNRate: 12}
\hat{F} = F(H^{\prime})^{ -1} - F(H^{\prime})^{ -1}\frac{1}{N}(\hat{B}- BH)^{\prime}\hat{B} + \frac{1}{N}1_{T} a^{\prime}(\hat{B}-BH) + \frac{1}{N}(\hat{\Pi}-\Pi)^{\prime}\hat{B},
\end{align}
where we have used $a^{\prime}B=0$ and $\Pi = a 1_{T}^{\prime} + BF^{\prime}$. Thus, the last result follows from \eqref{Eqn: Thm: NuclearNRate: 12}, the second result, Lemma \ref{Lem: TechA1}(ii) and Theorem \ref{Thm: NuclearNRate}(i).\qed

\subsection{Technical Lemmas}
\begin{lem}\label{Lem: TechA1}
For any $Np \times T$ matrix $\Delta$, let $\mathcal{P}(\Delta)$ and $\mathcal{M}(\Delta)$ be given in Section \ref{Sec4}. Assume $0<K<\min\{Np,T\}-1$. For any $Np \times T$ matrix $\Delta$, the followings are true.\\
(i) $\|\Pi +\mathcal{P}(\Delta)\|_{\ast} = \|\Pi\|_{\ast} +\|\mathcal{P}(\Delta)\|_{\ast}$.\\
(ii) The rank of $\mathcal{M}(\Delta)$ is no greater than $2(K+1)$. \\
(iii) $\|\Delta\|_{F}^2 = \|\mathcal{P}(\Delta)\|_{F}^{2} + \|\mathcal{M}(\Delta)\|_{F}^2 $.
\end{lem} 
\noindent{\sc Proof:} (i) Since $\mathcal{P}(\Delta)= U_2U_2^{\prime}\Delta  V_2V_2^{\prime}$ and $\Pi = U_{1} \Sigma_{11} V_{1}^{\prime}$ where $\Sigma_{11}$ is square diagonal matrix with nonzero singular values of $\Pi$ in the diagonal in descending order, the result follows by Lemma 2.3 of \citet{Rechtetal_Guaranteed_2010}.

(ii) We have the following decomposition:
\begin{align}\label{Eqn: Lem: TechA1: 1}
\Delta &= U(U_1,U_2)^{\prime}\Delta(V_1,V_2)V^{\prime} \notag\\
&= U\left(
                                                           \begin{array}{cc}
                                                             U_{1}^{\prime}\Delta V_{1} & U_{1}^{\prime}\Delta V_{2} \\
                                                             U_{2}^{\prime}\Delta V_{1} & U_{2}^{\prime}\Delta V_{2} \\
                                                           \end{array}
                                                         \right)
V^{\prime}\notag\\
& = U\left(
                                                           \begin{array}{cc}
                                                             0 & 0 \\
                                                             0 & U_{2}^{\prime}\Delta V_{2} \\
                                                           \end{array}
                                                         \right)
V^{\prime}
+U\left(
                                                           \begin{array}{cc}
                                                             U_{1}^{\prime}\Delta V_{1} & U_{1}^{\prime}\Delta V_{2} \\
                                                             U_{2}^{\prime}\Delta V_{1} & 0 \\
                                                           \end{array}
                                                         \right)
V^{\prime}\notag\\
& = \mathcal{P}(\Delta) + U\left(
                                                           \begin{array}{cc}
                                                             U_{1}^{\prime}\Delta V_{1} & U_{1}^{\prime}\Delta V_{2} \\
                                                             U_{2}^{\prime}\Delta V_{1} & 0 \\
                                                           \end{array}
                                                         \right)
V^{\prime}.
\end{align}
Therefore, by \eqref{Eqn: Lem: TechA1: 1} we obtain
\begin{align}\label{Eqn: Lem: TechA1: 2}
\mathcal{M}(\Delta) =  U\left(
                                                           \begin{array}{cc}
                                                             U_{1}^{\prime}\Delta V_{1} & U_{1}^{\prime}\Delta V_{2} \\
                                                             U_{2}^{\prime}\Delta V_{1} & 0 \\
                                                           \end{array}
                                                         \right)
V^{\prime}.
\end{align}
Thus, by \eqref{Eqn: Lem: TechA1: 2} it follows that
\begin{align}\label{Eqn: Lem: TechA1: 3}
\mathrm{rank}(\mathcal{M}(\Delta)) &=  \mathrm{rank} \left(\left(
                                                           \begin{array}{cc}
                                                             U_{1}^{\prime}\Delta V_{1} & U_{1}^{\prime}\Delta V_{2} \\
                                                             U_{2}^{\prime}\Delta V_{1} & 0 \\
                                                           \end{array}
                                                         \right)\right)\notag\\
                                  &\leq \mathrm{rank} \left(\left(
                                                           \begin{array}{cc}
                                                             U_{1}^{\prime}\Delta V_{1} & U_{1}^{\prime}\Delta V_{2} \\
                                                             0 & 0 \\
                                                           \end{array}
                                                         \right)\right)
                                                         +\mathrm{rank} \left(\left(
                                                           \begin{array}{cc}
                                                             0& 0 \\
                                                             U_{2}^{\prime}\Delta V_{1} & 0 \\
                                                           \end{array}
                                                         \right)\right)\notag\\
                                                         &\leq 2(K+1),
\end{align}
where the first inequality follows by the fact that $\mathrm{rank}(C+D)\leq \mathrm{rank}(C) +\mathrm{rank}(D)$ (see, for example, Fact 2.10.17 in \citet{Bernstein_Scalar_2018}) and  the second inequality follows since $\Pi$ has at most rank $K+1$.

(iii) By \eqref{Eqn: Lem: TechA1: 1} and \eqref{Eqn: Lem: TechA1: 2}, we obtain
\begin{align}\label{Eqn: Lem: TechA1: 4}
\|\mathcal{P}(\Delta)\|_{F}^{2} + \|\mathcal{M}(\Delta)\|_{F}^2 &= \left\|\left(
                                                           \begin{array}{cc}
                                                             0 & 0 \\
                                                             0 & U_{2}^{\prime}\Delta V_{2} \\
                                                           \end{array}
                                                         \right)\right\|_{F}^{2} + \left\|\left(
                                                           \begin{array}{cc}
                                                             U_{1}^{\prime}\Delta V_{1} & U_{1}^{\prime}\Delta V_{2} \\
                                                             U_{2}^{\prime}\Delta V_{1} & 0 \\
                                                           \end{array}
                                                         \right)\right\|_{F}^{2}\notag\\
                                                         &=\|\Delta\|_{F}^{2},
\end{align}
where the second equality follows by the first two equalities in \eqref{Eqn: Lem: TechA1: 1}. This completes of the proof of the lemma.\qed
\begin{lem}\label{Lem: TechA2}
Suppose Assumption \ref{Ass: NuclearKaBF} holds. Let $V$ be a $K\times K$ diagonal matrix of the first $K$ largest eigenvalues of $\hat{\Pi}M_T\hat{\Pi}^{\prime}/(NT)$.  Assume that $\|\hat{\Pi}-\Pi\|_{F}=o_{p}(\sqrt{NT})$ and $P(\hat{K}= K) \to 1$. Then (i) $\|V\|_{2} = O_{p}(1)$, $\|V^{-1}\|_{2}=O_{p}(1)$, and $\|H\|_{2}=O_{p}(1)$; (ii) $\|H^{-1}\|_{2}=O_{p}(1)$, if $\|\hat{B} - BH\|_{F}=o_{p}(\sqrt{N})$.
\end{lem}
\noindent{\sc Proof:} (i) Let $\sigma_{j}(A)$ be the $j$th largest singular value of $A$. We have $\lambda_{j}(\hat{\Pi}M_{T}\hat{\Pi}^{\prime}/(NT)) = \sigma^{2}_{j}(\hat{\Pi}M_{T}/\sqrt{NT})$ and $\lambda_{j}(\Pi M_{T}\Pi^{\prime}/(NT)) = \sigma^{2}_{j}(\Pi M_{T}/\sqrt{NT})$.
By the triangle inequality, it follows from \eqref{Eqn: Thm: NuclearNRate: 7} that
\begin{align}\label{Eqn: Lem: TechA2: 1}
\hspace{-0.4cm}\sqrt{\|V\|_{2}} = \sigma_{1}(\hat{\Pi}M_{T}/\sqrt{NT})\leq \sigma_{1}(\Pi M_{T}/\sqrt{NT}) + \|\hat{\Pi}-\Pi\|_{F}/\sqrt{NT}=O_{p}(1),
\end{align}
where the last equality follows since $\sigma_{1}(\Pi M_{T}/\sqrt{NT})$ is bounded. Similarly,
\begin{align}\label{Eqn: Lem: TechA2: 2}
\sqrt{\|V^{-1}\|_{2}} = \sigma^{-1}_{K}(\hat{\Pi}M_{T}/\sqrt{NT})\leq \sigma^{-1}_{K}(\Pi M_{T}/\sqrt{NT}) +o_{p}(1) =O_{p}(1),
\end{align}
where the last equality follows since $\sigma^{2}_{K}(\Pi M_{T}/\sqrt{NT})\hspace{-0.1cm}=\hspace{-0.1cm}\lambda_{\min}((B^{\prime}B/N)(F^{\prime} M_{T}F/T))$ $>d_{\min}^{2}$. Let $H^{\diamond}\equiv(F^{\prime}M_TF/T)(B^{\prime}\hat{B}/N)V^{-1}$. Recall that $H = (F^{\prime}M_T\hat{\Pi}^{\prime}\hat{B}/T)V^{-1}$. Then,
\begin{align}\label{Eqn: Lem: TechA2: 3}
\|H-H^{\diamond}\|_{2}\leq\frac{1}{NT}\|F\|_{2}\|\hat{\Pi}-\Pi\|_{F}\|\hat{B}\|_2\|V^{-1}\|_2=o_{p}(1),
\end{align}
where the equality follows by Assumption \ref{Ass: NuclearKaBF}(ii). Since $\|H^{\diamond}\|_{2}=O_{p}(1)$, it follows from \eqref{Eqn: Lem: TechA2: 3} that $\|H\|_{2}=O_{p}(1)$.

(ii) Since $\|\hat{B} - BH\|_{F}=o_{p}(\sqrt{N})$, we have $\|\hat{B}^{\prime}\hat{B}/(N)-H^{\prime}(B^{\prime}B/N)H\|_{F} = o_{p}(1)$ by the triangle inequality. This implies that $I_{K}-\lambda_{\max}(B^{\prime}B/N)H^{\prime}H$ is negative semidefinite with probability approaching one. Therefore, the eigenvalues of $H^{\prime}H$ are no smaller than $\lambda^{-1}_{\max}(B^{\prime}B/N)$ with probability approaching one. Thus, the result of the lemma follows from Assumption \ref{Ass: NuclearKaBF}(i). This completes the proof of the lemma. \qed

\section{Proof of Corollary \ref{Cor: Ex135: 1}}\label{App: B}
\renewcommand{\theequation}{B.\arabic{equation}}
\setcounter{equation}{0}
\noindent{\sc Proof of Corollary \ref{Cor: Ex135: 1}:} We have $\mathcal{S}\ominus \mathcal{S}=\mathbf{R}^{Np\times T}$. Utilizing the fact that $|\mathrm{tr}(C^{\prime}D)|\leq \|C\|_{2}\|D\|_{\ast}$\footnote{See, for example, Fact 11.14.1 in \citet{Bernstein_Scalar_2018}.}, we obtain that for any $\Delta\in \mathcal{S}\ominus \mathcal{S}$,
\begin{align}\label{Eqn: Ex135: 1}
\left|\sum_{i=1}^{N}\sum_{t=1}^{T}\mathrm{tr}(\varepsilon_{it} X_{it}^{\prime}\Delta)\right|\leq \left\|\sum_{i=1}^{N}\sum_{t=1}^{T} X_{it}\varepsilon_{it}\right\|_{2}\|\Delta\|_{\ast}.
\end{align}
Thus, by Assumption \ref{Ass: Ex135: 1}(ii) and Lemma \ref{Lem: UseD1}(i), Assumption \ref{Ass: NuclearN}(ii) is satisfied with $r_{NT} = O_{p}(\max\{\sqrt{Np},\sqrt{T}\})$ as $(N,T)\to\infty$. When $x_{it}=1$, Assumption \ref{Ass: NuclearN}(i) is trivially satisfied with $\mathcal{L}_{NT}(\cdot) = 0$ and $\kappa =1$. Otherwise, by Assumption \ref{Ass: Ex135: 1}(i), Assumption \ref{Ass: NuclearN}(i) is satisfied with $\mathcal{L}_{NT}(\cdot) = 0$. When $a=0$, Assumptions \ref{Ass: NuclearKaBF}(iv) and (v) are trivially satisfied. \qed

\subsection{Technical Lemmas}
Recall that $X_{it}= (e_{N,i}\otimes x_{it}) e_{T,t}^{\prime}$ be an $Np\times T$ matrix of $x_{it}$, where $e_{N,i}$ is the $i$th column of $I_N$ and $e_{T,t}$ is the $t$th column of $I_T$.
\begin{lem}\label{Lem: UseD1}
(i) Let $\{\xi_{Nt}\}_{t\leq T}$ be a sequence of independent $Np\times 1$ sub-Gaussian vectors with $\lambda_{\max}(E[\xi_{Nt}\xi_{Nt}^{\prime}])$ bounded. Assume that $(x_{1t}^{\prime}e_{1t},x_{2t}^{\prime}e_{2t},\ldots,x_{Nt}^{\prime}e_{Nt})^{\prime}$ is the $t$th column of $\Xi_{NT}\Omega_{NT}$, where $\Xi_{NT} = (\xi_{N1},\xi_{N2},\ldots, \xi_{NT})$ and $\Omega_{NT}$ is a $T\times T$ deterministic (possibly non-diagonal) matrix with $\|\Omega_{NT}\|_{2}$ bounded. Then as $(N,T)\to\infty$,
\[\left\|\sum_{i=1}^{N}\sum_{t=1}^{T} X_{it}\varepsilon_{it}\right\|_{2} = O_{p}(\max\{\sqrt{Np},\sqrt{T}\}).\]
(ii) Let $\{\nu_{Nt}\}_{t\leq T}$ be a sequence of independent $Np\times 1$ sub-Gaussian vectors with bounded $\lambda_{\max}(E[\nu_{Nt}\nu_{Nt}^{\prime}])$. Assume that $(x_{1t}^{\prime},x_{2t}^{\prime},\ldots,x_{Nt}^{\prime})^{\prime}$ is the $t$th column of $\mathcal{V}_{NT}\Omega_{NT}$, where $\mathcal{V}_{NT} = (\nu_{N1},\nu_{N2},\ldots, \nu_{NT})$ and $\Omega_{NT}$ is a $T\times T$ deterministic (possibly non-diagonal) matrix with $\|\Omega_{NT}\|_{2}$ bounded. Then as $(N,T)\to\infty$,
\[\left\|\sum_{i=1}^{N}\sum_{t=1}^{T} X_{it}\right\|_{2} = O_{p}(\max\{\sqrt{Np},\sqrt{T}\}).\]
(iii) Let $\{\eta_{Nt}\}_{t\leq T}$ be a sequence of independent $p\times 1$ sub-Gaussian vectors with bounded $\lambda_{\max}(E[\eta_{Nt}\eta_{Nt}^{\prime}])$. Assume that $\sum_{i=1}^{N}x_{it}e_{it}/\sqrt{N}$ is the $t$th column of $\Upsilon_{NT}\Omega_{NT}$, where $\Upsilon_{NT}\equiv (\eta_{N1},\eta_{N2}, \ldots, \eta_{NT})$ and $\Omega_{NT}$ is a $T\times T$ deterministic (possibly non-diagonal) matrix with $\|\Omega_{NT}\|_{2}$ bounded. Then as $(N,T)\to\infty$,
\[\left\|\left(\frac{1}{\sqrt{N}}\sum_{i=1}^{N}x_{i1}e_{i1},\frac{1}{\sqrt{N}}\sum_{i=1}^{N}x_{i2}e_{i2},\ldots, \frac{1}{\sqrt{N}}\sum_{i=1}^{N}x_{iT}e_{iT}\right)\right\|_{2} = O_{p}(\max\{\sqrt{p},\sqrt{T}\}).\]
\end{lem}
\noindent{\sc Proof:} (i) Since $(x_{1t}^{\prime}e_{1t},x_{2t}^{\prime}e_{2t},\ldots,x_{Nt}^{\prime}e_{Nt})^{\prime}$ is the $t$th column of $\sum_{i=1}^{N}\sum_{t=1}^{T}X_{it}e_{it}$,
\begin{align}\label{Eqn: Lem: UseD1: 1}
\sum_{i=1}^{N}\sum_{t=1}^{T}X_{it}\varepsilon_{it} =  \Xi_{NT}\Omega_{NT}.
\end{align}
Applying Theorem 5.39 and Remark 5.40 in \citet{Vershynin_nonasymptotic_2010} on $\Xi_{NT}^{\prime}$, we obtain $\|\Xi_{NT}\|_{2} = O_{p}(\max\{\sqrt{Np},\sqrt{T}\})$ as $(N,T)\to\infty$. Thus, the result follows by \eqref{Eqn: Lem: UseD1: 1} since $\|\Omega_{NT}\|_{2}$ is bounded and $\|CD\|_{2}\leq \|C\|_{2}\|D\|_{2}$.

(ii) and (iii) The proof is similar to the proof of (i), thus omitted. \qed

\section{Proof of Corollary \ref{Cor: Ex2: 1}}\label{App: C}
\renewcommand{\theequation}{C.\arabic{equation}}
\setcounter{equation}{0}
\noindent{\sc Proof of Corollary \ref{Cor: Ex2: 1}:} Clearly, $\mathcal{S} = \mathcal{D}_M$ is convex in $\mathbf{R}^{Np\times T}$, and $\mathcal{S}\ominus \mathcal{S} =\mathcal{D}_{2M}$. We verify Assumptions \ref{Ass: NuclearN} and \ref{Ass: NuclearKaBF}. By Assumption \ref{Ass: Eqn: Ex2: 1}(ii), $\Pi\in \mathcal{S}$. By Lemma \ref{Lem: UseD4}, for any $\Delta\in \mathcal{S}\ominus \mathcal{S}$, there exists $\mathcal{R}_{NT}(\cdot)$ such that
\begin{align}\label{Eqn: Ex2: 2}
\sum_{i=1}^{N}\sum_{t=1}^{T}|\mathrm{tr}(X_{it}^{\prime}\Delta)|^2\geq \min\left\{1,\min_{t\leq T}\lambda_{\min}\left(\frac{\sum_{i=1}^{N}x^{\ast}_{it}x_{it}^{\ast\prime}}{N}\right) \right\}\|\Delta\|_{F}^{2}+2\mathcal{R}_{NT}(\Delta),
\end{align}
\begin{align}\label{Eqn: Ex2: 3}
\left|\mathcal{R}_{NT}(\Delta)\right|\leq 2M\sqrt{p-1}\left\|\left(
                                                              \begin{array}{cccc}
                                                                x_{11}^{\ast} & x_{12}^{\ast} & \cdots & x_{1T}^{\ast} \\
                                                                x_{21}^{\ast} & x_{22}^{\ast} & \cdots & x_{2T}^{\ast} \\
                                                                \vdots & \vdots & \vdots & \vdots \\
                                                                x_{N1}^{\ast} & x_{N2}^{\ast} & \cdots & x_{NT}^{\ast} \\
                                                              \end{array}
                                                            \right)
\right\|_{2}\|\Delta\|_{\ast},
\end{align}
and
\begin{align}\label{Eqn: Ex2: 4}
\left|\sum_{i=1}^{N}\sum_{t=1}^{T}\mathrm{tr}(\varepsilon_{it} X_{it}^{\prime}\Delta)\right|&\leq \left(\hspace{-0.2cm}
                                                                                                          \begin{array}{c}
                                                                                                             \vspace{0.35cm}\\
                                                                                                            \left\|\left(\frac{1}{\sqrt{N}}\sum_{i=1}^{N}x^{\ast}_{i1}\varepsilon_{i1},\frac{1}{\sqrt{N}}\sum_{i=1}^{N}x^{\ast}_{i2}\varepsilon_{i2},\ldots, \frac{1}{\sqrt{N}}\sum_{i=1}^{N}x^{\ast}_{iT}\varepsilon_{iT}\right)\right\|_{2} \\
                                                                                                             \vspace{0.35cm}\\
                                                                                                          \end{array}
                                                                                                        \right.\notag\\
&\hspace{0.5cm} + \left.\left\|\left(
                                                              \begin{array}{cccc}
                                                                \varepsilon_{11} & \varepsilon_{12} & \cdots & \varepsilon_{1T}\\
                                                                \varepsilon_{21} & \varepsilon_{22} & \cdots & \varepsilon_{2T}\\
                                                                \vdots & \vdots & \vdots & \vdots \\
                                                                \varepsilon_{N1} & \varepsilon_{N2} & \cdots & \varepsilon_{NT}\\
                                                              \end{array}
                                                            \right)
\right\|_{2}\right)\|\Delta\|_{\ast}.
\end{align}
By \eqref{Eqn: Ex2: 2}, \eqref{Eqn: Ex2: 3}, Assumption \ref{Ass: Eqn: Ex2: 1}(iv), and Lemma \ref{Lem: UseD1}(ii), if $\min_{t\leq T}\lambda_{\min}(\sum_{i=1}^{N}x^{\ast}_{it}x_{it}^{\ast\prime}/N)\geq c_{\min}$ for some constant $0<c_{\min}<\infty$, then Assumption \ref{Ass: NuclearN}(i) is satisfied with $\mathcal{L}_{NT}(\cdot) = 2\mathcal{R}_{NT}(\cdot)$, $\kappa = \min\{1,c_{\min}\}$, and $r_{NT} = O_{p}(M\sqrt{p}\max\{\sqrt{Np}, \sqrt{T}\})$ as $(N,T)\to\infty$. By Assumption \ref{Ass: Eqn: Ex2: 1}(i), the condition holds with probability approaching one as $N\to\infty$. As discussed below Theorem \ref{Thm: NuclearNRate}, this is sufficient for us to establish a rate of convergence of $\hat{\Pi}$. Note that  $\varepsilon_{it} = e_{it} + d_{it}$ where $d_{it} = \delta(z_{it}) + \Delta(z_{it})^{\prime}f_t$, it follows that
\begin{align}\label{App: Eqn: Ex2: 3}
&\left\|\left(\frac{1}{\sqrt{N}}\sum_{i=1}^{N}x^{\ast}_{i1}\varepsilon_{i1},\frac{1}{\sqrt{N}}\sum_{i=1}^{N}x^{\ast}_{i2}\varepsilon_{i2},\ldots, \frac{1}{\sqrt{N}}\sum_{i=1}^{N}x^{\ast}_{iT}\varepsilon_{iT}\right)\right\|_{2}\notag\\
&\hspace{-0.5cm}\leq \left\|\left(\frac{1}{\sqrt{N}}\sum_{i=1}^{N}x^{\ast}_{i1}e_{i1},\frac{1}{\sqrt{N}}\sum_{i=1}^{N}x^{\ast}_{i2}e_{i2},\ldots, \frac{1}{\sqrt{N}}\sum_{i=1}^{N}x^{\ast}_{iT}e_{iT}\right)\right\|_{2} + \sqrt{c_{\max}\sum_{t=1}^{T}\sum_{i=1}^{N}|d_{it}|^{2}},
\end{align}
where the last inequality holds with probability approaching one by Assumption \ref{Ass: Eqn: Ex2: 1}(i) and the fact that $\|A\|_{2}\leq \|A\|_{F}$. Similarly,
\begin{align}\label{App: Eqn: Ex2: 4}
\left\|\left(
                                                              \begin{array}{cccc}
                                                                \varepsilon_{11} & \varepsilon_{12} & \cdots & \varepsilon_{1T}\\
                                                                \varepsilon_{21} & \varepsilon_{22} & \cdots & \varepsilon_{2T}\\
                                                                \vdots & \vdots & \vdots & \vdots \\
                                                                \varepsilon_{N1} & \varepsilon_{N2} & \cdots & \varepsilon_{NT}\\
                                                              \end{array}
                                                            \right)
\right\|_{2}
\leq \left\|\left(
                                                              \begin{array}{cccc}
                                                                e_{11} & e_{12} & \cdots & e_{1T}\\
                                                                e_{21} & e_{22} & \cdots & e_{2T}\\
                                                                \vdots & \vdots & \vdots & \vdots \\
                                                                e_{N1} & e_{N2} & \cdots & e_{NT}\\
                                                              \end{array}
                                                            \right)
\right\|_{2} +  \sqrt{\sum_{t=1}^{T}\sum_{i=1}^{N}|d_{it}|^{2}}.
\end{align}
By \eqref{Eqn: Ex2: 4}-\eqref{App: Eqn: Ex2: 4}, Assumptions \ref{Ass: Eqn: Ex2: 1}(iii), (v), (vi), \ref{Ass: Eqn: Ex2: 2}(iv), and Lemmas \ref{Lem: UseD1}(i) and (iii), Assumption \ref{Ass: NuclearN}(ii) is satisfied with $r_{NT} =O_{p}(\max\{\sqrt{N+p},\sqrt{T}\} + \sqrt{NT}p^{-s})$ as $(N,T)\to\infty$. It is easy to see that Assumption \ref{Ass: NuclearKaBF} holds by Assumption \ref{Ass: Eqn: Ex2: 2}.\qed

\subsection{Technical Lemmas}
Recall that $x_{it} = (1,x_{it}^{\ast\prime})^{\prime}$ and $X_{it}= (e_{N,i}\otimes x_{it}) e_{T,t}^{\prime}$ be an $Np\times T$ matrix of $x_{it}$, where $e_{N,i}$ is the $i$th column of $I_N$ and $e_{T,t}$ is the $t$th column of $I_T$.
\begin{lem}\label{Lem: UseD4}
Let $\mathcal{X}^{\ast}$ be an $N\times T$ block matrix with the $it$th block $x_{it}^{\ast}$, $\mathcal{E}$ be an $N\times T$ matrix with the $it$th entry $\varepsilon_{it}$, and $\mathcal{F}^{\ast}\equiv (\sum_{i=1}^{N}x^{\ast}_{i1}\varepsilon_{i1}/\sqrt{N},$ $\sum_{i=1}^{N}x^{\ast}_{i2}\varepsilon_{i2}/\sqrt{N}, \ldots, \sum_{i=1}^{N}x^{\ast}_{iT}\varepsilon_{iT}/\sqrt{N})$. For any $\Delta\in\mathcal{D}_{M}$ given in \eqref{Eqn: Ex2: 1}, we have
\[\sum_{i=1}^{N}\sum_{t=1}^{T}|\mathrm{tr}(X_{it}^{\prime}\Delta)|^2\geq \min\left\{1,\min_{t\leq T}\lambda_{\min}\left(\frac{\sum_{i=1}^{N}x^{\ast}_{it}x_{it}^{\ast\prime}}{N}\right) \right\}\|\Delta\|_{F}^{2}+2\mathcal{R}_{NT}(\Delta)\]
for some $\mathcal{R}_{NT}(\Delta)$ such that $|\mathcal{R}_{NT}(\Delta)|\leq M\sqrt{p-1}\|\mathcal{X}^{\ast}\|_{2}\|\Delta\|_{\ast}$, and
\[\left|\sum_{i=1}^{N}\sum_{t=1}^{T}\mathrm{tr}(\varepsilon_{it} X_{it}^{\prime}\Delta)\right|\leq (\|\mathcal{E}\|_{2}+\|\mathcal{F}^{\ast}\|_{2})\|\Delta\|_{\ast}.\]
\end{lem}
\noindent{\sc Proof:} Fix $\Delta = ((\gamma_1,\Gamma^{\ast\prime}),(\gamma_2,\Gamma^{\ast\prime}), (\gamma_N,\Gamma^{\ast\prime}))^{\prime}\in\mathcal{D}_{M}$ for some $(\gamma_1,\gamma_2,\ldots,\gamma_N)^{\prime}\in\mathbf{R}^{N\times T}$ and $\Gamma^{\ast}\in\mathbf{R}^{(p-1)\times T}$. Write  $\gamma_i = (\gamma_{i1},\gamma_{i2},\ldots, \gamma_{iT})^{\prime}$ and $\Gamma^{\ast} = (\gamma^{\ast}_1,\gamma^{\ast}_2,\ldots, \gamma^{\ast}_{T})$, where $\gamma_{it}$ is a scalar and $\gamma^{\ast}_{t}$ is a $(p-1)\times 1$ vector. Since $x_{it} = (1,x_{it}^{\ast})^{\prime}$, it follows that $\mathrm{tr}(X_{it}^{\prime}\Delta) = \gamma_{it} + x_{it}^{\ast\prime}\gamma_{t}^{\ast}$ and then
\begin{align}\label{Eqn: Lem: UseD4: 1}
&\hspace{0.5cm}\sum_{i=1}^{N}\sum_{t=1}^{T}|\mathrm{tr}(X_{it}^{\prime}\Delta)|^2= \sum_{i=1}^{N}\sum_{t=1}^{T}(\gamma_{it} + x_{it}^{\ast\prime}\gamma_{t}^{\ast})^2\notag\\
&=\sum_{i=1}^{N}\sum_{t=1}^{T} \gamma_{it}^{2} + N\sum_{t=1}^{T}\gamma_t^{\ast\prime}\left(\frac{\sum_{i=1}^{N}x^{\ast}_{it}x_{it}^{\ast\prime}}{N}\right)\gamma^{\ast}_t + 2\sum_{i=1}^{N}\sum_{t=1}^{T} \gamma_{it} x_{it}^{\ast\prime}\gamma_{t}^{\ast}\notag\\
&\geq \min\left\{1,\min_{t\leq T}\lambda_{\min}\left(\frac{\sum_{i=1}^{N}x^{\ast}_{it}x_{it}^{\ast\prime}}{N}\right) \right\}\left(\sum_{i=1}^{N}\sum_{t=1}^{T} \gamma_{it}^{2}  + N \|\Gamma^{\ast}\|_{F}^{2}\right)+ 2\sum_{i=1}^{N}\sum_{t=1}^{T} \gamma_{it} x_{it}^{\ast\prime}\gamma_{t}^{\ast}\notag\\
&= \min\left\{1,\min_{t\leq T}\lambda_{\min}\left(\frac{\sum_{i=1}^{N}x^{\ast}_{it}x_{it}^{\ast\prime}}{N}\right) \right\}\|\Delta\|_{F}^{2}+ 2\sum_{i=1}^{N}\sum_{t=1}^{T} \gamma_{it} x_{it}^{\ast\prime}\gamma_{t}^{\ast},
\end{align}
where the last equality holds since $\|\Delta\|_{F}^{2} = \sum_{i=1}^{N}\sum_{t=1}^{T} \gamma_{it}^{2} + N\|\Gamma^{\ast}\|_{F}^{2}$. Write $x_{it}^{\ast} = (x_{it,1}^{\ast},x_{it,2}^{\ast},$ $\ldots,x_{it,p-1}^{\ast})^{\prime}$ and $\gamma_{t}^{\ast} = (\gamma_{1t}^{\ast},\gamma_{2t}^{\ast},\ldots, \gamma_{(p-1)t}^{\ast})$. Let $\Gamma^{\diamond}\equiv (\gamma_1,\gamma_2,\ldots,\gamma_N)^{\prime}$,  $\Gamma_{j}^{\dagger}\equiv \Gamma^{\diamond} \mathrm{diag}(\gamma_{j1}^{\ast},\gamma_{j2}^{\ast},\ldots, \gamma_{jT}^{\ast})$, and $X_{j}^{\ast}$ be an $N\times T$ matrix with the $it$th entry $x_{it,j}^{\ast}$. Write $\Gamma^{\diamond} = (\zeta_1, \zeta_2, \ldots, \zeta_T)$. It follows that
\begin{align}\label{Eqn: Lem: UseD4: 2}
\sum_{i=1}^{N}\sum_{t=1}^{T} \gamma_{it} x_{it}^{\ast\prime}\gamma_{t}^{\ast} &= \sum_{j=1}^{p-1}\sum_{i=1}^{N}\sum_{t=1}^{T}\gamma_{it} x_{it,j}^{\ast}\gamma_{jt}^{\ast}\notag\\
& = \sum_{j=1}^{p-1} \mathrm{tr}(X_{j}^{\ast\prime}\Gamma_{j}^{\dagger})\notag\\
& = \mathrm{tr}\left(\left(
                                        \begin{array}{c}
                                          X_{1}^{\ast}\\
                                          X_{2}^{\ast} \\
                                          \vdots \\
                                          X_{p-1}^{\ast} \\
                                        \end{array}
                                      \right)^{\prime}
                                      \left(
                                        \begin{array}{c}
                                          \Gamma_{1}^{\dagger}\\
                                          \Gamma_{2}^{\dagger} \\
                                          \vdots \\
                                          \Gamma_{p-1}^{\dagger} \\
                                        \end{array}
                                      \right)
\right)\notag\\
&\leq \left\|\left(
                                        \begin{array}{c}
                                          X_{1}^{\ast}\\
                                          X_{2}^{\ast} \\
                                          \vdots \\
                                          X_{p-1}^{\ast} \\
                                        \end{array}
                                      \right)\right\|_{2}\left\|\left(
                                        \begin{array}{c}
                                          \Gamma_{1}^{\dagger}\\
                                          \Gamma_{2}^{\dagger} \\
                                          \vdots \\
                                          \Gamma_{p-1}^{\dagger} \\
                                        \end{array}
                                      \right)\right\|_{\ast}\notag\\
&= \|\mathcal{X}^{\ast}\|_{2}\left\|\left(
                                        \begin{array}{c}
                                          \Gamma_{1}^{\dagger}\\
                                          \Gamma_{2}^{\dagger} \\
                                          \vdots \\
                                          \Gamma_{p-1}^{\dagger} \\
                                        \end{array}
                                      \right)\right\|_{\ast}\notag\\
& \leq \max_{j\leq p-1,t\leq T}|\gamma_{jt}^{\ast}|\sum_{j=1}^{p-1}\sqrt{p-1}\|\mathcal{X}^{\ast}\|_{2} \|\Gamma_{j}^{\dagger}\|_{\ast},
\end{align}
where the first inequality holds by the fact that $|\mathrm{tr}(C^{\prime}D)|\leq \|C\|_{2}\|D\|_{\ast}$, the fourth equality holds since $\mathcal{X}^{\ast}$ and $(X_{1}^{\ast\prime},X_{2}^{\ast\prime},\ldots, X_{p-1}^{\ast\prime})^{\prime}$ share a common set of nonzero singular values, the last inequality follows since the nonzero singular values of $(\Gamma_{1}^{\dagger\prime}, \Gamma_{2}^{\dagger\prime}, \ldots, \Gamma_{p-1}^{\dagger\prime})^{\prime}$ are given by the square root of the nonzero eigenvalues of
\begin{align}\label{Eqn: Lem: UseD4: 3}
&\hspace{0.5cm}(\Gamma_{1}^{\dagger\prime}, \Gamma_{2}^{\dagger\prime}, \ldots, \Gamma_{p-1}^{\dagger\prime}) \left(
                                        \begin{array}{c}
                                          \Gamma_{1}^{\dagger}\\
                                          \Gamma_{2}^{\dagger} \\
                                          \vdots \\
                                          \Gamma_{p-1}^{\dagger} \\
                                        \end{array}
                                      \right)= \sum_{j=1}^{p-1}\Gamma_{j}^{\dagger\prime}\Gamma_{j}^{\dagger}\notag\\
                                      & = \sum_{j=1}^{p-1}\left(
                                                                                 \begin{array}{cccc}
                                                                                   \gamma_{j1}^{\ast} & 0 & \cdots & 0 \\
                                                                                   0 & \gamma_{j2}^{\ast} & \cdots & 0 \\
                                                                                   \vdots & \vdots & \vdots & \vdots \\
                                                                                   0 & 0 & \cdots & \gamma_{jT}^{\ast} \\
                                                                                 \end{array}
                                                                               \right)\Gamma^{\diamond\prime}\Gamma^{\diamond}\left(
                                                                                 \begin{array}{cccc}
                                                                                   \gamma_{j1}^{\ast} & 0 & \cdots & 0 \\
                                                                                   0 & \gamma_{j2}^{\ast} & \cdots & 0 \\
                                                                                   \vdots & \vdots & \vdots & \vdots \\
                                                                                   0 & 0 & \cdots & \gamma_{jT}^{\ast} \\
                                                                                 \end{array}
                                                                               \right)\notag\\
                                &  = \sum_{j=1}^{p-1}\sum_{t=1}^{T}\gamma^{\ast2}_{jt}\zeta_{t}\zeta_{t}^{\prime}\preceq  \max_{j\leq p-1,t\leq T}|\gamma_{jt}^{\ast}|^2 \sum_{j=1}^{p-1}\sum_{t=1}^{T}\zeta_{t}\zeta_{t}^{\prime} = (p-1)\max_{j\leq p-1,t\leq T}|\gamma_{jt}^{\ast}|^2 \Gamma^{\diamond\prime}\Gamma^{\diamond},
\end{align}
and ``$C\preceq D$'' means that $D-C$ is positive semi-definite. Thus, the first result of the lemma follows from \eqref{Eqn: Lem: UseD4: 1} and \eqref{Eqn: Lem: UseD4: 2} by letting $\mathcal{R}_{NT}(\Delta) = \sum_{i=1}^{N}\sum_{t=1}^{T} \gamma_{it} x_{it}^{\ast\prime}\gamma_{t}^{\ast}$. Since $\mathrm{tr}(X_{it}^{\prime}\Delta) = \gamma_{it} + x_{it}^{\ast\prime}\gamma_{t}^{\ast}$,
\begin{align}\label{Eqn: Lem: UseD4: 4}
\sum_{i=1}^{N}\sum_{t=1}^{T}\mathrm{tr}(\varepsilon_{it} X_{it}^{\prime}\Delta)&= \sum_{i=1}^{N}\sum_{t=1}^{T}\varepsilon_{it} \gamma_{it} + \sum_{i=1}^{N}\sum_{t=1}^{T}\varepsilon_{it}x_{it}^{\ast\prime}\gamma^{\ast}_{t} \notag\\
&= \mathrm{tr}( \mathcal{E}^{\prime} \Gamma ) + \mathrm{tr}\left( \mathcal{F}^{\ast\prime} \sqrt{N}\Gamma^{\ast} \right)\notag\\
& \leq \|\mathcal{E}\|_{2}\|\Gamma\|_{\ast} + \|\mathcal{F}^{\ast}\|_{2}\sqrt{N}\|\Gamma^{\ast}\|_{\ast}\notag\\
&\leq (\|\mathcal{E}\|_{2}+\|\mathcal{F}^{\ast}\|_{2})\left\|\left(
                                                         \begin{array}{c}
                                                           \Gamma \\
                                                           \sqrt{N}\Gamma^{\ast} \\
                                                         \end{array}
                                                       \right)
\right\|\notag\\
& = (\|\mathcal{E}\|_{2}+\|\mathcal{F}^{\ast}\|_{2})\|\Delta\|_{\ast},
\end{align}
where the first inequality holds by the fact that $|\mathrm{tr}(C^{\prime}D)|\leq \|C\|_{2}\|D\|_{\ast}$, the second inequality follows since $\|\Gamma\|_{\ast}\leq \|(\Gamma^{\prime}, \sqrt{N}\Gamma^{\ast\prime})^{\prime}\|_{\ast}$ and $\sqrt{N}\|\Gamma^{\ast}\|_{\ast}\leq \|(\Gamma^{\prime}, \sqrt{N}\Gamma^{\ast\prime})^{\prime}\|_{\ast}$, and the last equality follows by Lemma \ref{Lem: TechD2}(iii). This completes the proof of the lemma. \qed

\section{Proof of Corollary \ref{Cor: Ex4: 1}}\label{App: D}
\renewcommand{\theequation}{E.\arabic{equation}}
\setcounter{equation}{0}
\noindent{\sc Proof of Corollary \ref{Cor: Ex4: 1}:} Clearly, $\mathcal{S} = \{1_{N}\otimes \Gamma: \Gamma\in \mathbf{R}^{p\times T}\}$ is convex in $\mathbf{R}^{Np\times T}$, and $\mathcal{S}\ominus \mathcal{S} =\mathcal{S}$. We verify Assumptions \ref{Ass: NuclearN} and \ref{Ass: NuclearKaBF}. By Lemma \ref{Lem: UseD2}, for any $\Delta\in \mathcal{S}\ominus \mathcal{S}$,
\begin{align}\label{Eqn: Ex4: 1}
\sum_{i=1}^{N}\sum_{t=1}^{T}|\mathrm{tr}(X_{it}^{\prime}\Delta)|^2\leq \min_{t\leq T}\lambda_{\min}\left(\frac{\sum_{i=1}^{N}x_{it}x_{it}^{\prime}}{N}\right)\|\Delta\|_{F}^{2}
\end{align}
and
\begin{align}\label{Eqn: Ex4: 2}
\hspace{-0.3cm}\left|\sum_{i=1}^{N}\sum_{t=1}^{T}\hspace{-0.1cm}\mathrm{tr}(\varepsilon_{it} X_{it}^{\prime}\Delta)\right|\hspace{-0.05cm}\leq \hspace{-0.05cm}\left\|\hspace{-0.05cm}\left(\hspace{-0.05cm}\frac{1}{\sqrt{N}}\hspace{-0.1cm}\sum_{i=1}^{N}x_{i1}\varepsilon_{i1},\frac{1}{\sqrt{N}}\hspace{-0.1cm}\sum_{i=1}^{N}x_{i2}\varepsilon_{i2},\ldots, \frac{1}{\sqrt{N}}\hspace{-0.1cm}\sum_{i=1}^{N}x_{iT}\varepsilon_{iT}\hspace{-0.05cm}\right)\hspace{-0.05cm}\right\|_{2}\hspace{-0.2cm}\|\Delta\|_{\ast}.
\end{align}
In view of \eqref{Eqn: Ex4: 1}, if $\min_{t\leq T}\lambda_{\min}(\sum_{i=1}^{N}x_{it}x_{it}^{\prime}/N)\geq c_{\min}$ for some constant $0<c_{\min}<\infty$, then Assumption \ref{Ass: NuclearN}(i) is satisfied with $\mathcal{L}_{NT}(\cdot) = 0$ and $\kappa = c_{\min}$. By Assumption \ref{Ass: Eqn: Ex4: 1}(i), the condition holds with probability approaching one as $N\to\infty$ with fixed $T$ or as $(N,T)\to\infty$. As discussed below Theorem \ref{Thm: NuclearNRate}, this is sufficient for us to establish a rate of convergence of $\hat{\Pi}$. Note that  $\varepsilon_{it} = e_{it} + d_{it}$ where $d_{it} = \delta(z_{it}) + \Delta(z_{it})^{\prime}f_t$, it follows that
\begin{align}\label{App: Eqn: Ex4: 3}
&\left\|\left(\frac{1}{\sqrt{N}}\sum_{i=1}^{N}x_{i1}\varepsilon_{i1},\frac{1}{\sqrt{N}}\sum_{i=1}^{N}x_{i2}\varepsilon_{i2},\ldots, \frac{1}{\sqrt{N}}\sum_{i=1}^{N}x_{iT}\varepsilon_{iT}\right)\right\|_{2}\notag\\
&\hspace{-0.5cm}\leq \left\|\left(\frac{1}{\sqrt{N}}\sum_{i=1}^{N}x_{i1}e_{i1},\frac{1}{\sqrt{N}}\sum_{i=1}^{N}x_{i2}e_{i2},\ldots, \frac{1}{\sqrt{N}}\sum_{i=1}^{N}x_{iT}e_{iT}\right)\right\|_{2} + \sqrt{c_{\max}\sum_{t=1}^{T}\sum_{i=1}^{N}|d_{it}|^{2}},
\end{align}
where the last inequality holds with probability approaching one by Assumption \ref{Ass: Eqn: Ex4: 1}(i) and the fact that $\|A\|_{2}\leq \|A\|_{F}$. By \eqref{Eqn: Ex4: 2}, \eqref{App: Eqn: Ex4: 3}, Assumption \ref{Ass: Eqn: Ex4: 1} (ii), (iv), and \ref{Ass: Eqn: Ex4: 2}(ii), Assumption \ref{Ass: NuclearN}(ii) is trivially satisfied with $r_{NT} = O_{p}(\sqrt{p}+\sqrt{N}p^{-s})$ as $N\to\infty$ with fixed $T$. Alternatively, by \eqref{Eqn: Ex4: 2}, \eqref{App: Eqn: Ex4: 3}, Assumption \ref{Ass: Eqn: Ex4: 1}(iii), (iv), and \ref{Ass: Eqn: Ex4: 2}(ii), Assumption \ref{Ass: NuclearN}(ii) is satisfied with $r_{NT} =O_{p}(\max\{\sqrt{p},\sqrt{T}\} + \sqrt{NT}p^{-s})$ as $(N,T)\to\infty$; see Lemma \ref{Lem: UseD1}(iii). It is easy to see that Assumption \ref{Ass: NuclearKaBF} holds by Assumption \ref{Ass: Eqn: Ex4: 2}. \qed

\subsection{Technical Lemmas}
Recall that $X_{it}= (e_{N,i}\otimes x_{it}) e_{T,t}^{\prime}$ be an $Np\times T$ matrix of $x_{it}$, where $e_{N,i}$ is the $i$th column of $I_N$ and $e_{T,t}$ is the $t$th column of $I_T$.
\begin{lem}\label{Lem: UseD2}
Let $\mathcal{F}\equiv (\sum_{i=1}^{N}x_{i1}\varepsilon_{i1}/\sqrt{N},\sum_{i=1}^{N}x_{i2}\varepsilon_{i2}/\sqrt{N}, \ldots, \sum_{i=1}^{N}x_{iT}\varepsilon_{iT}/\sqrt{N})$. For any $\Delta\in\{1_{N}\otimes \Gamma:  \Gamma\in\mathbf{R}^{p\times T}\}$, we have
\[\sum_{i=1}^{N}\sum_{t=1}^{T}|\mathrm{tr}(X_{it}^{\prime}\Delta)|^2\geq \min_{t\leq T}\lambda_{\min}\left(\frac{\sum_{i=1}^{N}x_{it}x_{it}^{\prime}}{N}\right)\|\Delta\|_{F}^{2}\]
and
\[\left|\sum_{i=1}^{N}\sum_{t=1}^{T}\mathrm{tr}(\varepsilon_{it} X_{it}^{\prime}\Delta)\right|\leq \|\mathcal{F}\|_{2}\|\Delta\|_{\ast}.\]
\end{lem}
\noindent{\sc Proof:} Fix $\Delta = 1_{N}\otimes \Gamma$ for some $\Gamma\in\mathbf{R}^{p\times T}$. Write $\Gamma = (\gamma_1,\gamma_2,\ldots, \gamma_{T})$, where $\gamma_{t}$ is a $p\times 1$ vector. Since $\mathrm{tr}(X_{it}^{\prime}\Delta) = x_{it}^{\prime}\gamma_{t}$, it follows that
\begin{align}\label{Eqn: Lem: UseD2: 1}
\sum_{i=1}^{N}\sum_{t=1}^{T}|\mathrm{tr}(X_{it}^{\prime}\Delta)|^2 &= \sum_{i=1}^{N}\sum_{t=1}^{T}|x_{it}^{\prime}\gamma_{t}|^2\notag\\ &=N\sum_{t=1}^{T}\gamma_t^{\prime}\left(\frac{\sum_{i=1}^{N}x_{it}x_{it}^{\prime}}{N}\right)\gamma_t\notag\\
&\geq \min_{t\leq T}\lambda_{\min}\left(\frac{\sum_{i=1}^{N}x_{it}x_{it}^{\prime}}{N}\right)N\|\Gamma\|_{F}^{2}\notag\\
&= \min_{t\leq T}\lambda_{\min}\left(\frac{\sum_{i=1}^{N}x_{it}x_{it}^{\prime}}{N}\right)\|\Delta\|_{F}^{2},
\end{align}
where the last equality holds since $\|\Delta\|_{F}^{2} = N\|\Gamma\|_{F}^{2}$. For the same reason, we have
\begin{align}\label{Eqn: Lem: UseD2: 2}
\sum_{i=1}^{N}\sum_{t=1}^{T}\mathrm{tr}(\varepsilon_{it} X_{it}^{\prime}\Delta)&= \sum_{i=1}^{N}\sum_{t=1}^{T}\varepsilon_{it} x_{it}^{\prime}\gamma_{t}\notag\\
&= \mathrm{tr}(\mathcal{F}^{\prime}\sqrt{N}\Gamma)\notag\\
&\leq \|\mathcal{F}\|_{2}\sqrt{N}\|\Gamma\|_{\ast}\notag\\
&= \|\mathcal{F}\|_{2}\|\Delta\|_{\ast},
\end{align}
where the inequality holds by the fact that $|\mathrm{tr}(C^{\prime}D)|\leq \|C\|_{2}\|D\|_{\ast}$, and the last equality follows by Lemma \ref{Lem: TechD1}(iii). This completes the proof of the lemma. \qed

\section{Computing Algorithms}\label{App: E}
\renewcommand{\theequation}{E.\arabic{equation}}
\setcounter{equation}{0}
In this appendix, we present computing algorithms for finding the nuclear norm regularized estimators in Examples \ref{Ex: Classical}-\ref{Ex: ConitionalObs}. Specifically, we use the accelerated proximal gradient algorithm by \citet{JiYe_AcceleratedGradient_2009} and \citet{TohYun_Accelerated_2010}. The algorithm solves the following general nonsmooth convex minimization problem:
\begin{align}\label{Eqn: Alg: NuclearNEst: 1}
\min_{\Gamma\in\mathbf{R}^{m\times T}}F(\Gamma)\equiv f(\Gamma) + \varphi_{NT}\|\Gamma\|_{\ast},
\end{align}
where $\Gamma\in\mathbf{R}^{m\times T}$ is the decision matrix, $f:\mathbf{R}^{m\times T}\mapsto [0,\infty)$ is a smooth loss function with the gradient $\nabla f(\Gamma)$ being Lipschitz continuous with constant $L_f$ (namely, $\|\nabla f(\Gamma^{(1)}) - \nabla f(\Gamma^{(2)})\|_{F}\leq L_f\|\Gamma^{(1)}-\Gamma^{(2)}\|_{F}$ for any $\Gamma^{(1)},\Gamma^{(2)}\in \mathbf{R}^{m\times T}$), $\|\Gamma\|_{\ast}$ is the nuclear norm of $\Gamma$, $\varphi_{NT}>0$ is a regularization parameter.
The algorithm consists of recursively solving a sequence of minimizations of linear approximations of $f(\Gamma)$ regularized by a quadratic proximal term and the nuclear norm, which is given by
\begin{align}\label{Eqn: Alg: NuclearNEst: 2}
&\hspace{0.5cm}\min_{\Gamma\in\mathbf{R}^{m\times T}}Q_{\tau_k}(\Gamma,\Gamma_{k})\equiv f(\Gamma_{k}) + \mathrm{tr}((\Gamma - \Gamma_{k})^{\prime}\nabla f(\Gamma_{k})) + \frac{\tau_k}{2}\|\Gamma - \Gamma_{k}\|_{F}^{2} +  \varphi_{NT}\|\Gamma\|_{\ast},\notag\\
&:= \min_{\Gamma\in\mathbf{R}^{m\times T}} \frac{\tau_k}{2}\left\|\Gamma - \left(\Gamma_{k} - \frac{1}{\tau_k}\nabla f(\Gamma_{k})\right)\right\|_{F}^2+\varphi_{NT}\|\Gamma\|_{\ast} + f(\Gamma_{k}) - \frac{1}{2\tau_k}\|\nabla f(\Gamma_{k})\|_{F}^{2}
\end{align}
for $k\in\mathbf{Z}^{+}$, where $\tau_k>0$ and $\Gamma_{k}$ are recursively updated. The algorithm is attractive in two aspects. First, the problem in \eqref{Eqn: Alg: NuclearNEst: 2} can be explicitly solved via the singular value decomposition of $\Gamma_{k} - \frac{1}{\tau_k}\nabla f(\Gamma_{k})$ and then applying some soft-thresholding on the singular values. This is because $ f(\Gamma_{k}) - \frac{1}{2\tau_k}\|\nabla f(\Gamma_{k})\|_{F}^{2}$ does not depend on $\Gamma$ and $\min_{\Gamma\in\mathbf{R}^{m\times T}} \frac{\tau_k}{2}\|\Gamma - [\Gamma_{k} -\frac{1}{\tau_k}\nabla f(\Gamma_{k})]\|_{F}^2+\varphi_{NT}\|\Gamma\|_{\ast}$ can be explicitly solved by the technique; see, for example, \citet{Caietal_SVD_2010} and \citet{Maetal_Fixedpoint_2011}. For $A\in\mathbf{R}^{m\times T}$, let $A = U\Sigma V^{\prime}$ be a singular value decomposition of $A$, where $U\in\mathbf{R}^{m\times m}$ with $U^{\prime}U=I_{m}$, $V\in\mathbf{R}^{T\times T}$ with $V^{\prime}V=I_{T}$, and $\Sigma\in\mathbf{R}^{m\times T}$ is a diagonal matrix with singular values in the diagonal in descending order. For $x>0$, define $\mathcal{S}_{x}(A) \equiv U\Sigma_{x} V^{\prime}$, where $\Sigma_{x}$ is diagonal with the $jj$th entry equal to $ \max\{0,\Sigma_{jj}-x\}$ for all $j$ and $\Sigma_{jj}$ denotes the $jj$th entry of $\Sigma$. The solution to \eqref{Eqn: Alg: NuclearNEst: 2} is given by
\begin{align}\label{Eqn: Alg: NuclearNEst: 3}
\mathcal{S}_{\tau_{k}^{-1}\varphi_{NT}}\left(\Gamma_{k} - \frac{1}{\tau_k}\nabla f(\Gamma_{k})\right).
\end{align}
Second, \citet{JiYe_AcceleratedGradient_2009} and \citet{TohYun_Accelerated_2010} show that if $\tau_k>0$ and $\Gamma_{k}$ are updated properly, the algorithm can achieve the optimal convergence rate of $O(1/k^2)$.

Let $\eta \in (0,1)$ be a given constant. Choose $\Gamma^{\ast}_0=\Gamma^{\ast}_1 \in \mathbf{R}^{m \times T}$. Set $w_0=w_1=1$ and $\tau_0=L_{f}$. Set $k=1$. The algorithm is given as follows.
\begin{enumerate}
\item[]\textbf{Step 1.} Set $\Gamma_{k}=\Gamma_{k}^{\ast}+\frac{w_{k-1}-1}{w_k}(\Gamma^{\ast}_k-\Gamma^{\ast}_{k-1})$.
\item[]\textbf{Step 2.} Set $\hat{\tau}_0=\eta \tau_{k-1}$. Set $j=0$ and execute the following step:
	\begin{itemize}
	\item  Compute $A_j = \mathcal{S}_{\hat\tau_{j}^{-1}\varphi_{NT}}(\Gamma_{k} - \hat\tau_{j}^{-1}\nabla f(\Gamma_{k}))$. If $F(A_j)\leq Q_{\hat{\tau}_j}(A_j,\Gamma_k)$, set $\tau_k=\hat{\tau}_j$ and proceed to \textbf{Step 3}; Otherwise, set $\hat{\tau}_{j+1}=\min\{\eta^{-1}\hat{\tau}_j, \tau_0\}$ and $j=j+1$, and return to the beginning of this step.
	\end{itemize}
\item[]\textbf{Step 3.} Set $\Gamma^{\ast}_{k+1}= \mathcal{S}_{\tau_{k}^{-1}\varphi_{NT}}(\Gamma_{k} - \tau_{k}^{-1}\nabla f(\Gamma_{k}))$.
\item[]\textbf{Step 4.} Set $w_{k+1}=(1+\sqrt{1+4w_k^2})/{2}$.
\item[]\textbf{Step 5.} Compute $D_{k+1}=\tau_{k}(\Gamma_{k}-\Gamma^{\ast}_{k+1}) + \nabla f(\Gamma^{\ast}_{k+1}) - \nabla f(\Gamma_{k})$.
If ${\|D_{k+1}\|_F}/$ $[{\tau_k \max\{1,\|\Gamma^{\ast}_{k+1}\|_F\}}] \leq \epsilon$ where $\epsilon$ is a pre-specified tolerance level, set the output $\hat{\Pi}=\Gamma^{\ast}_{k+1}$. Otherwise, set $k=k+1$ and return to \textbf{Step 1}.
\end{enumerate}
Step 2 is to ensure that the objective value generated at the $k$th iteration is bounded by the minimum of the approximating function, that is, $F(\Gamma_{k+1}^{\ast})\leq Q_{\tau_{k}}(\Gamma_{k+1}^{\ast},\Gamma_k)$, which is crucial to the algorithm. Alternatively, we may fix $\tau_{k} = L_{f}$ to meet the requirement; see, for example, Lemma 1.2.3 of \citet{Nesterov_ConvexOptimization_2003}. By shrinking $\tau_{k}$, the resulting solution tends to have lower rank than the one generated by setting $\tau_{k} = L_{f}$, since smaller value of $\tau_k$ may lead to fewer nonzero singular values in $\mathcal{S}_{\tau_{k}^{-1}\varphi_{NT}}(\Gamma_{k} - \tau_{k}^{-1}\nabla f(\Gamma_{k}))$. Steps 1 and 4 are key steps for the convergence rate of $O(1/k^2)$. Rather than fixing the search point (i.e.,$\Gamma_{k}$) at the solution from the previous iteration (i.e., $\Gamma^{\ast}_{k}$), the algorithm constructs the search point as a linear combination of the solutions from the latest two iterations. This may accelerate the convergence rate from $O(1/k)$ to $O(1/k^2)$ \citep{Nesterov_Convex_1983,Nesterov_ConvexOptimization_2003}; see \citet{JiYe_AcceleratedGradient_2009} and \citet{TohYun_Accelerated_2010} for the proofs. The sequence $w_{k}$ is generated in the manner in Step 4 to satisfy the constraint $w_{k+1}^2-w_{k+1}\leq w_k^2$. In Step 5, $D_{k+1}$ is a subgradient of $F(\Gamma)$ at $\Gamma = \Gamma^{\ast}_{k+1}$, see \citet{TohYun_Accelerated_2010}. In simulations and real data applications, we set $\eta = 0.8$, $\Gamma^{\ast}_0=\Gamma^{\ast}_1=0$ and $\epsilon = 10^{-5}$.

We next show how the problems in \eqref{Eqn: NuclearNEst} with $\mathcal{S} = \mathbf{R}^{Np\times T}$, $\mathcal{S}=\mathcal{D}_{M}$, and $\mathcal{S} = \{1_{N}\otimes \Gamma: \Gamma\in \mathbf{R}^{p\times T}\}$, which respectively define our estimators in Examples \ref{Ex: Classical}, \ref{Ex: Statevarying} and \ref{Ex: ConitionalObs}, Example \ref{Ex: SemiTin}, and Example \ref{Ex: HomoConditional}, can fit into the general framework in \eqref{Eqn: Alg: NuclearNEst: 1}. In all cases, the algorithms can be easily adapted to allow for the presence of missing values. In both Examples \ref{Ex: Classical}, \ref{Ex: Statevarying} and \ref{Ex: ConitionalObs} and Example \ref{Ex: HomoConditional}, we can simply replace the observations with $y_{it}m_{it}$ and $x_{it}m_{it}$, where $m_{it}$ is a dummy variable of missing status defined in Section \ref{Sec3}. It is straightforward to modify the algorithm to accommodate the presence of missing values in Example \ref{Ex: SemiTin}. Below we focus on the case without missing values.

\subsection{Examples \ref{Ex: Classical}, \ref{Ex: Statevarying}, and \ref{Ex: ConitionalObs}}\label{App: E1}
For \eqref{Eqn: NuclearNEst} with $\mathcal{S} = \mathbf{R}^{Np\times T}$, to use the algorithm, we set $m=Np$, $\varphi_{NT} = \lambda_{NT}$ and
\begin{align}\label{Eqn: Alg: NuclearNEst: 4}
f(\Gamma)= \frac{1}{2}\sum_{i=1}^{N}\sum_{t=1}^{T}(y_{it}-x_{it}^{\prime}\gamma_{it})^{2} \text{ for } \Gamma \equiv \left(
                                                              \begin{array}{cccc}
                                                                \gamma_{11} &\gamma_{12} & \cdots &\gamma_{1T} \\
                                                                \gamma_{21} & \gamma_{22} & \cdots & \gamma_{2T} \\
                                                                \vdots & \vdots & \vdots & \vdots \\
                                                                \gamma_{N1} & \gamma_{N2} & \cdots & \gamma_{NT} \\
                                                              \end{array}
                                                            \right)\in \mathbf{R}^{Np\times T}.
\end{align}
We need to show that the gradient $\nabla f(\Gamma)$ is Lipschitz continuous. It follows that
\begin{align}\label{Eqn: Alg: NuclearNEst: 5}
\nabla f(\Gamma)\hspace{-0.1cm}=\hspace{-0.1cm}\left(\hspace{-0.2cm}
                                                              \begin{array}{cccc}
                                                                x_{11}(x_{11}^{\prime}\gamma_{11} - y_{11}) &x_{12}(x_{12}^{\prime}\gamma_{11} - y_{12}) & \cdots &x_{1T}(x_{1T}^{\prime}\gamma_{1T} - y_{1T}) \\
                                                                x_{21}(x_{21}^{\prime}\gamma_{21} - y_{21}) & x_{22}(x_{22}^{\prime}\gamma_{22} - y_{22}) & \cdots & x_{2T}(x_{2T}^{\prime}\gamma_{2T} - y_{2T}) \\
                                                                \vdots & \vdots & \vdots & \vdots \\
                                                                x_{N1}(x_{N1}^{\prime}\gamma_{N1} - y_{N1}) & x_{N2}(x_{N2}^{\prime}\gamma_{N2} - y_{N2}) & \cdots & x_{NT}(x_{NT}^{\prime}\gamma_{NT} - y_{NT}) \\
                                                              \end{array}
                                                            \hspace{-0.2cm}\right).
\end{align}
Indeed,$\nabla f(\Gamma)$ is Lipschitz continuous with constant $L_f = \max_{i\leq N, t\leq T}\|x_{it}\|^2$, because for $\Gamma^{(1)}\equiv (\gamma_{it}^{(1)})\in \mathbf{R}^{Np\times T}$ and $\Gamma^{(2)}\equiv  (\gamma_{it}^{(2)})\in \mathbf{R}^{Np\times T}$,
\begin{align}\label{Eqn: Alg: NuclearNEst: 6}
&\hspace{0.5cm}\|\nabla f(\Gamma^{(1)})-\nabla f(\Gamma^{(2)})\|^{2}_F\notag\\
&= \left\|\left(
                                                              \begin{array}{cccc}
                                                                x_{11}x_{11}^{\prime}(\gamma^{(1)}_{11} - \gamma^{(2)}_{11}) &x_{12}x_{12}^{\prime}(\gamma^{(1)}_{12} - \gamma^{(2)}_{12}) & \cdots &x_{1T}x_{1T}^{\prime}(\gamma^{(1)}_{1T} - \gamma^{(2)}_{1T}) \\
                                                                x_{21}x_{21}^{\prime}(\gamma^{(1)}_{21} - \gamma^{(2)}_{21}) & x_{22}x_{22}^{\prime}(\gamma^{(1)}_{22} - \gamma^{(2)}_{22}) & \cdots & x_{2T}x_{2T}^{\prime}(\gamma^{(1)}_{2T} - \gamma^{(2)}_{2T}) \\
                                                                \vdots & \vdots & \vdots & \vdots \\
                                                                x_{N1}x_{N1}^{\prime}(\gamma^{(1)}_{N1} - \gamma^{(2)}_{N1}) & x_{N2}x_{N2}^{\prime}(\gamma^{(1)}_{N2} - \gamma^{(2)}_{N2}) & \cdots & x_{NT}x_{NT}^{\prime}(\gamma^{(1)}_{NT} - \gamma^{(2)}_{NT}) \\
                                                              \end{array}
                                                            \right)\right\|^{2}_{F}\notag\\
& = \sum_{i=1}^{N}\sum_{t=1}^{T}\|x_{it}x_{it}^{\prime}(\gamma^{(1)}_{it} - \gamma^{(2)}_{it})\|^{2}\notag\\
&\leq \max_{i\leq N, t\leq T}\|x_{it}\|^{4}\|\Gamma^{(1)}-\Gamma^{(2)}\|^{2}_{F}.
\end{align}

\subsection{Example \ref{Ex: SemiTin}}\label{App: E2}
We transform the problem in \eqref{Eqn: NuclearNEst} with $\mathcal{S} = \mathcal{D}_{M}$ to an unconstrained problem by plugging in the homogeneity restriction from $\mathcal{D}_{M}$. As discussed in Section \ref{Sec52}, finding $\hat{\Pi}$ reduces to finding $\hat{\Pi}^{\diamond}$ and $\hat{\Pi}^{\ast}$. By Lemma \ref{Lem: UseD5}, $\hat{\Pi}^{\diamond}$ and $\hat{\Pi}^{\ast}$ can be equivalently obtained as follows:
\begin{align}\label{Eqn: Ex2: 5}
\hspace{-0.3cm}\{\hat{\Pi}^{\diamond},\hat{\Pi}^{\ast}\}=\hspace{-0.4cm}\amin_{\substack{\Gamma^{\diamond} = (\gamma_{it})_{i\leq N, t\leq T}\in\mathbf{R}^{N\times T}\\\Gamma^{\ast}= (\gamma^{\ast}_1,\ldots,\gamma^{\ast}_T)\in\mathbf{R}^{(p-1)\times T}\\ \|\Gamma^{\ast}\|_{\max}\leq M}}\frac{1}{2}\sum_{i=1}^{N}\sum_{t=1}^{T}(y_{it}-\gamma_{it}- x_{it}^{\ast\prime}\gamma^{\ast}_{t})^{2} +\lambda_{NT}\left\|\left(\hspace{-0.2cm}
     \begin{array}{c}
       \Gamma^{\diamond} \\
       \sqrt{N}\Gamma^{\ast} \\
     \end{array}
   \hspace{-0.2cm}\right)
\right\|_{\ast}.
\end{align}
By changing values, we may equivalently rewrite \eqref{Eqn: Ex2: 5} as
\begin{align}\label{Eqn: Alg: NuclearNEst: 70}
\left(
     \begin{array}{c}
       \hat{\Pi}^{\diamond} \\
       \sqrt{N}\hat{\Pi}^{\ast} \\
     \end{array}
   \right) =\amin_{\substack{\Gamma^{\diamond} = (\gamma_{it})_{i\leq N, t\leq T}\in\mathbf{R}^{N\times T}\\\Gamma^{\ast}= (\gamma^{\ast}_1,\ldots,\gamma^{\ast}_T)\in\mathbf{R}^{(p-1)\times T}\\ \|\Gamma^{\ast}\|_{\max}\leq \sqrt{N}M}}\frac{1}{2}\sum_{i=1}^{N}\sum_{t=1}^{T}(y_{it}-\gamma_{it}- w_{it}^{\ast\prime}\gamma^{\ast}_{t})^{2} +\lambda_{NT}\left\|\left(
     \begin{array}{c}
       \Gamma^{\diamond} \\
       \Gamma^{\ast} \\
     \end{array}
   \right)
\right\|_{\ast},
\end{align}
where $w_{it}^{\ast} = x_{it}^{\ast}/\sqrt{N}$. Here, we consider the problem by dropping the constraint that $\|\Gamma^{\ast}\|\leq \sqrt{N} M$. First, as noted in Footnote \ref{foot7}, the constraint is only a technical condition that simplifies the proof, so may not be necessary. Second, in practice, the constraint is not binding for a sufficiently large value of $M$, thus can be dropped. Therefore, the problem in \eqref{Eqn: Alg: NuclearNEst: 70} falls into the general framework in \eqref{Eqn: Alg: NuclearNEst: 1}. To use the algorithm, we set $m=N+p-1$, $\varphi_{NT} = \lambda_{NT}$ and
\begin{align}\label{Eqn: Alg: NuclearNEst: 7}
f(\Gamma)\hspace{-0.1cm}= \hspace{-0.1cm}\frac{1}{2}\sum_{i=1}^{N}\sum_{t=1}^{T}\left(y_{it}-\gamma_{it}- w_{it}^{\ast\prime}\gamma^{\ast}_{t}\right)^{2} \hspace{-0.1cm}\text{ for } \hspace{-0.1cm}\Gamma \hspace{-0.1cm}\equiv \hspace{-0.1cm}\left(\hspace{-0.2cm}
                                                              \begin{array}{cccc}
                                                                \gamma_{11} &\gamma_{12} & \cdots &\gamma_{1T} \\
                                                                \gamma_{21} & \gamma_{22} & \cdots & \gamma_{2T} \\
                                                                \vdots & \vdots & \vdots & \vdots \\
                                                                \gamma_{N1} & \gamma_{N2} & \cdots & \gamma_{NT} \\
                                                                \gamma^{\ast}_{1} &\gamma^{\ast}_{2} & \cdots &\gamma^{\ast}_{T} \\
                                                              \end{array}
                                                            \hspace{-0.2cm}\right)\in \mathbf{R}^{(N+p-1)\times T}.
\end{align}
We need to show that the gradient $\nabla f(\Gamma)$ is Lipschitz continuous. It follows that
\begin{align}\label{Eqn: Alg: NuclearNEst: 8}
\nabla f(\Gamma)&=\left(
                                                              \begin{array}{cc}
                                                                \gamma_{11} + w_{11}^{\ast\prime}\gamma^{\ast}_{1} - y_{11} &\gamma_{12} + w_{12}^{\ast\prime}\gamma^{\ast}_{2} - y_{12} \\
                                                               \gamma_{21} + w_{21}^{\ast\prime}\gamma^{\ast}_{1} - y_{21} & \gamma_{22} + w_{22}^{\ast\prime}\gamma^{\ast}_{2} - y_{22} \\
                                                                \vdots & \vdots\\
                                                                \gamma_{N1} + w_{N1}^{\ast\prime}\gamma^{\ast}_{1} - y_{N1} & \gamma_{N2} + w_{N2}^{\ast\prime}\gamma^{\ast}_{2} - y_{N2} \\
                                                                \sum_{i=1}^{N}w^{\ast}_{i1}(\gamma_{i1} + w_{i1}^{\ast\prime}\gamma^{\ast}_{1} - y_{i1}) & \sum_{i=1}^{N}w^{\ast}_{i2}(\gamma_{i2} + w_{i2}^{\ast\prime}\gamma^{\ast}_{2} - y_{i2})\\
                                                              \end{array}
                                                            \right.\notag\\
&                                                            \notag\\
                                                            &\hspace{4cm}
\left.
                                                              \begin{array}{cc}
                                                               \cdots &(\gamma_{1T} + w_{1T}^{\ast\prime}\gamma^{\ast}_{T} - y_{1T}) \\
                                                             \cdots & (\gamma_{2T} + w_{2T}^{\ast\prime}\gamma^{\ast}_{T} - y_{2T}) \\
                                                               \vdots & \vdots \\
                                                                \cdots & (\gamma_{NT} + w_{NT}^{\ast\prime}\gamma^{\ast}_{T} - y_{NT}) \\
                                                                \cdots & \sum_{i=1}^{N}w^{\ast}_{iT}(\gamma_{iT} + w_{iT}^{\ast\prime}\gamma^{\ast}_{T} - y_{iT}) \\
                                                              \end{array}
                                                            \right),
\end{align}
and for $\Gamma^{(1)}\hspace{-0.1cm}\equiv\hspace{-0.1cm} (\gamma_{it}^{(1)},\gamma_{t}^{\ast(1)})\hspace{-0.05cm}\in \hspace{-0.05cm}\mathbf{R}^{(N+p-1)\times T}$ and $\Gamma^{(2)}\hspace{-0.1cm}\equiv \hspace{-0.1cm} (\gamma_{it}^{(2)},\gamma_{t}^{\ast(2)})\hspace{-0.05cm}\in\hspace{-0.05cm}\mathbf{R}^{(N+p-1)\times T}$,
\begin{align}\label{Eqn: Alg: NuclearNEst: 9}
&\hspace{0.5cm}\|\nabla f(\Gamma^{(1)})-\nabla f(\Gamma^{(2)})\|^{2}_F\notag\\
&= \left\|\left(
                                                              \begin{array}{c}
                                                                \gamma^{(1)}_{11}-\gamma^{(2)}_{11} + w_{11}^{\ast\prime}(\gamma^{\ast(1)}_{1}-\gamma^{\ast(2)}_{1}) \\
                                                                \gamma^{(1)}_{21}-\gamma^{(2)}_{21} + w_{21}^{\ast\prime}(\gamma^{\ast(1)}_{1}-\gamma^{\ast(2)}_{1}) \\
                                                                \vdots\\
                                                                \gamma^{(1)}_{N1}-\gamma^{(2)}_{N1} + w_{N1}^{\ast\prime}(\gamma^{\ast(1)}_{1}-\gamma^{\ast(2)}_{1}) \\
                                                                 \sum_{i=1}^{N}w^{\ast}_{i1}(\gamma^{(1)}_{i1}-\gamma^{(2)}_{i1}) +\sum_{i=1}^{N}w^{\ast}_{i1}w_{i1}^{\ast\prime}(\gamma^{\ast(1)}_{1} - \gamma^{\ast(2)}_{1})\\
                                                              \end{array}
                                                            \right.\right.\notag\\
&                                                            \notag\\
&\hspace{2cm}
\left.
                                                              \begin{array}{c}
                                                                \gamma^{(1)}_{12}-\gamma^{(2)}_{12} + w_{12}^{\ast\prime}(\gamma^{\ast(1)}_{2}-\gamma^{\ast(2)}_{2}) \\
                                                                 \gamma^{(1)}_{22}-\gamma^{(2)}_{22} + w_{22}^{\ast\prime}(\gamma^{\ast(1)}_{2}-\gamma^{\ast(2)}_{2}) \\
                                                                \vdots\\
                                                               \gamma^{(1)}_{N2}-\gamma^{(2)}_{N2} + w_{N2}^{\ast\prime}(\gamma^{\ast(1)}_{2}-\gamma^{\ast(2)}_{2})\\
                                                                 \sum_{i=1}^{N}w^{\ast}_{i2}(\gamma^{(1)}_{i2}-\gamma^{(2)}_{i2}) +\sum_{i=1}^{N}w^{\ast}_{i2}w_{i2}^{\ast\prime}(\gamma^{\ast(1)}_{2} - \gamma^{\ast(2)}_{2})\\
                                                              \end{array}
                                                            \right.\notag\\
&                                                            \notag\\
&
                                                            \hspace{1.5cm}
\left.\left.
                                                              \begin{array}{cc}
                                                               \cdots &\gamma^{(1)}_{1T}-\gamma^{(2)}_{1T} + w_{1T}^{\ast\prime}(\gamma^{\ast(1)}_{T}-\gamma^{\ast(2)}_{T}) \\
                                                             \cdots & \gamma^{(1)}_{2T}-\gamma^{(2)}_{2T} + w_{2T}^{\ast\prime}(\gamma^{\ast(1)}_{T}-\gamma^{\ast(2)}_{T}) \\
                                                               \vdots & \vdots \\
                                                                \cdots & \gamma^{(1)}_{NT}-\gamma^{(2)}_{NT} + w_{NT}^{\ast\prime}(\gamma^{\ast(1)}_{T}-\gamma^{\ast(2)}_{T}) \\
                                                                \cdots & \sum_{i=1}^{N}w^{\ast}_{iT}(\gamma^{(1)}_{iT}-\gamma^{(2)}_{iT}) +\sum_{i=1}^{N}w^{\ast}_{iT}w_{iT}^{\ast\prime}(\gamma^{\ast(1)}_{T} - \gamma^{\ast(2)}_{T}) \\
                                                              \end{array}
                                                            \right)\right\|^{2}_{F}\notag\\
& = \sum_{i=1}^{N}\sum_{t=1}^{T}\left[\gamma^{(1)}_{it}-\gamma^{(2)}_{it} + w_{it}^{\ast\prime}(\gamma^{\ast(1)}_{t}-\gamma^{\ast(2)}_{t})\right]^{2}\notag\\
&\hspace{0.5cm}+ \sum_{t=1}^{T}\left\|\sum_{i=1}^{N}w^{\ast}_{it}(\gamma^{(1)}_{it}-\gamma^{(2)}_{it}) +\sum_{i=1}^{N}w^{\ast}_{iT}w_{it}^{\ast\prime}(\gamma^{\ast(1)}_{t} - \gamma^{\ast(2)}_{t})\right\|^2\notag\\
&\leq 2\sum_{i=1}^{N}\sum_{t=1}^{T}(\gamma^{(1)}_{it}-\gamma^{(2)}_{it})^2 + 2\max_{t\leq T}\lambda_{\max}\left(\sum_{i=1}^{N} w_{it}^{\ast}w_{it}^{\ast\prime}\right)\sum_{t=1}^{T}\|\gamma^{\ast(1)}_{t}-\gamma^{\ast(2)}_{t}\|^{2}\notag\\
&\hspace{0.2cm}+2N\hspace{-0.2cm}\max_{i\leq N,t\leq N}\hspace{-0.1cm}\|w_{it}^{\ast}\|^2\sum_{i=1}^{N}\sum_{t=1}^{T}(\gamma^{(1)}_{it}-\gamma^{(2)}_{it})^2 + 2\max_{t\leq T}\lambda^{2}_{\max}\left(\sum_{i=1}^{N} w_{it}^{\ast}w_{it}^{\ast\prime}\right)\hspace{-0.1cm}\sum_{t=1}^{T}\|\gamma^{\ast(1)}_{t}-\gamma^{\ast(2)}_{t}\|^{2}\notag\\
&\leq 2\max\left\{1+N\max_{i\leq N,t\leq N}\|w_{it}^{\ast}\|^2, \max_{t\leq T}\lambda_{\max}\left(\sum_{i=1}^{N} w_{it}^{\ast}w_{it}^{\ast\prime}\right)+\max_{t\leq T}\lambda^{2}_{\max}\left(\sum_{i=1}^{N} w_{it}^{\ast}w_{it}^{\ast\prime}\right)\right\}\notag\\
&\hspace{1.5cm}\times\|\Gamma^{(1)}-\Gamma^{(2)}\|^{2}_{F}\notag\\
&= 2\max\left\{1+\max_{i\leq N,t\leq N}\|x_{it}^{\ast}\|^2, \max_{t\leq T}\lambda_{\max}\left(\frac{1}{N}\sum_{i=1}^{N} x_{it}^{\ast}x_{it}^{\ast\prime}\right)+\max_{t\leq T}\lambda^{2}_{\max}\left(\frac{1}{N}\sum_{i=1}^{N} x_{it}^{\ast}x_{it}^{\ast\prime}\right)\right\}\notag\\
&\hspace{1.5cm}\times \|\Gamma^{(1)}-\Gamma^{(2)}\|^{2}_{F},
\end{align}
where the first inequality follows due to the Cauchy Schwartz inequality together with the triangle inequality. Thus, $\nabla f(\Gamma)$ is Lipschitz continuous with constant $L_f = \sqrt{2}[\max\{1+\max_{i\leq N,t\leq N}\|x_{it}^{\ast}\|^2,\max_{t\leq T}\lambda_{\max}(\sum_{i=1}^{N} x_{it}^{\ast}x_{it}^{\ast\prime}/N)+\max_{t\leq T}\lambda^{2}_{\max}(\sum_{i=1}^{N} x_{it}^{\ast}x_{it}^{\ast\prime}/N)\}]^{1/2}$.

\begin{rem}
The equivalence in \eqref{Eqn: Ex2: 5} has greatly simplified the computation of $\hat{\Pi}$, since \eqref{Eqn: NuclearNEst} involves an $Np\times T$ unknown matrix while \eqref{Eqn: Ex2: 5} involves two unknown matrices with relatively smaller sizes. By Lemma \ref{Lem: TechD2}(ii) and (iv), $\hat{K}$ can be equivalently obtained as
\begin{align}\label{Eqn: Ex2: 6}
\hat{K} = \sum_{j=1}^{T}1\{\lambda_{j}(M_{T}(\hat{\Pi}^{\diamond\prime}\hat{\Pi}^{\diamond} + N\hat{\Pi}^{\ast\prime}\hat{\Pi}^{\ast})M_{T})\geq \delta_{NT}\},
\end{align}
and $(\hat{\Lambda}^{\prime}/\sqrt{N}, \hat{\Phi}^{\prime})^{\prime}$ as the left singular vector of  $(\hat{\Pi}^{\diamond\prime}, \sqrt{N}\hat{\Pi}^{\ast\prime})^{\prime}M_{T}$ corresponding to its largest $\hat{K}$ singular values.
Moreover, it is straightforward to show that
\begin{align}\label{Eqn: Ex2: 7}
\hat{\mu} &= \left(I_N - \frac{\hat{\Lambda}\hat{\Lambda}^{\prime}}{N}\right)\frac{\hat{\Pi}^{\diamond\prime}1_{T}}{T} -\hat{\Lambda}\hat{\Phi}^{\prime}\frac{\hat{\Pi}^{\ast\prime}1_{T}}{T},\notag\\
\hat{\phi}& = (I_{p-1} - \hat{\Phi}\hat{\Phi}^{\prime})\frac{\hat{\Pi}^{\ast\prime}1_{T}}{T} -\frac{\hat{\Phi}\hat{\Lambda}^{\prime}}{N}\frac{\hat{\Pi}^{\diamond}1_{T}}{T},\\
\hat{F} &= \frac{\hat{\Pi}^{\diamond\prime}\hat{\Lambda}}{N} + \hat{\Pi}^{\ast\prime}\hat{\Phi}.\notag
\end{align}
\end{rem}

\begin{rem}
We can extract additional estimators for $K$, $\mu$, $\Lambda$, $\phi$, $\Phi$, and $F$ from $\hat{\Pi}^{\diamond}$ and $\hat{\Pi}^{\ast}$ separately. First, since $\Pi^{\diamond} M_{T} = \Lambda F^{\prime}M_{T}$, we may extract estimators for $K$, $\mu$, $\Lambda$, and $F$ from $\hat{\Pi}^{\diamond}$ analogously to \eqref{Eqn: KEst} and \eqref{Eqn: aFEst}. Second, similarly, since $\Pi^{\ast} M_{T} = \Phi F^{\prime}M_{T}$, we can derive estimators for $K$, $\phi$, $\Phi$, and $F$ from $\hat{\Pi}^{\ast}$. These estimators differ from $\hat{K}$, $\hat{\mu}$, $\hat\Lambda$, $\hat{\phi}, \hat{\Phi}$, and $\hat{F}$ in Corollary \ref{Cor: Ex2: 1}. However, following the arguments in the proof of Theorem \ref{Thm: NuclearNRate}(ii), we can establish the consistency and the same convergence rate for the estimators; the details are omitted.
\end{rem}

\subsubsection{Technical Lemmas}
Recall that $x_{it} = (1,x_{it}^{\ast\prime})^{\prime}$ and $X_{it}= (e_{N,i}\otimes x_{it}) e_{T,t}^{\prime}$ be an $Np\times T$ matrix of $x_{it}$, where $e_{N,i}$ is the $i$th column of $I_N$ and $e_{T,t}$ is the $t$th column of $I_T$.
\begin{lem}\label{Lem: UseD5}
For any $\Gamma^{\diamond}=(\gamma_1,\gamma_2,\ldots,\gamma_N)^{\prime}\in\mathbf{R}^{N\times T}$ and $\Gamma^{\ast}=(\gamma^{\ast}_1,\gamma^{\ast}_2,\ldots,\gamma^{\ast}_T)^{\prime}\in\mathbf{R}^{(p-1)\times T}$, we have
\begin{align*}
&\hspace{0.5cm}\frac{1}{2}\sum_{i=1}^{N}\sum_{t=1}^{T}\left(y_{it}-\mathrm{tr}\left(X_{it}^{\prime}\left(
                    \begin{array}{c}
                      \gamma_1^{\prime} \\
                      \Gamma^{\ast} \\
                      \gamma_2^{\prime} \\
                      \Gamma^{\ast} \\
                      \vdots \\
                      \gamma_N^{\prime} \\
                      \Gamma^{\ast} \\
                    \end{array}
                  \right)\right)\right)^{2} +\lambda_{NT} \left\|\left(
                    \begin{array}{c}
                      \gamma_1^{\prime} \\
                      \Gamma^{\ast} \\
                      \gamma_2^{\prime} \\
                      \Gamma^{\ast} \\
                      \vdots \\
                      \gamma_N^{\prime} \\
                      \Gamma^{\ast} \\
                    \end{array}
                  \right)\right\|_{\ast}\notag\\
&= \frac{1}{2}\sum_{i=1}^{N}\sum_{t=1}^{T}(y_{it}-\gamma_{it} - x_{it}^{\ast\prime}\gamma^{\ast}_{t})^{2} +\lambda_{NT}\left\|\left(
              \begin{array}{c}
                \Gamma^{\diamond} \\
                \sqrt{N}\Gamma^{\ast} \\
              \end{array}
            \right)
\right\|_{\ast},
\end{align*}
where $\gamma_{i} = (\gamma_{i1}, \gamma_{i2}, \ldots, \gamma_{iT})^{\prime}$.
\end{lem}
\noindent{\sc Proof:} Fix $\Gamma^{\diamond}=(\gamma_1,\gamma_2,\ldots,\gamma_N)^{\prime}\in\mathbf{R}^{N\times T}$ and $\Gamma^{\ast}=(\gamma^{\ast}_1,\gamma^{\ast}_2,\ldots,\gamma^{\ast}_T)^{\prime}\in\mathbf{R}^{(p-1)\times T}$. It is easy to see that $\mathrm{tr}(X_{it}^{\prime}((\gamma_1,\Gamma^{\ast\prime}),(\gamma_2,\Gamma^{\ast\prime}), (\gamma_N,\Gamma^{\ast\prime}))^{\prime}) = \gamma_{it} + x_{it}^{\ast\prime}\gamma^{\ast}_{t}$. By Lemma \ref{Lem: TechD2}(iii), $\|((\gamma_1,\Gamma^{\ast\prime}),(\gamma_2,\Gamma^{\ast\prime}), (\gamma_N,\Gamma^{\ast\prime}))^{\prime}\|_{\ast}=\|(\Gamma^{\diamond\prime}, \sqrt{N}\Gamma^{\ast})^{\prime}\|_{\ast}$. Thus, the result follows. \qed
\begin{lem}\label{Lem: TechD2}
For any matrices $C=(c_1,c_2,\ldots,c_k)^{\prime}$ and $D$ with the same number of columns where $c_{j}$'s are column vectors, (i) the rank of $(c_1,D^{\prime},c_2,D^{\prime},\ldots, c_k,D^{\prime})$ is equal to the rank of $(C^{\prime}, \sqrt{k}D^{\prime})$; (ii) the nonzero singular values of $(c_1,D^{\prime},c_2,D^{\prime},$ $\ldots,c_k,D^{\prime})$ are equal to the nonzero singular values of $(C^{\prime}, \sqrt{k}D^{\prime})$; (iii) $\|(c_1,D^{\prime},c_2,D^{\prime},$ $\ldots, c_k,D^{\prime})\|_{\ast} =\|(C^{\prime},\sqrt{k}D^{\prime})\|_{\ast}$; (iv) the left singular vector matrix of nonzero matrix $(c_1,D^{\prime},c_2,D^{\prime},\ldots, c_k,D^{\prime})^{\prime}$ corresponding to its nonzero singular values have the form of $(u_1,V^{\prime},u_2,V^{\prime},\ldots, u_k,V^{\prime})^{\prime}$, where $U = (u_1,u_2,\ldots, u_{k})^{\prime}$ and $V$ have the same number of rows with $C$ and $D$, respectively. Moreover, $(U^{\prime},\sqrt{k} V^{\prime})^{\prime}$ is the left singular vector matrix of $(C^{\prime},\sqrt{k} D^{\prime})^{\prime}$ corresponding to its nonzero singular values.
\end{lem}
\noindent{\sc Proof:} It is without loss of generality to assume that $C$ or $D$ is nonzero. Let $d>0$ be the rank of $(c_1,D^{\prime},c_2,D^{\prime},\ldots, c_k,D^{\prime})$ and $\sigma_1\geq \sigma_2\geq \ldots\geq \sigma_{d}>0$ be the nonzero singular values of $(c_1,D^{\prime},c_2,D^{\prime},\ldots, c_k,D^{\prime})$. It follows that $\sigma^{2}_1\geq \sigma^{2}_2\geq \ldots\geq \sigma^{2}_{d}>0$ are the nonzero eigenvalues of\vspace{-0.2cm}
\begin{align}\label{Eqn: Lem: TechD2: 1}
(c_1,D^{\prime},c_2,D^{\prime},\ldots, c_k,D^{\prime})\left(
                                               \begin{array}{c}
                                                 c_1^{\prime} \\
                                                 D \\
                                                 c_2^{\prime} \\
                                                 D \\
                                                 \vdots\\
                                                 c_k^{\prime} \\
                                                 D \\
                                               \end{array}
                                             \right)
 = C^{\prime}C +k D^{\prime}D = (C^{\prime}, \sqrt{k}D^{\prime})\left(
                                               \begin{array}{c}
                                                 C \\
                                                 \sqrt{k} D \\
                                               \end{array}
                                             \right).
\end{align}
Thus, the nonzero singular values of $(C^{\prime}, \sqrt{k}D^{\prime})$ are $\sigma_1\geq \sigma_2\geq \ldots\geq \sigma_{d}>0$ and the rank of $(C^{\prime}, \sqrt{k}D^{\prime})$ is equal to $d$. Let $(c_1,D^{\prime},c_2,D^{\prime},\ldots, c_k,D^{\prime})^{\prime} = U^{\ast} \Sigma V^{\ast\prime}$ be a singular value decomposition of $(c_1,D^{\prime},c_2,D^{\prime},\ldots, c_k,D^{\prime})^{\prime}$, where $\Sigma$ is a $d\times d$ diagonal matrix with $\sigma_{j}$'s in the diagonal in descending order. It follows that
\begin{align}\label{Eqn: Lem: TechD2: 2}
U^{\ast} = \left(
                                               \begin{array}{c}
                                                 c_1^{\prime} \\
                                                 D \\
                                                 c_2^{\prime} \\
                                                 D \\
                                                 \vdots\\
                                                 c_k^{\prime} \\
                                                 D \\
                                               \end{array}
                                             \right) V^{\ast}\Sigma^{-1} = \left(
                                               \begin{array}{c}
                                                 c_1^{\prime}V^{\ast}\Sigma^{-1} \\
                                                 DV^{\ast}\Sigma^{-1} \\
                                                 c_2^{\prime}V^{\ast}\Sigma^{-1} \\
                                                 DV^{\ast}\Sigma^{-1} \\
                                                 \vdots\\
                                                 c_k^{\prime}V^{\ast}\Sigma^{-1} \\
                                                 DV^{\ast}\Sigma^{-1} \\
                                               \end{array}
                                             \right)=\left(
                                               \begin{array}{c}
                                                 u_1^\prime \\
                                                 V \\
                                                 u_2^{\prime} \\
                                                 V \\
                                                 \vdots\\
                                                 u_k^{\prime} \\
                                                 V \\
                                               \end{array}
                                             \right),
\end{align}
where $u_j = \Sigma^{-1}V^{\ast\prime}c_{j}$ and $V = DV^{\ast}\Sigma^{-1}$. In view of \eqref{Eqn: Lem: TechD2: 1}, $V^{\ast}$ is also the right singular vector matrix of $(C^{\prime},\sqrt{k} D^{\prime})^{\prime}$. Thus, the left singular vector matrix of $(C^{\prime},\sqrt{k} D^{\prime})^{\prime}$ corresponding to its nonzero singular values is given by
\begin{align}\label{Eqn: Lem: TechD2: 3}
\left(
                                               \begin{array}{c}
                                                 C \\
                                                 \sqrt{k} D \\
                                               \end{array}
                                             \right) V^{\ast}\Sigma^{-1} = \left(
                                               \begin{array}{c}
                                                 C V^{\ast}\Sigma^{-1} \\
                                                 \sqrt{k} DV^{\ast}\Sigma^{-1} \\
                                               \end{array}
                                             \right) =\left(
                                               \begin{array}{c}
                                                 U \\
                                                 \sqrt{k} V \\
                                               \end{array}
                                             \right).
\end{align}
 This completes the proof of the lemma. \qed

\subsection{Example \ref{Ex: HomoConditional}}\label{App: E3}
We transform the problem in \eqref{Eqn: NuclearNEst} with $\mathcal{S} = \{1_{N}\otimes \Gamma: \Gamma\in \mathbf{R}^{p\times T}\}$ to an unconstrained problem by plugging in the homogeneity restriction from $\{1_{N}\otimes \Gamma: \Gamma\in \mathbf{R}^{p\times T}\}$. As discussed in Section \ref{Sec53}, finding $\hat{\Pi}$ reduces to finding $\hat{\Pi}_0$. By Lemma \ref{Lem: UseD3}, $\hat{\Pi}_0$ can be equivalently obtained as follows:
\begin{align}\label{Eqn: Ex4: 3}
\hat{\Pi}_0 =\amin_{\Gamma= (\gamma_1,\ldots,\gamma_T)\in\mathbf{R}^{p\times T}}\frac{1}{2}\sum_{i=1}^{N}\sum_{t=1}^{T}(y_{it}-x_{it}^{\prime}\gamma_{t})^{2} +\sqrt{N}\lambda_{NT} \|\Gamma\|_{\ast},
\end{align}
The problem in \eqref{Eqn: Ex4: 3} fall into the general framework in \eqref{Eqn: Alg: NuclearNEst: 1}.  To use the algorithm, we set $m=p$, $\varphi_{NT} = \sqrt{N}\lambda_{NT}$ and
\begin{align}\label{Eqn: Alg: NuclearNEst: 10}
f(\Gamma)= \frac{1}{2}\sum_{i=1}^{N}\sum_{t=1}^{T}(y_{it} - x_{it}^{\prime}\gamma_t)^{2}\text{ for } \Gamma \equiv (\gamma_1,\gamma_2,\ldots, \gamma_T)\in \mathbf{R}^{p\times T}.
\end{align}
We need to show that the gradient $\nabla f(\Gamma)$ is Lipschitz continuous. It follows that
\begin{align}\label{Eqn: Alg: NuclearNEst: 11}
\nabla f(\Gamma)= \left(\sum_{i=1}^{N}x_{i1}(x_{i1}^{\prime} \gamma_1 -y_{i1}) , \sum_{i=1}^{N}x_{i2}(x_{i2}^{\prime} \gamma_2-y_{i2}), \ldots, \sum_{i=1}^{N}x_{iT}(x_{iT}^{\prime} \gamma_T -y_{iT})\right).
\end{align}
For $\Gamma^{(1)}\equiv (\gamma^{(1)}_1,\gamma^{(1)}_2,\ldots, \gamma^{(1)}_T)\in \mathbf{R}^{p\times T}$ and $\Gamma^{(2)}\equiv  (\gamma^{(2)}_1,\gamma^{(2)}_2,\ldots, \gamma^{(2)}_T)\in \mathbf{R}^{p\times T}$,
\begin{align}\label{Eqn: Alg: NuclearNEst: 12}
&\hspace{0.5cm}\|\nabla f(\Gamma^{(1)})-\nabla f(\Gamma^{(2)})\|^{2}_F\notag\\
&= \left\|\sum_{i=1}^{N}x_{i1}x_{i1}^{\prime}(\gamma^{(1)}_1-\gamma^{(2)}_1),\sum_{i=1}^{N}x_{i2}x_{i2}^{\prime}(\gamma^{(1)}_2-\gamma^{(2)}_2) \ldots, \sum_{i=1}^{N}x_{iT}x_{iT}^{\prime}(\gamma^{(1)}_T-\gamma^{(2)}_T)\right\|^{2}_{F}\notag\\
&=\sum_{t=1}^{T}\left\|\sum_{i=1}^{N}x_{it}x_{it}^{\prime}(\gamma^{(1)}_t-\gamma^{(2)}_t)\right\|^{2}\notag\\
&\leq\max_{t\leq T}\lambda^{2}_{\max}\left(\sum_{i=1}^{N}x_{it}x_{it}^{\prime}\right)\|\Gamma^{(1)}-\Gamma^{(2)}\|^{2}_{F}.
\end{align}
Thus, $\nabla f(\Gamma)$ is Lipschitz continuous with constant $L_f = \max_{t\leq T}\lambda_{\max}(\sum_{i=1}^{N}x_{it}x_{it}^{\prime})$.

\begin{rem}
The equivalence in \eqref{Eqn: Ex4: 3} has greatly simplified the computation of $\hat{\Pi}$, since \eqref{Eqn: NuclearNEst} involves an $Np\times T$ matrix with constraints while \eqref{Eqn: Ex4: 3} involves a matrix of much smaller size. By Lemma \ref{Lem: TechD1}(ii) and (iv), $\hat{K}$ can be equivalently obtained as
\begin{align}\label{Eqn: Ex4: 4}
\hat{K} = \sum_{j=1}^{p}1\{\lambda_{j}(\hat{\Pi}_0 M_{T}\hat{\Pi}_0^{\prime})\geq \delta_{NT}/N\},
\end{align}
and $\hat{\Phi}_0$ as the left singular vector matrix of $\hat{\Pi}_{0}M_{T}$ corresponding to its largest $\hat{K}$ singular values. Moreover, it is straightforward to show that
\begin{align}\label{Eqn: Ex4: 5}
\hat{\phi}_0 = (I_{p}-\hat{\Phi}_0\hat{\Phi}_0^{\prime})\frac{\hat{\Pi}_01_{T}}{T} \text{ and } \hat{F} =  \hat{\Pi}_0^{\prime}\hat{\Phi}_0.
\end{align}
\end{rem}
\begin{rem}
The model in \eqref{Eqn: ModelRef} with $\Pi = 1_{N} \otimes \Pi_0$ can be alternatively viewed as a multivariate linear regression model with reduced rank coefficient matrix $\Pi_0$, which has rank at most $K+1$. Therefore, our result extends Example 1 of \citet{NegahbanWainwright_Scaling_2011} by allowing $x_{it}$ to change over $t$.
\end{rem}

\subsubsection{Technical Lemmas}
Recall that $X_{it}= (e_{N,i}\otimes x_{it}) e_{T,t}^{\prime}$ be an $Np\times T$ matrix of $x_{it}$, where $e_{N,i}$ is the $i$th column of $I_N$ and $e_{T,t}$ is the $t$th column of $I_T$.

\begin{lem}\label{Lem: UseD3}
For any $\Gamma=(\gamma_1,\gamma_2,\ldots,\gamma_T)\in\mathbf{R}^{p\times T}$, we have
\[\frac{1}{2}\sum_{i=1}^{N}\sum_{t=1}^{T}(y_{it}-\mathrm{tr}(X_{it}^{\prime}(1_{N} \otimes \Gamma)))^{2} +\lambda_{NT} \|1_{N} \otimes \Gamma\|_{\ast} \hspace{-0.1cm}=\hspace{-0.1cm} \frac{1}{2}\sum_{i=1}^{N}\sum_{t=1}^{T}(y_{it}-x_{it}^{\prime}\gamma_{t})^{2} +\sqrt{N}\lambda_{NT}\|\Gamma\|_{\ast}.\]
\end{lem}
\noindent{\sc Proof:} Fix $\Gamma=(\gamma_1,\gamma_2,\ldots,\gamma_T)\in\mathbf{R}^{p\times T}$. It is easy to see that $\mathrm{tr}(X_{it}^{\prime}(1_{N} \otimes \Gamma)) = x_{it}^{\prime}\gamma_{t}$. By Lemma \ref{Lem: TechD1}(iii), $\|1_{N} \otimes \Gamma\|_{\ast}=\sqrt{N} \|\Gamma\|_{\ast}$. Thus, the result follows. \qed
\begin{lem}\label{Lem: TechD1}
For any matrix $A$, (i) the rank of $1_{k}\otimes A$ is equal to the rank of $A$; (ii) the nonzero singular values of $1_{k}\otimes A$ are equal to the nonzero singular values of $A$ multiplied by $\sqrt{k}$; (iii) $\|1_{k}\otimes A\|_{\ast} =\sqrt{k} \|A\|_{\ast}$; (iv) the left singular vector matrix of nonzero matrix $1_{k}\otimes A$ corresponding to its nonzero singular values are given by $1_{k}\otimes U/\sqrt{k}$, where $U$ is the left singular vector matrix of $A$ corresponding to its nonzero singular values.
\end{lem}
\noindent{\sc Proof:} It is without loss of generality to assume that $A$ is nonzero. Let $d>0$ be the rank of $A$ and $\sigma_1\geq \sigma_2\geq \ldots\geq \sigma_{d}>0$ be nonzero singular values of $A$. Let $A = U \Sigma V^{\prime}$ be a singular value decomposition of $A$, where $\Sigma$ is a $d\times d$ diagonal matrix with $\sigma_{j}$'s in the diagonal in descending order. It follows that
\begin{align}\label{Eqn: Lem: TechD1: 1}
1_{k}\otimes A = \frac{1}{\sqrt{k}}(1_{k}\otimes U)\sqrt{k}\Sigma V^{\prime},
\end{align}
which gives a singular value decomposition of $1_{k}\otimes A$. Thus, the rank of $1_{k}\otimes A$ is equal to $d$, the nonzero singular values of $1_{k}\otimes A$ given by $\sqrt{k}\sigma_1\geq \sqrt{k}\sigma_2\geq \ldots\geq \sqrt{k}\sigma_{d}>0$, and the left singular vector matrix of $1_{k}\otimes A$ corresponding to its nonzero singular values is $1_{k}\otimes U/\sqrt{k}$. This completes the proof of the lemma. \qed

\section{Additional Discussions}\label{App: F}
\renewcommand{\theequation}{F.\arabic{equation}}
\setcounter{equation}{0}

\subsection{Estimation under \texorpdfstring{$a=0$}{a=0}}\label{App: F: 1}
In the case where $a=0$, we can still utilize the available information to derive estimators for $K$, $B$, and $F$ from $\hat{\Pi}$ in a similar manner. Denote the estimators by $\tilde{K}$, $\tilde{B}$, and $\tilde{F}$. Since $\Pi = BF^{\prime}$, we can obtain $\tilde{K}$ and $\tilde{B}$ from the eigenvalues and eigenvectors of $\hat{\Pi}\hat{\Pi}^{\prime}$. Specifically, $\tilde{K}$ is given by
\begin{align}
\tilde{K} = \sum_{j=1}^{Np}1\{\lambda_{j}(\hat{\Pi}\hat{\Pi}^{\prime})\geq \delta_{NT}\}.
\end{align}
If $\tilde{K} = 0$, $\tilde{B} =0$ and $\tilde{F}=0$; otherwise we proceed as follows. To estimate $B$, we use the following normalization: $B^{\prime}B/N=I_K$ and $F^{\prime}F/T$ being diagonal with diagonal entries in descending order. Then the columns of $\tilde{B}/\sqrt{N}$ are given by the eigenvectors of $\hat{\Pi}\hat{\Pi}^{\prime}$ corresponding to its largest $\tilde{K}$ eigenvalues. Since $F = \Pi^{\prime}B(B^{\prime}B)^{-1}$, we thus obtain
\begin{align}
 \tilde{F} =  \frac{\hat{\Pi}^{\prime}\tilde{B}}{N}.
\end{align}
We can also establish the same convergence rate for the restricted estimators $\tilde{K}$, $\tilde{B}$, and $\tilde{F}$ as in Theorem \ref{Thm: NuclearNRate}(ii). Let $G\equiv (F^{\prime}\tilde{F})(\tilde{F}^{\prime}\tilde{F})^{-1}$. Under the same conditions as in Theorem \ref{Thm: NuclearNRate}(ii), following similar arguments as in its proof, we can establish the following:
\begin{align}
P(\tilde{K}= K) &\to 1,\\
\|\tilde{B} - B G\|_{F}&=O_{p}\left(\frac{\sqrt{K}\lambda_{NT}}{\sqrt{T}}\right),\\
\|\tilde{F}-F(G^{\prime})^{-1}\|_{F}&=O_{p}\left(\frac{\sqrt{K}\lambda_{NT}}{\sqrt{N}}\right).
\end{align}

\subsection{Estimation with Errors \texorpdfstring{in $\alpha_{it}$ and $\beta_{it}$ }{ }}\label{App: F: 2}
Our estimation procedure continues to be effective even when the pricing errors and risk exposures are not fully explained by $x_{it}$. Let $e_{\alpha,it}$ and $e_{\beta,it}$ be the error terms in the pricing errors and the risk exposures, respectively, which are orthogonal to $x_{it}$. In this case, the model becomes:
\begin{align}
y_{it} = [a_i^{\prime}x_{it}+e_{\alpha,it}] + [B_i^{\prime}x_{it}+e_{\beta,it}]^{\prime}f_{t} + \varepsilon_{it} = a_i^{\prime}x_{it} + x_{it}B_if_{t} + \varepsilon^{\ast}_{it},
\end{align}
where $\varepsilon^{\ast}_{it} = \varepsilon_{it} + e_{\alpha,it} + e_{\beta,it}^{\prime}f_{t}$. Since we are not interested in estimating $e_{\alpha,it}$ and $e_{\beta,it}$, our asymptotic results remain valid if we replace $\varepsilon_{it}$ in the original model with $\varepsilon^{\ast}_{it}$.

It is worth to note that the orthogonality between pricing errors ($a_i^{\prime}x_{it}+e_{\alpha,it}$) and risk exposures ($B_i^{\prime}x_{it}+e_{\beta,it}$) cannot used for identification. The orthogonality implies
\begin{align}
\sum_{i=1}^{N}[a_i^{\prime}x_{it}+e_{\alpha,it}][x_{it}^{\prime}B_i+e_{\beta,it}^{\prime}] = \sum_{i=1}^{N}a_i^{\prime}x_{it}x_{it}^{\prime}B_i + \sum_{i=1}^{N}e_{\alpha,it}e_{\beta,it}^{\prime}=0,
\end{align}
which cannot be used for identification, since $e_{\alpha,it}$ and $e_{\beta,it}$ are unobserved. Therefore, we impose $a'B=0$ in Assumption \ref{Ass: NuclearKaBF}(v) for identification of $a$, which is not contradicting with the orthogonality between pricing errors and risk exposures.

\section{Additional Simulations}\label{App: G}
\renewcommand{\theequation}{G.\arabic{equation}}
\setcounter{equation}{0}

\subsection{Sparse \texorpdfstring{$a$ and $B$}{a and B} with \texorpdfstring{$p=37$}{p=37}}\label{App: G: sub1}
We consider the same settings as in Section \ref{Sec6}, but with sparse $a$ and $B$. We also consider three DGPs: DGP4, DGP5, and DGP6. In DGP4,
\begin{align}\label{Eqn: alphabeta1aa}
a_i &= \left(
         \begin{array}{cccccccc}
             0 & 1 & \theta_i & 0 & 0 & 0 &\cdots & 0 \\
         \end{array}
       \right)^{\prime} \text{ and }
\notag\\
B_i &=\left(
       \begin{array}{cccccccc}
         0 & 0 & 0 & 2 & 0 & 0 & \cdots & 0\\
         \varrho_{i} & 0 & 0 &  0 & 0 & 0 & \cdots & 0\\
       \end{array}
     \right)^{\prime},
\end{align}
where $\theta_i$'s are i.i.d. $N(0,1)$ across $i$ and $\varrho_{i}$'s are i.i.d. $U(1,3)$ across $i$. This setup corresponds to Example \ref{Ex: ConitionalObs}. In DGP5,
\begin{align}\label{Eqn: alphabeta2aa}
a_i &= \left(
         \begin{array}{cc}
            \mu_i & \phi^{\prime} \\
         \end{array}
       \right)^{\prime} = \left(
         \begin{array}{cccccccc}
            0 & 1 & 1 & 0 & 0 & 0 & \cdots & 0 \\
         \end{array}
       \right)^{\prime} \text{ and }
\notag\\
B_i &=\left(
         \begin{array}{cc}
            \lambda_i & \Phi^{\prime} \\
         \end{array}
       \right)^{\prime} = \left(
       \begin{array}{cccccccc}
         0 & 0 & 0 & 2 & 0 & 0 & \cdots & 0\\
         \vartheta_{i} &  0 & 0 & 0 & 0 & 0 & \cdots &  0 \\
       \end{array}
     \right)^{\prime},
\end{align}
where $\vartheta_{i}$'s are i.i.d. $U(1,3)$ across $i$. This setup corresponds to Example \ref{Ex: SemiTin}. In DGP6,
\begin{align}\label{Eqn: alphabeta3aa}
a_i &=\phi_0= \left(
         \begin{array}{cccccccc}
          0 & 1 & 1 & 0 & 0 & 0 & \cdots & 0 \\
         \end{array}
       \right)^{\prime} \text{ and }
\notag\\
B_i &=\Phi_0 = \left(
       \begin{array}{cccccccc}
         0 & 0 & 0 & 2 & 0 & 0 & \cdots & 0 \\
         2 & 0 & 0 & 0 & 0 & 0 & \cdots & 0 \\
       \end{array}
     \right)^{\prime}.
\end{align}
This setup corresponds to Example \ref{Ex: HomoConditional}. We implement the same estimation as in Section \ref{Sec6} and observe similar findings, as summarized in Tables \ref{Tab: MSEDGP1aa}-\ref{Tab: MSEDGP3aa}.

\pgfplotstableread{
c	d1	d2	d3	d4	d5	d6	d10	d20	d30	d40	d50	d60
0	30.22761208	29.77666673	30.22475973	28.715897	28.74124076	29.31353064	3.443759291	3.295780703	3.16618448	1.58289552	1.45366112	1.37131363
0.05	21.19279486	20.60484645	20.9649548	18.78968628	18.64704062	18.8861139	3.443759291	3.295780703	3.16618448	1.58289552	1.45366112	1.37131363
0.1	14.73413684	14.12955068	14.45017682	11.96491376	11.71747439	11.74232972	3.443759291	3.295780703	3.16618448	1.58289552	1.45366112	1.37131363
0.2	7.175532034	6.72392872	6.911086142	4.511137487	4.30418674	4.218173227	3.443759291	3.295780703	3.16618448	1.58289552	1.45366112	1.37131363
0.3	4.137187992	3.867193078	3.931490898	2.007572624	1.884976152	1.814948189	3.443759291	3.295780703	3.16618448	1.58289552	1.45366112	1.37131363
0.4	3.410560121	3.218164512	3.129753187	1.537702695	1.416646117	1.345108401	3.443759291	3.295780703	3.16618448	1.58289552	1.45366112	1.37131363
0.5	3.750927188	3.596893203	3.38909462	1.87947374	1.683047919	1.56763262	3.443759291	3.295780703	3.16618448	1.58289552	1.45366112	1.37131363
0.6	4.39085557	4.259758672	4.113113244	2.318197332	2.168252657	2.069558154	3.443759291	3.295780703	3.16618448	1.58289552	1.45366112	1.37131363
0.7	4.639537328	4.389417748	4.291091057	2.178812166	2.020543045	2.014058565	3.443759291	3.295780703	3.16618448	1.58289552	1.45366112	1.37131363
0.8	4.469881884	4.199146491	4.17328277	1.927928778	1.787761081	1.802759553	3.443759291	3.295780703	3.16618448	1.58289552	1.45366112	1.37131363
0.9	4.282451753	4.12118054	4.163596025	1.903793947	1.86382216	1.939205168	3.443759291	3.295780703	3.16618448	1.58289552	1.45366112	1.37131363
1	4.346925019	4.436652187	4.708445557	2.091099364	2.159444215	2.282051519	3.443759291	3.295780703	3.16618448	1.58289552	1.45366112	1.37131363
1.5	6.563254413	6.878033713	7.270513612	3.443583981	3.641980242	3.889317679	3.443759291	3.295780703	3.16618448	1.58289552	1.45366112	1.37131363
2	9.091198839	9.54631569	10.11680493	5.10218996	5.4102677	5.787996694	3.443759291	3.295780703	3.16618448	1.58289552	1.45366112	1.37131363
}\dgpaaa

\begin{figure}[htbp]
\centering
\begin{subfigure}[b]{0.32\textwidth}
\centering
\resizebox{\linewidth}{!}{
\begin{tikzpicture}
\begin{axis}[
    legend style={draw=none},
    grid = minor,
    xmax=2,xmin=0,
    ymax=30,ymin=0,
    xtick={0,1,2},
    ytick={15,30},
    title={$N=500,T=250$},
    tick label style={/pgf/number format/fixed},
legend style={at={(0.2,0.9)},anchor=north,
    row sep = 3pt}]
\addplot[smooth,tension=0.5,color=black, line width=0.75pt,dashdotted] table[x = c,y=d1] from \dgpaaa;
\addplot[smooth,tension=0.5,no markers, color=blue, line width=0.75pt] table[x = c,y=d10] from \dgpaaa;
\legend{\footnotesize Fixed $c$, \footnotesize CV}
\end{axis}
\end{tikzpicture}}
\end{subfigure}
\begin{subfigure}[b]{0.32\textwidth}
\centering
\resizebox{\linewidth}{!}{
\begin{tikzpicture}
\begin{axis}[
    legend style={draw=none},
    grid = minor,
    xmax=2,xmin=0,
    ymax=30,ymin=0,
    xtick={0,1,2},
    ytick={15,30},
    title={$N=1000,T=250$},
    tick label style={/pgf/number format/fixed},
legend style={at={(0.2,0.9)},anchor=north,
    row sep = 3pt}]
\addplot[smooth,tension=0.5,color=black, line width=0.75pt,dashdotted] table[x = c,y=d2] from \dgpaaa;
\addplot[smooth,tension=0.5,no markers, color=blue, line width=0.75pt] table[x = c,y=d20] from \dgpaaa;
\legend{\footnotesize Fixed $c$, \footnotesize CV}
\end{axis}
\end{tikzpicture}}
\end{subfigure}
\begin{subfigure}[b]{0.32\textwidth}
\centering
\resizebox{\linewidth}{!}{
\begin{tikzpicture}
\begin{axis}[
    legend style={draw=none},
    grid = minor,
    xmax=2,xmin=0,
    ymax=30,ymin=0,
    xtick={0,1,2},
    ytick={15,30},
    title={$N=2000,T=250$},
    tick label style={/pgf/number format/fixed},
legend style={at={(0.2,0.9)},anchor=north,
    row sep = 3pt}]
\addplot[smooth,tension=0.5,color=black, line width=0.75pt,dashdotted] table[x = c,y=d3] from \dgpaaa;
\addplot[smooth,tension=0.5,no markers, color=blue, line width=0.75pt] table[x = c,y=d30] from \dgpaaa;
\legend{\footnotesize Fixed $c$, \footnotesize CV}
\end{axis}
\end{tikzpicture}}
\end{subfigure}

\begin{subfigure}[b]{0.32\textwidth}
\centering
\resizebox{\linewidth}{!}{
\begin{tikzpicture}
\begin{axis}[
    legend style={draw=none},
    grid = minor,
    xmax=2,xmin=0,
    ymax=30,ymin=0,
    xtick={0,1,2},
    ytick={15,30},
    title={$N=500,T=500$},
    tick label style={/pgf/number format/fixed},
legend style={at={(0.2,0.9)},anchor=north,
    row sep = 3pt}]
\addplot[smooth,tension=0.5,color=black, line width=0.75pt,dashdotted] table[x = c,y=d4] from \dgpaaa;
\addplot[smooth,tension=0.5,no markers, color=blue, line width=0.75pt] table[x = c,y=d40] from \dgpaaa;
\legend{\footnotesize Fixed $c$, \footnotesize CV}
\end{axis}
\end{tikzpicture}}
\end{subfigure}
\begin{subfigure}[b]{0.32\textwidth}
\centering
\resizebox{\linewidth}{!}{
\begin{tikzpicture}
\begin{axis}[
    legend style={draw=none},
    grid = minor,
    xmax=2,xmin=0,
    ymax=30,ymin=0,
    xtick={0,1,2},
    ytick={15,30},
    title={$N=1000,T=500$},
    tick label style={/pgf/number format/fixed},
legend style={at={(0.2,0.9)},anchor=north,
    row sep = 3pt}]
\addplot[smooth,tension=0.5,color=black, line width=0.75pt,dashdotted] table[x = c,y=d5] from \dgpaaa;
\addplot[smooth,tension=0.5,no markers, color=blue, line width=0.75pt] table[x = c,y=d50] from \dgpaaa;
\legend{\footnotesize Fixed $c$, \footnotesize CV}
\end{axis}
\end{tikzpicture}}
\end{subfigure}
\begin{subfigure}[b]{0.32\textwidth}
\centering
\resizebox{\linewidth}{!}{
\begin{tikzpicture}
\begin{axis}[
    legend style={draw=none},
    grid = minor,
    xmax=2,xmin=0,
    ymax=30,ymin=0,
    xtick={0,1,2},
    ytick={15,30},
    title={$N=2000,T=500$},
    tick label style={/pgf/number format/fixed},
legend style={at={(0.2,0.9)},anchor=north,
    row sep = 3pt}]
\addplot[smooth,tension=0.5,color=black, line width=0.75pt,dashdotted] table[x = c,y=d6] from \dgpaaa;
\addplot[smooth,tension=0.5,no markers, color=blue, line width=0.75pt] table[x = c,y=d60] from \dgpaaa;
\legend{\footnotesize Fixed $c$, \footnotesize CV}
\end{axis}
\end{tikzpicture}}
\end{subfigure}
\caption{Mean square errors of $\hat{\Pi}$ when using fixed $c$ and CV: DGP4}\label{Fig: DGP1aa}
\end{figure}

\pgfplotstableread{
c	d1	d2	d3	d4	d5	d6  d10	d20	d30	d40	d50	d60
0	11.98743856	11.71774021	11.57510246	12.2040618	11.94973465	11.8149917	0.184171971	0.134655727	0.136331175	0.114290618	0.100318881	0.070148928
0.05	3.452883971	3.478778684	3.490763641	3.434093416	3.422261833	3.436710259	0.184171971	0.134655727	0.136331175	0.114290618	0.100318881	0.070148928
0.1	2.937568322	2.964313648	2.97343518	2.916294724	2.899162847	2.912415302	0.184171971	0.134655727	0.136331175	0.114290618	0.100318881	0.070148928
0.2	2.108361147	2.09808112	2.083687229	2.075154038	2.020675338	2.007441343	0.184171971	0.134655727	0.136331175	0.114290618	0.100318881	0.070148928
0.3	1.459771783	1.409398998	1.370567692	1.427640737	1.347171489	1.298698531	0.184171971	0.134655727	0.136331175	0.114290618	0.100318881	0.070148928
0.4	0.974401561	0.893892831	0.829380825	0.945623805	0.853385982	0.779017504	0.184171971	0.134655727	0.136331175	0.114290618	0.100318881	0.070148928
0.5	0.628633991	0.536783144	0.457560943	0.601527118	0.510768654	0.430144056	0.184171971	0.134655727	0.136331175	0.114290618	0.100318881	0.070148928
0.6	0.397724807	0.311613039	0.239336911	0.369489558	0.28989678	0.219610061	0.184171971	0.134655727	0.136331175	0.114290618	0.100318881	0.070148928
0.7	0.257159487	0.187748521	0.136850676	0.225243921	0.161957882	0.111199096	0.184171971	0.134655727	0.136331175	0.114290618	0.100318881	0.070148928
0.8	0.183782565	0.134896073	0.107835127	0.146371565	0.100100982	0.069966891	0.184171971	0.134655727	0.136331175	0.114290618	0.100318881	0.070148928
0.9	0.156973948	0.125703054	0.114315524	0.113016827	0.080538416	0.065180872	0.184171971	0.134655727	0.136331175	0.114290618	0.100318881	0.070148928
1	0.159009697	0.1375479	0.129741872	0.108098213	0.083313477	0.073088166	0.184171971	0.134655727	0.136331175	0.114290618	0.100318881	0.070148928
1.5	0.271873582	0.24330213	0.232773823	0.182439102	0.14660263	0.13088878	0.184171971	0.134655727	0.136331175	0.114290618	0.100318881	0.070148928
2	0.433818075	0.391528792	0.377153291	0.291624772	0.235867947	0.211908686	0.184171971	0.134655727	0.136331175	0.114290618	0.100318881	0.070148928
}\dgpbaa

\begin{figure}[htbp]
\centering
\begin{subfigure}[b]{0.32\textwidth}
\centering
\resizebox{\linewidth}{!}{
\begin{tikzpicture}
\begin{axis}[
    grid = minor,
    xmax=2,xmin=0,
    ymax=10,ymin=0,
    xtick={0,1,2},
    ytick={5,10},
    title={$N=500,T=250$},
    tick label style={/pgf/number format/fixed},
    legend style={draw=none, at={(0.2,0.9)},anchor=north,
    row sep = 3pt}]
\addplot[smooth,tension=0.1,color=black, line width=0.75pt,dashdotted] table[x = c,y=d1] from \dgpbaa;
\addplot[smooth,tension=0.1,no markers, color=blue, line width=0.75pt] table[x = c,y=d10] from \dgpbaa;
\legend{\footnotesize Fixed $c$, \footnotesize CV}
\end{axis}
\end{tikzpicture}}
\end{subfigure}
\begin{subfigure}[b]{0.32\textwidth}
\centering
\resizebox{\linewidth}{!}{
\begin{tikzpicture}
\begin{axis}[
    grid = minor,
    xmax=2,xmin=0,
    ymax=10,ymin=0,
    xtick={0,1,2},
    ytick={5,10},
    title={$N=1000,T=250$},
    tick label style={/pgf/number format/fixed},
    legend style={draw=none, at={(0.2,0.9)},anchor=north,
    row sep = 3pt}]
\addplot[smooth,tension=0.1,color=black, line width=0.75pt,dashdotted] table[x = c,y=d2] from \dgpbaa;
\addplot[smooth,tension=0.1,no markers, color=blue, line width=0.75pt] table[x = c,y=d20] from \dgpbaa;
\legend{\footnotesize Fixed $c$, \footnotesize CV}
\end{axis}
\end{tikzpicture}}
\end{subfigure}
\begin{subfigure}[b]{0.32\textwidth}
\centering
\resizebox{\linewidth}{!}{
\begin{tikzpicture}
\begin{axis}[
    grid = minor,
    xmax=2,xmin=0,
    ymax=10,ymin=0,
    xtick={0,1,2},
    ytick={5,10},
    title={$N=2000,T=250$},
    tick label style={/pgf/number format/fixed},
    legend style={draw=none, at={(0.2,0.9)},anchor=north,
    row sep = 3pt}]
\addplot[smooth,tension=0.1,color=black, line width=0.75pt,dashdotted] table[x = c,y=d3] from \dgpbaa;
\addplot[smooth,tension=0.1,no markers, color=blue, line width=0.75pt] table[x = c,y=d30] from \dgpbaa;
\legend{\footnotesize Fixed $c$, \footnotesize CV}
\end{axis}
\end{tikzpicture}}
\end{subfigure}

\begin{subfigure}[b]{0.32\textwidth}
\centering
\resizebox{\linewidth}{!}{
\begin{tikzpicture}
\begin{axis}[
    grid = minor,
    xmax=2,xmin=0,
    ymax=10,ymin=0,
    xtick={0,1,2},
    ytick={5,10},
    title={$N=500,T=500$},
    tick label style={/pgf/number format/fixed},
    legend style={draw=none, at={(0.2,0.9)},anchor=north,
    row sep = 3pt}]
\addplot[smooth,tension=0.1,color=black, line width=0.75pt,dashdotted] table[x = c,y=d4] from \dgpbaa;
\addplot[smooth,tension=0.1,no markers, color=blue, line width=0.75pt] table[x = c,y=d40] from \dgpbaa;
\legend{\footnotesize Fixed $c$, \footnotesize CV}
\end{axis}
\end{tikzpicture}}
\end{subfigure}
\begin{subfigure}[b]{0.32\textwidth}
\centering
\resizebox{\linewidth}{!}{
\begin{tikzpicture}
\begin{axis}[
    grid = minor,
    xmax=2,xmin=0,
    ymax=10,ymin=0,
    xtick={0,1,2},
    ytick={5,10},
    title={$N=1000,T=500$},
    tick label style={/pgf/number format/fixed},
    legend style={draw=none, at={(0.2,0.9)},anchor=north,
    row sep = 3pt}]
\addplot[smooth,tension=0.1,color=black, line width=0.75pt,dashdotted] table[x = c,y=d5] from \dgpbaa;
\addplot[smooth,tension=0.1,no markers, color=blue, line width=0.75pt] table[x = c,y=d50] from \dgpbaa;
\legend{\footnotesize Fixed $c$, \footnotesize CV}
\end{axis}
\end{tikzpicture}}
\end{subfigure}
\begin{subfigure}[b]{0.32\textwidth}
\centering
\resizebox{\linewidth}{!}{
\begin{tikzpicture}
\begin{axis}[
    grid = minor,
    xmax=2,xmin=0,
    ymax=10,ymin=0,
    xtick={0,1,2},
    ytick={5,10},
    title={$N=2000,T=500$},
    tick label style={/pgf/number format/fixed},
    legend style={draw=none, at={(0.2,0.9)},anchor=north,
    row sep = 3pt}]
\addplot[smooth,tension=0.1,color=black, line width=0.75pt,dashdotted] table[x = c,y=d6] from \dgpbaa;
\addplot[smooth,tension=0.1,no markers, color=blue, line width=0.75pt] table[x = c,y=d60] from \dgpbaa;
\legend{\footnotesize Fixed $c$, \footnotesize CV}
\end{axis}
\end{tikzpicture}}
\end{subfigure}
\caption{Mean square errors of $(\hat{\Pi}^{\diamond\prime}, \sqrt{N}\hat{\Pi}^{\ast\prime})$ when using fixed $c$ and CV: DGP5}\label{Fig: DGP2aa}
\end{figure}

\pgfplotstableread{
c	d1	d2	d3	d4	d5	d6	d10	d20	d30	d40	d50	d60
0	0.315224246	0.151241167	0.074309309	0.315433208	0.151431044	0.074182378	0.047333759	0.025184553	0.012207841	0.043707479	0.023818231	0.011558331
0.05	0.274302167	0.131693064	0.064534478	0.275261265	0.132249652	0.064606905	0.047333759	0.025184553	0.012207841	0.043707479	0.023818231	0.011558331
0.1	0.237354963	0.11392085	0.055637458	0.238840599	0.114738446	0.055854143	0.047333759	0.025184553	0.012207841	0.043707479	0.023818231	0.011558331
0.2	0.174725513	0.083548763	0.040435153	0.176641265	0.084584999	0.040780164	0.047333759	0.025184553	0.012207841	0.043707479	0.023818231	0.011558331
0.3	0.126090623	0.059825397	0.028623864	0.127681548	0.060689312	0.028888506	0.047333759	0.025184553	0.012207841	0.043707479	0.023818231	0.011558331
0.4	0.090307512	0.042465027	0.020123394	0.090897844	0.042792434	0.020104171	0.047333759	0.025184553	0.012207841	0.043707479	0.023818231	0.011558331
0.5	0.066324258	0.031166711	0.014827492	0.065332325	0.030639069	0.014353458	0.047333759	0.025184553	0.012207841	0.043707479	0.023818231	0.011558331
0.6	0.052643249	0.025221019	0.012323139	0.050100608	0.023954614	0.011528317	0.047333759	0.025184553	0.012207841	0.043707479	0.023818231	0.011558331
0.7	0.047011134	0.023400466	0.011910032	0.04377299	0.021896525	0.011054209	0.047333759	0.025184553	0.012207841	0.043707479	0.023818231	0.011558331
0.8	0.047172171	0.024437577	0.012868311	0.043906063	0.022994599	0.012047541	0.047333759	0.025184553	0.012207841	0.043707479	0.023818231	0.011558331
0.9	0.051140884	0.027198791	0.014569705	0.04801865	0.025786651	0.013709213	0.047333759	0.025184553	0.012207841	0.043707479	0.023818231	0.011558331
1	0.057287572	0.030816854	0.016609468	0.054031292	0.029249762	0.015633604	0.047333759	0.025184553	0.012207841	0.043707479	0.023818231	0.011558331
1.5	0.100554471	0.055054675	0.030084377	0.094839103	0.05218589	0.028327156	0.047333759	0.025184553	0.012207841	0.043707479	0.023818231	0.011558331
2	0.161179156	0.089001279	0.048945648	0.15200433	0.084288656	0.046097871	0.047333759	0.025184553	0.012207841	0.043707479	0.023818231	0.011558331
}\dgpcaa

\begin{figure}[htbp]
\centering
\begin{subfigure}[b]{0.32\textwidth}
\centering
\resizebox{\linewidth}{!}{
\begin{tikzpicture}
\begin{axis}[
    legend style={draw=none},
    grid = minor,
    xmax=2,xmin=0,
    ymax=0.5,ymin=0,
    xtick={0,1,2},
    ytick={0.25,0.5},
    title={$N=500,T=250$},
    tick label style={/pgf/number format/fixed},
legend style={at={(0.2,0.9)},anchor=north,
    row sep = 3pt}]
\addplot[smooth,tension=0.5,color=black, line width=0.75pt,dashdotted] table[x = c,y=d1] from \dgpcaa;
\addplot[smooth,tension=0.5,no markers, color=blue, line width=0.75pt] table[x = c,y=d10] from \dgpcaa;
\legend{\footnotesize Fixed $c$, \footnotesize CV}
\end{axis}
\end{tikzpicture}}
\end{subfigure}
\begin{subfigure}[b]{0.32\textwidth}
\centering
\resizebox{\linewidth}{!}{
\begin{tikzpicture}
\begin{axis}[
    legend style={draw=none},
    grid = minor,
    xmax=2,xmin=0,
    ymax=0.25,ymin=0,
    xtick={0,1,2},
    ytick={0.12,0.25},
    title={$N=1000,T=250$},
    tick label style={/pgf/number format/fixed},
legend style={at={(0.2,0.9)},anchor=north,
    row sep = 3pt}]
\addplot[smooth,tension=0.5,color=black, line width=0.75pt,dashdotted] table[x = c,y=d2] from \dgpcaa;
\addplot[smooth,tension=0.5,no markers, color=blue, line width=0.75pt] table[x = c,y=d20] from \dgpcaa;
\legend{\footnotesize Fixed $c$, \footnotesize CV}
\end{axis}
\end{tikzpicture}}
\end{subfigure}
\begin{subfigure}[b]{0.32\textwidth}
\centering
\resizebox{\linewidth}{!}{
\begin{tikzpicture}
\begin{axis}[
    legend style={draw=none},
    grid = minor,
    xmax=2,xmin=0,
    ymax=0.12,ymin=0,
    xtick={0,1,2},
    ytick={0.06,0.12},
    title={$N=2000,T=250$},
    tick label style={/pgf/number format/fixed},
legend style={at={(0.2,0.9)},anchor=north,
    row sep = 3pt}]
\addplot[smooth,tension=0.5,color=black, line width=0.75pt,dashdotted] table[x = c,y=d3] from \dgpcaa;
\addplot[smooth,tension=0.5,no markers, color=blue, line width=0.75pt] table[x = c,y=d30] from \dgpcaa;
\legend{\footnotesize Fixed $c$, \footnotesize CV}
\end{axis}
\end{tikzpicture}}
\end{subfigure}

\begin{subfigure}[b]{0.32\textwidth}
\centering
\resizebox{\linewidth}{!}{
\begin{tikzpicture}
\begin{axis}[
    legend style={draw=none},
    grid = minor,
    xmax=2,xmin=0,
    ymax=0.5,ymin=0,
    xtick={0,1,2},
    ytick={0.25,0.5},
    title={$N=500,T=500$},
    tick label style={/pgf/number format/fixed},
legend style={at={(0.2,0.9)},anchor=north,
    row sep = 3pt}]
\addplot[smooth,tension=0.5,color=black, line width=0.75pt,dashdotted] table[x = c,y=d4] from \dgpcaa;
\addplot[smooth,tension=0.5,no markers, color=blue, line width=0.75pt] table[x = c,y=d40] from \dgpcaa;
\legend{\footnotesize Fixed $c$, \footnotesize CV}
\end{axis}
\end{tikzpicture}}
\end{subfigure}
\begin{subfigure}[b]{0.32\textwidth}
\centering
\resizebox{\linewidth}{!}{
\begin{tikzpicture}
\begin{axis}[
    legend style={draw=none},
    grid = minor,
    xmax=2,xmin=0,
    ymax=0.25,ymin=0,
    xtick={0,1,2},
    ytick={0.12,0.25},
    title={$N=1000,T=500$},
    tick label style={/pgf/number format/fixed},
legend style={at={(0.2,0.9)},anchor=north,
    row sep = 3pt}]
\addplot[smooth,tension=0.5,color=black, line width=0.75pt,dashdotted] table[x = c,y=d5] from \dgpcaa;
\addplot[smooth,tension=0.5,no markers, color=blue, line width=0.75pt] table[x = c,y=d50] from \dgpcaa;
\legend{\footnotesize Fixed $c$, \footnotesize CV}
\end{axis}
\end{tikzpicture}}
\end{subfigure}
\begin{subfigure}[b]{0.32\textwidth}
\centering
\resizebox{\linewidth}{!}{
\begin{tikzpicture}
\begin{axis}[
    legend style={draw=none},
    grid = minor,
    xmax=2,xmin=0,
    ymax=0.12,ymin=0,
    xtick={0,1,2},
    ytick={0.06,0.12},
    title={$N=2000,T=500$},
    tick label style={/pgf/number format/fixed},
legend style={at={(0.2,0.9)},anchor=north,
    row sep = 3pt}]
\addplot[smooth,tension=0.5,color=black, line width=0.75pt,dashdotted] table[x = c,y=d6] from \dgpcaa;
\addplot[smooth,tension=0.5,no markers, color=blue, line width=0.75pt] table[x = c,y=d60] from \dgpcaa;
\legend{\footnotesize Fixed $c$, \footnotesize CV}
\end{axis}
\end{tikzpicture}}
\end{subfigure}
\caption{Mean square errors of $\hat{\Pi}_0$ when using fixed $c$ and CV: DGP6}\label{Fig: DGP3aa}
\end{figure}

\setlength{\tabcolsep}{18pt}
\begin{table}[htbp]
\centering
\resizebox{0.99\textwidth}{!}{
\begin{threeparttable}
\renewcommand{\arraystretch}{1.35}
\caption{Mean square errors of $\hat{\Pi}$, $\hat{a}$, $\hat{B}$, and $\hat{F}$, and correct rates of $\hat{K}$: DGP4\tnote{\dag}}\label{Tab: MSEDGP1aa}
\begin{tabular}{cccccccccc}
\hline\hline
&(N,T)&&$\hat{\Pi}$&$\hat{a}$&$\hat{B}$&$\hat{F}$&&$\hat{K}$&\\
\cline{2-9}
&$(500,250)$    &&3.444&1.299&1.095&0.157&&0.000&\\
&$(1000,250)$   &&3.296&1.352&1.029&0.148&&0.000&\\
&$(2000,250)$   &&3.166&1.316&0.975&0.138&&0.000&\\
\cline{2-9}
&$(500,500)$   &&1.583&1.012&0.315&0.074&&1.000&\\
&$(1000,500)$  &&1.454&0.941&0.292&0.052&&1.000&\\
&$(2000,500)$  &&1.371&0.904&0.273&0.039&&1.000&\\
\hline\hline
\end{tabular}
\begin{tablenotes}
      \small
      \item[\dag] The mean square errors of $\hat{\Pi}$, $\hat{a}$ , $\hat{B}$, and $\hat{F}$ are given by $\sum_{\ell=1}^{200}\|\hat{\Pi}^{(\ell)}-\Pi\|_{F}^2/200NT$, $\sum_{\ell=1}^{200}\|\hat{a}^{(\ell)}-a\|^2/200N$, $\sum_{\ell=1}^{200}\|\hat{B}^{(\ell)}-BH^{(\ell)}\|_{F}^2/200N$ and $\sum_{\ell=1}^{200}\|\hat{F}^{(\ell)}- F(H^{{(\ell)}\prime})^{-1}\|_{F}^2/{200T}$, where $\hat{\Pi}^{(\ell)}$, $\hat{a}^{(\ell)}$, $\hat{B}^{(\ell)}$, and $\hat{F}^{(\ell)}$ are estimates in the $\ell$th simulation replication, and $H^{(\ell)}\equiv (F^{\prime}M_T\hat{F}^{(\ell)})(\hat{F}^{{(\ell)}\prime}M_T\hat{F}^{(\ell)})^{-1}$ is a rotational transformation matrix. The value of $c$ is chosen from $\{0, 0.05, 0.1, 0.2, \ldots, 0.9,1,1.5,2\}$ by using the 5-fold CV method as outlined in Section \ref{Sec3}.
    \end{tablenotes}
\end{threeparttable}
}
\end{table}

\setlength{\tabcolsep}{12pt}
\begin{table}[htbp]
\centering
\resizebox{0.99\textwidth}{!}{
\begin{threeparttable}
\renewcommand{\arraystretch}{1.35}
\caption{Mean square errors of $\hat{\Pi}^{\diamond}$, $\hat{\Pi}^{\ast}$, $\hat{\mu}$, $\hat\Lambda$, $\hat{\phi}, \hat{\Phi}$, and $\hat{F}$, and correct rates of $\hat{K}$: DGP5\tnote{\dag}}\label{Tab: MSEDGP2aa}
\begin{tabular}{ccccccccccccc}
\hline\hline
&(N,T)&&$\hat{\Pi}^{\diamond}$&$\hat{\Pi}^{\ast}$&$\hat{\mu}$&$\hat\Lambda$&$\hat{\phi}$&$\hat{\Phi}$&$\hat{F}$&&$\hat{K}$&\\
\cline{2-12}
&$(500,250)$     &&0.123&0.062&0.056&0.010&0.083&0.009&0.032&&1.000&\\
&$(1000,250)$    &&0.088&0.046&0.057&0.010&0.071&0.008&0.023&&1.000&\\
&$(2000,250)$    &&0.102&0.034&0.061&0.009&0.054&0.006&0.016&&1.000&\\
\cline{2-12}
&$(500,500)$    &&0.067&0.048&0.029&0.006&0.060&0.006&0.028&&1.000&\\
&$(1000,500)$   &&0.070&0.031&0.031&0.006&0.042&0.004&0.017&&1.000&\\
&$(2000,500)$   &&0.047&0.023&0.031&0.005&0.034&0.004&0.012&&1.000&\\
\hline\hline
\end{tabular}
\begin{tablenotes}
      \small
      \item[\dag] The mean square errors of $\hat{\Pi}^{\diamond}$, $\hat{\Pi}^{\ast}$, $\hat{\mu}$, $\hat\Lambda$, $\hat{\phi}, \hat{\Phi}$, and $\hat{F}$ are given by $\sum_{\ell=1}^{200}\|\hat{\Pi}^{\diamond(\ell)}-\Pi^{\diamond}\|_{F}^2/200NT$, $\sum_{\ell=1}^{200}\|\hat{\Pi}^{\ast(\ell)}-\Pi^{\ast}\|_{F}^2/200T$,$\sum_{\ell=1}^{200}\|\hat{\mu}^{(\ell)}-\mu\|^2/200N$, $\sum_{\ell=1}^{200}\|\hat{\Lambda}^{(\ell)}-\Lambda H^{(\ell)}\|_{F}^2/200N$, $\sum_{\ell=1}^{200}\|\hat{\phi}^{(\ell)}-\phi\|^2/200$, $\sum_{\ell=1}^{200}\|\hat{\Phi}^{(\ell)}-\Phi H^{(\ell)}\|^2/200$ and $\sum_{\ell=1}^{200}\|\hat{F}^{(\ell)}- F(H^{{(\ell)}\prime})^{-1}\|_{F}^2/{200T}$, where $\hat{\Pi}^{\diamond(\ell)}$, $\hat{\Pi}^{\ast(\ell)}$, $\hat{\mu}^{(\ell)}$, $\hat{\Lambda}^{(\ell)}$, $\hat{\phi}^{(\ell)}$, $\hat{\Phi}^{(\ell)}$, and $\hat{F}^{(\ell)}$ are estimates in the $\ell$th simulation replication, and $H^{(\ell)}\equiv (F^{\prime}M_T\hat{F}^{(\ell)})(\hat{F}^{{(\ell)}\prime}M_T\hat{F}^{(\ell)})^{-1}$ is a rotational transformation matrix. The value of $c$ is chosen from $\{0, 0.05, 0.1, 0.2, \ldots, 0.9,1,1.5,2\}$ by using the 5-fold CV method as outlined in Section \ref{Sec3}.
    \end{tablenotes}
\end{threeparttable}
}
\end{table}

\setlength{\tabcolsep}{18pt}
\begin{table}[htbp]
\centering
\resizebox{0.99\textwidth}{!}{
\begin{threeparttable}
\renewcommand{\arraystretch}{1.35}
\caption{Mean square errors of $\hat{\Pi}_0$, $\hat{\phi}_0$, $\hat{\Phi}_0$, and $\hat{F}$ ($\times 10^{-2}$), and correct rates of $\hat{K}$: DGP6\tnote{\dag}}\label{Tab: MSEDGP3aa}
\begin{tabular}{cccccccccc}
\hline\hline
&(N,T)&&$\hat{\Pi}_0$&$\hat{\phi}_0$&$\hat{\Phi}_0$&$\hat{F}$&&$\hat{K}$&\\
\cline{2-9}
&$(500,250)$    &&4.733&2.552&0.342&2.033&&1.000&\\
&$(1000,250)$   &&2.519&1.112&0.146&0.998&&1.000&\\
&$(2000,250)$   &&1.221&0.600&0.079&0.523&&1.000&\\
\cline{2-9}
&$(500,500)$   &&4.371&2.093&0.288&1.967&&1.000&\\
&$(1000,500)$  &&2.382&0.870&0.118&0.960&&1.000&\\
&$(2000,500)$  &&1.156&0.475&0.065&0.510&&1.000&\\
\hline\hline
\end{tabular}
\begin{tablenotes}
      \small
      \item[\dag] The mean square errors of $\hat{\Pi}_0$, $\hat{\phi}_0$ , $\hat{\Phi}_0$, and $\hat{F}$ are given by $\sum_{\ell=1}^{200}\|\hat{\Pi}_0^{(\ell)}-\Pi_0\|_{F}^2/200T$, $\sum_{\ell=1}^{200}\|\hat{\phi}_0^{(\ell)}-\phi\|^2/200$, $\sum_{\ell=1}^{200}\|\hat{\Phi}_0^{(\ell)}-\Phi H^{(\ell)}\|_{F}^2/200$ and $\sum_{\ell=1}^{200}\|\hat{F}^{(\ell)}- F(H^{{(\ell)}\prime})^{-1}\|_{F}^2/{200T}$, where $\hat{\Pi}_0^{(\ell)}$, $\hat{\phi}_0^{(\ell)}$, $\hat{\Phi}_0^{(\ell)}$, and $\hat{F}^{(\ell)}$ are estimates in the $\ell$th simulation replication, and $H^{(\ell)}\equiv (F^{\prime}M_T\hat{F}^{(\ell)})(\hat{F}^{{(\ell)}\prime}M_T\hat{F}^{(\ell)})^{-1}$ is a rotational transformation matrix. The value of $c$ is chosen from $\{0, 0.05, 0.1, 0.2, \ldots, 0.9,1,1.5,2\}$ by using the 5-fold CV method as outlined in Section \ref{Sec3}.
    \end{tablenotes}
\end{threeparttable}
}
\end{table}

\subsection{Settings with \texorpdfstring{$p=4$}{p=4}}\label{App: G: sub2}
We consider settings with a small number of covariates in $x_{it}$. Specially, let $x_{it} = (x_{it,1},x_{it,2},$ $x_{it,3},x_{it,4})^{\prime}$, which consist of the first four covariates from Section \ref{Sec6}. We also consider three DGPs: DGP7, DGP8, and DGP9, corresponding to the settings described in Examples \ref{Ex: ConitionalObs}, \ref{Ex: SemiTin}, and \ref{Ex: HomoConditional}, respectively. For each DGP, let $a_i$ be the vector containing the first four elements of $a_i$ in Section \ref{App: G: sub1} and $B_i$ be the matrix consisting of the first four rows of $B_i$. The error terms $\varepsilon_{it}$'s and latent factors $f_t$'s are generated as described in Section \ref{Sec6}. Given $p=4$, we investigate cases with small values of $N$ and $T$, specifically  $N=50, 100, 200$ and $T=50, 100, 200$. Our estimators demonstrate the same promising performance in these settings, as summarized in Tables \ref{Tab: MSEDGP1app}-\ref{Tab: MSEDGP3app}.

\pgfplotstableread{
c	d1	d2	d3	d4	d5	d6	d7	d8	d9	d10	d20	d30	d40	d50	d60	d70	d80	d90
0	20.2515	20.3613	22.2782	19.8983	21.2673	22.0438	20.3043	21.1392	22.6044	2.6072	1.6100	1.1760	2.3234	1.3322	0.8765	2.1105	1.1550	0.7074
0.05	8.7773	8.3534	9.3906	8.4439	8.4450	8.7753	8.2772	8.4190	8.2813	2.6072	1.6100	1.1760	2.3234	1.3322	0.8765	2.1105	1.1550	0.7074
0.1	4.8728	4.2295	4.5268	4.5266	3.9720	3.9323	4.2380	3.7640	3.3861	2.6072	1.6100	1.1760	2.3234	1.3322	0.8765	2.1105	1.1550	0.7074
0.2	2.9968	2.3158	2.0729	2.5915	1.9794	1.6963	2.3065	1.6608	1.4407	2.6072	1.6100	1.1760	2.3234	1.3322	0.8765	2.1105	1.1550	0.7074
0.3	2.7785	2.0626	1.6828	2.3868	1.7198	1.3679	2.1259	1.4096	1.1746	2.6072	1.6100	1.1760	2.3234	1.3322	0.8765	2.1105	1.1550	0.7074
0.4	2.8048	2.0152	1.5928	2.4496	1.6854	1.2816	2.2280	1.4076	1.0861	2.6072	1.6100	1.1760	2.3234	1.3322	0.8765	2.1105	1.1550	0.7074
0.5	2.7593	1.9066	1.5089	2.2915	1.5907	1.1819	2.2430	1.3607	0.9655	2.6072	1.6100	1.1760	2.3234	1.3322	0.8765	2.1105	1.1550	0.7074
0.6	2.6275	1.7441	1.3733	2.3145	1.4361	1.0288	2.1341	1.2219	0.8129	2.6072	1.6100	1.1760	2.3234	1.3322	0.8765	2.1105	1.1550	0.7074
0.7	2.5277	1.6357	1.2444	2.2199	1.3363	0.9145	2.0451	1.1166	0.7247	2.6072	1.6100	1.1760	2.3234	1.3322	0.8765	2.1105	1.1550	0.7074
0.8	2.5188	1.6105	1.1789	2.2222	1.3207	0.8765	2.0630	1.1130	0.7048	2.6072	1.6100	1.1760	2.3234	1.3322	0.8765	2.1105	1.1550	0.7074
0.9	2.5699	1.6228	1.1653	2.2915	1.3426	0.8749	2.1552	1.1530	0.6976	2.6072	1.6100	1.1760	2.3234	1.3322	0.8765	2.1105	1.1550	0.7074
1	2.6356	1.6385	1.1739	2.3769	1.3665	0.8720	2.2572	1.1855	0.6990	2.6072	1.6100	1.1760	2.3234	1.3322	0.8765	2.1105	1.1550	0.7074
1.5	3.2421	1.9945	1.3925	3.0918	1.7961	1.0990	3.1559	1.7093	0.9838	2.6072	1.6100	1.1760	2.3234	1.3322	0.8765	2.1105	1.1550	0.7074
2	4.3059	2.7448	1.9324	4.3002	2.6474	1.6361	4.5399	2.6242	1.5413	2.6072	1.6100	1.1760	2.3234	1.3322	0.8765	2.1105	1.1550	0.7074
}\dgpcaa

\begin{figure}[htbp]
\centering
\begin{subfigure}[b]{0.32\textwidth}
\centering
\resizebox{\linewidth}{!}{
\begin{tikzpicture}
\begin{axis}[
    legend style={draw=none},
    grid = minor,
    xmax=2,xmin=0,
    ymax=20,ymin=0,
    xtick={0,1,2},
    ytick={10,20},
    title={$N=50,T=50$},
    tick label style={/pgf/number format/fixed},
legend style={at={(0.2,0.9)},anchor=north,
    row sep = 3pt}]
\addplot[smooth,tension=0.5,color=black, line width=0.75pt,dashdotted] table[x = c,y=d1] from \dgpcaa;
\addplot[smooth,tension=0.5,no markers, color=blue, line width=0.75pt] table[x = c,y=d10] from \dgpcaa;
\legend{\footnotesize Fixed $c$, \footnotesize CV}
\end{axis}
\end{tikzpicture}}
\end{subfigure}
\begin{subfigure}[b]{0.32\textwidth}
\centering
\resizebox{\linewidth}{!}{
\begin{tikzpicture}
\begin{axis}[
    legend style={draw=none},
    grid = minor,
    xmax=2,xmin=0,
    ymax=20,ymin=0,
    xtick={0,1,2},
    ytick={10,20},
    title={$N=100,T=50$},
    tick label style={/pgf/number format/fixed},
legend style={at={(0.2,0.9)},anchor=north,
    row sep = 3pt}]
\addplot[smooth,tension=0.5,color=black, line width=0.75pt,dashdotted] table[x = c,y=d4] from \dgpcaa;
\addplot[smooth,tension=0.5,no markers, color=blue, line width=0.75pt] table[x = c,y=d40] from \dgpcaa;
\legend{\footnotesize Fixed $c$, \footnotesize CV}
\end{axis}
\end{tikzpicture}}
\end{subfigure}
\begin{subfigure}[b]{0.32\textwidth}
\centering
\resizebox{\linewidth}{!}{
\begin{tikzpicture}
\begin{axis}[
    legend style={draw=none},
    grid = minor,
    xmax=2,xmin=0,
    ymax=20,ymin=0,
    xtick={0,1,2},
    ytick={10,20},
    title={$N=200,T=50$},
    tick label style={/pgf/number format/fixed},
legend style={at={(0.2,0.9)},anchor=north,
    row sep = 3pt}]
\addplot[smooth,tension=0.5,color=black, line width=0.75pt,dashdotted] table[x = c,y=d7] from \dgpcaa;
\addplot[smooth,tension=0.5,no markers, color=blue, line width=0.75pt] table[x = c,y=d70] from \dgpcaa;
\legend{\footnotesize Fixed $c$, \footnotesize CV}
\end{axis}
\end{tikzpicture}}
\end{subfigure}

\begin{subfigure}[b]{0.32\textwidth}
\centering
\resizebox{\linewidth}{!}{
\begin{tikzpicture}
\begin{axis}[
    legend style={draw=none},
    grid = minor,
    xmax=2,xmin=0,
    ymax=20,ymin=0,
    xtick={0,1,2},
    ytick={10,20},
    title={$N=50,T=100$},
    tick label style={/pgf/number format/fixed},
legend style={at={(0.2,0.9)},anchor=north,
    row sep = 3pt}]
\addplot[smooth,tension=0.5,color=black, line width=0.75pt,dashdotted] table[x = c,y=d2] from \dgpcaa;
\addplot[smooth,tension=0.5,no markers, color=blue, line width=0.75pt] table[x = c,y=d20] from \dgpcaa;
\legend{\footnotesize Fixed $c$, \footnotesize CV}
\end{axis}
\end{tikzpicture}}
\end{subfigure}
\begin{subfigure}[b]{0.32\textwidth}
\centering
\resizebox{\linewidth}{!}{
\begin{tikzpicture}
\begin{axis}[
    legend style={draw=none},
    grid = minor,
    xmax=2,xmin=0,
    ymax=20,ymin=0,
    xtick={0,1,2},
    ytick={10,20},
    title={$N=100,T=100$},
    tick label style={/pgf/number format/fixed},
legend style={at={(0.2,0.9)},anchor=north,
    row sep = 3pt}]
\addplot[smooth,tension=0.5,color=black, line width=0.75pt,dashdotted] table[x = c,y=d5] from \dgpcaa;
\addplot[smooth,tension=0.5,no markers, color=blue, line width=0.75pt] table[x = c,y=d50] from \dgpcaa;
\legend{\footnotesize Fixed $c$, \footnotesize CV}
\end{axis}
\end{tikzpicture}}
\end{subfigure}
\begin{subfigure}[b]{0.32\textwidth}
\centering
\resizebox{\linewidth}{!}{
\begin{tikzpicture}
\begin{axis}[
    legend style={draw=none},
    grid = minor,
    xmax=2,xmin=0,
    ymax=20,ymin=0,
    xtick={0,1,2},
    ytick={10,20},
    title={$N=200,T=100$},
    tick label style={/pgf/number format/fixed},
legend style={at={(0.2,0.9)},anchor=north,
    row sep = 3pt}]
\addplot[smooth,tension=0.5,color=black, line width=0.75pt,dashdotted] table[x = c,y=d8] from \dgpcaa;
\addplot[smooth,tension=0.5,no markers, color=blue, line width=0.75pt] table[x = c,y=d80] from \dgpcaa;
\legend{\footnotesize Fixed $c$, \footnotesize CV}
\end{axis}
\end{tikzpicture}}
\end{subfigure}

\begin{subfigure}[b]{0.32\textwidth}
\centering
\resizebox{\linewidth}{!}{
\begin{tikzpicture}
\begin{axis}[
    legend style={draw=none},
    grid = minor,
    xmax=2,xmin=0,
    ymax=20,ymin=0,
    xtick={0,1,2},
    ytick={10,20},
    title={$N=50,T=200$},
    tick label style={/pgf/number format/fixed},
legend style={at={(0.2,0.9)},anchor=north,
    row sep = 3pt}]
\addplot[smooth,tension=0.5,color=black, line width=0.75pt,dashdotted] table[x = c,y=d3] from \dgpcaa;
\addplot[smooth,tension=0.5,no markers, color=blue, line width=0.75pt] table[x = c,y=d30] from \dgpcaa;
\legend{\footnotesize Fixed $c$, \footnotesize CV}
\end{axis}
\end{tikzpicture}}
\end{subfigure}
\begin{subfigure}[b]{0.32\textwidth}
\centering
\resizebox{\linewidth}{!}{
\begin{tikzpicture}
\begin{axis}[
    legend style={draw=none},
    grid = minor,
    xmax=2,xmin=0,
    ymax=20,ymin=0,
    xtick={0,1,2},
    ytick={10,20},
    title={$N=100,T=200$},
    tick label style={/pgf/number format/fixed},
legend style={at={(0.2,0.9)},anchor=north,
    row sep = 3pt}]
\addplot[smooth,tension=0.5,color=black, line width=0.75pt,dashdotted] table[x = c,y=d6] from \dgpcaa;
\addplot[smooth,tension=0.5,no markers, color=blue, line width=0.75pt] table[x = c,y=d60] from \dgpcaa;
\legend{\footnotesize Fixed $c$, \footnotesize CV}
\end{axis}
\end{tikzpicture}}
\end{subfigure}
\begin{subfigure}[b]{0.32\textwidth}
\centering
\resizebox{\linewidth}{!}{
\begin{tikzpicture}
\begin{axis}[
    legend style={draw=none},
    grid = minor,
    xmax=2,xmin=0,
    ymax=20,ymin=0,
    xtick={0,1,2},
    ytick={10,20},
    title={$N=200,T=200$},
    tick label style={/pgf/number format/fixed},
legend style={at={(0.2,0.9)},anchor=north,
    row sep = 3pt}]
\addplot[smooth,tension=0.5,color=black, line width=0.75pt,dashdotted] table[x = c,y=d9] from \dgpcaa;
\addplot[smooth,tension=0.5,no markers, color=blue, line width=0.75pt] table[x = c,y=d90] from \dgpcaa;
\legend{\footnotesize Fixed $c$, \footnotesize CV}
\end{axis}
\end{tikzpicture}}
\end{subfigure}
\caption{Mean square errors of $\hat{\Pi}$ when using fixed $c$ and CV: DGP7}\label{Fig: DGP1app}
\end{figure}

\pgfplotstableread{
c	d1	d2	d3	d4	d5	d6	d7	d8	d9	d10	d20	d30	d40	d50	d60	d70	d80	d90
0	12.67393356	11.41031947	11.84613728	12.41605098	11.10453489	11.60711175	12.30252656	10.93409589	11.46237063	0.91912723	0.707008528	0.601244938	0.7056535	0.498150481	0.378188106	0.595965384	0.386001153	0.264869037
0.05	3.974180755	3.890385144	3.857848232	3.887588752	3.772115676	3.73972767	3.888951947	3.708239015	3.619025818	0.91912723	0.707008528	0.601244938	0.7056535	0.498150481	0.378188106	0.595965384	0.386001153	0.264869037
0.1	3.562319958	3.457685376	3.457478626	3.38436899	3.265023221	3.225541933	3.341051748	3.176809903	3.099697029	0.91912723	0.707008528	0.601244938	0.7056535	0.498150481	0.378188106	0.595965384	0.386001153	0.264869037
0.2	2.865198839	2.770414841	2.78067724	2.66778519	2.518437369	2.481636924	2.595430828	2.391159606	2.299422484	0.91912723	0.707008528	0.601244938	0.7056535	0.498150481	0.378188106	0.595965384	0.386001153	0.264869037
0.3	2.288945784	2.182125013	2.183931617	2.060592045	1.912190772	1.86491144	1.960339389	1.754203542	1.66560621	0.91912723	0.707008528	0.601244938	0.7056535	0.498150481	0.378188106	0.595965384	0.386001153	0.264869037
0.4	1.838467404	1.713537044	1.693329825	1.583430258	1.436658028	1.375063277	1.450279675	1.26202103	1.179658486	0.91912723	0.707008528	0.601244938	0.7056535	0.498150481	0.378188106	0.595965384	0.386001153	0.264869037
0.5	1.490463681	1.34822532	1.302186617	1.224955763	1.077280914	1.002288305	1.069415201	0.898649638	0.819790995	0.91912723	0.707008528	0.601244938	0.7056535	0.498150481	0.378188106	0.595965384	0.386001153	0.264869037
0.6	1.233740319	1.075248235	1.0060592	0.968171756	0.817063432	0.730915405	0.806874023	0.645091233	0.566061355	0.91912723	0.707008528	0.601244938	0.7056535	0.498150481	0.378188106	0.595965384	0.386001153	0.264869037
0.7	1.055289556	0.882107359	0.79521517	0.797560983	0.640160994	0.544930456	0.644580437	0.481935543	0.39913223	0.91912723	0.707008528	0.601244938	0.7056535	0.498150481	0.378188106	0.595965384	0.386001153	0.264869037
0.8	0.943409586	0.756986451	0.657567409	0.697883637	0.531378195	0.428707365	0.56249628	0.390276695	0.300632788	0.91912723	0.707008528	0.601244938	0.7056535	0.498150481	0.378188106	0.595965384	0.386001153	0.264869037
0.9	0.887004153	0.6880828	0.580437203	0.655339227	0.476592464	0.367269025	0.541057947	0.352391152	0.25361088	0.91912723	0.707008528	0.601244938	0.7056535	0.498150481	0.378188106	0.595965384	0.386001153	0.264869037
1	0.875927292	0.664356329	0.551409012	0.657070279	0.463094973	0.34697989	0.56279377	0.352612391	0.242965263	0.91912723	0.707008528	0.601244938	0.7056535	0.498150481	0.378188106	0.595965384	0.386001153	0.264869037
1.5	1.230847573	0.923309396	0.772658161	1.003546576	0.69188202	0.514859816	0.931775045	0.569992522	0.38425521	0.91912723	0.707008528	0.601244938	0.7056535	0.498150481	0.378188106	0.595965384	0.386001153	0.264869037
2	1.867370686	1.406632301	1.177084609	1.557110728	1.074870816	0.801240017	1.468821575	0.897859244	0.605712548	0.91912723	0.707008528	0.601244938	0.7056535	0.498150481	0.378188106	0.595965384	0.386001153	0.264869037
}\dgpbapp

\begin{figure}[htbp]
\centering
\begin{subfigure}[b]{0.32\textwidth}
\centering
\resizebox{\linewidth}{!}{
\begin{tikzpicture}
\begin{axis}[
    grid = minor,
    xmax=2,xmin=0,
    ymax=10,ymin=0,
    xtick={0,1,2},
    ytick={5,10},
    title={$N=50,T=50$},
    tick label style={/pgf/number format/fixed},
    legend style={draw=none, at={(0.2,0.9)},anchor=north,
    row sep = 3pt}]
\addplot[smooth,tension=0.1,color=black, line width=0.75pt,dashdotted] table[x = c,y=d1] from \dgpbapp;
\addplot[smooth,tension=0.1,no markers, color=blue, line width=0.75pt] table[x = c,y=d10] from \dgpbapp;
\legend{\footnotesize Fixed $c$, \footnotesize CV}
\end{axis}
\end{tikzpicture}}
\end{subfigure}
\begin{subfigure}[b]{0.32\textwidth}
\centering
\resizebox{\linewidth}{!}{
\begin{tikzpicture}
\begin{axis}[
    grid = minor,
    xmax=2,xmin=0,
    ymax=10,ymin=0,
    xtick={0,1,2},
    ytick={5,10},
    title={$N=100,T=50$},
    tick label style={/pgf/number format/fixed},
    legend style={draw=none, at={(0.2,0.9)},anchor=north,
    row sep = 3pt}]
\addplot[smooth,tension=0.1,color=black, line width=0.75pt,dashdotted] table[x = c,y=d4] from \dgpbapp;
\addplot[smooth,tension=0.1,no markers, color=blue, line width=0.75pt] table[x = c,y=d40] from \dgpbapp;
\legend{\footnotesize Fixed $c$, \footnotesize CV}
\end{axis}
\end{tikzpicture}}
\end{subfigure}
\begin{subfigure}[b]{0.32\textwidth}
\centering
\resizebox{\linewidth}{!}{
\begin{tikzpicture}
\begin{axis}[
    grid = minor,
    xmax=2,xmin=0,
    ymax=10,ymin=0,
    xtick={0,1,2},
    ytick={5,10},
    title={$N=200,T=50$},
    tick label style={/pgf/number format/fixed},
    legend style={draw=none, at={(0.2,0.9)},anchor=north,
    row sep = 3pt}]
\addplot[smooth,tension=0.1,color=black, line width=0.75pt,dashdotted] table[x = c,y=d7] from \dgpbapp;
\addplot[smooth,tension=0.1,no markers, color=blue, line width=0.75pt] table[x = c,y=d70] from \dgpbapp;
\legend{\footnotesize Fixed $c$, \footnotesize CV}
\end{axis}
\end{tikzpicture}}
\end{subfigure}

\begin{subfigure}[b]{0.32\textwidth}
\centering
\resizebox{\linewidth}{!}{
\begin{tikzpicture}
\begin{axis}[
    grid = minor,
    xmax=2,xmin=0,
    ymax=10,ymin=0,
    xtick={0,1,2},
    ytick={5,10},
    title={$N=50,T=100$},
    tick label style={/pgf/number format/fixed},
    legend style={draw=none, at={(0.2,0.9)},anchor=north,
    row sep = 3pt}]
\addplot[smooth,tension=0.1,color=black, line width=0.75pt,dashdotted] table[x = c,y=d2] from \dgpbapp;
\addplot[smooth,tension=0.1,no markers, color=blue, line width=0.75pt] table[x = c,y=d20] from \dgpbapp;
\legend{\footnotesize Fixed $c$, \footnotesize CV}
\end{axis}
\end{tikzpicture}}
\end{subfigure}
\begin{subfigure}[b]{0.32\textwidth}
\centering
\resizebox{\linewidth}{!}{
\begin{tikzpicture}
\begin{axis}[
    grid = minor,
    xmax=2,xmin=0,
    ymax=10,ymin=0,
    xtick={0,1,2},
    ytick={5,10},
    title={$N=100,T=100$},
    tick label style={/pgf/number format/fixed},
    legend style={draw=none, at={(0.2,0.9)},anchor=north,
    row sep = 3pt}]
\addplot[smooth,tension=0.1,color=black, line width=0.75pt,dashdotted] table[x = c,y=d5] from \dgpbapp;
\addplot[smooth,tension=0.1,no markers, color=blue, line width=0.75pt] table[x = c,y=d50] from \dgpbapp;
\legend{\footnotesize Fixed $c$, \footnotesize CV}
\end{axis}
\end{tikzpicture}}
\end{subfigure}
\begin{subfigure}[b]{0.32\textwidth}
\centering
\resizebox{\linewidth}{!}{
\begin{tikzpicture}
\begin{axis}[
    grid = minor,
    xmax=2,xmin=0,
    ymax=10,ymin=0,
    xtick={0,1,2},
    ytick={5,10},
    title={$N=200,T=100$},
    tick label style={/pgf/number format/fixed},
    legend style={draw=none, at={(0.2,0.9)},anchor=north,
    row sep = 3pt}]
\addplot[smooth,tension=0.1,color=black, line width=0.75pt,dashdotted] table[x = c,y=d8] from \dgpbapp;
\addplot[smooth,tension=0.1,no markers, color=blue, line width=0.75pt] table[x = c,y=d80] from \dgpbapp;
\legend{\footnotesize Fixed $c$, \footnotesize CV}
\end{axis}
\end{tikzpicture}}
\end{subfigure}

\begin{subfigure}[b]{0.32\textwidth}
\centering
\resizebox{\linewidth}{!}{
\begin{tikzpicture}
\begin{axis}[
    grid = minor,
    xmax=2,xmin=0,
    ymax=10,ymin=0,
    xtick={0,1,2},
    ytick={5,10},
    title={$N=50,T=200$},
    tick label style={/pgf/number format/fixed},
    legend style={draw=none, at={(0.2,0.9)},anchor=north,
    row sep = 3pt}]
\addplot[smooth,tension=0.1,color=black, line width=0.75pt,dashdotted] table[x = c,y=d3] from \dgpbapp;
\addplot[smooth,tension=0.1,no markers, color=blue, line width=0.75pt] table[x = c,y=d30] from \dgpbapp;
\legend{\footnotesize Fixed $c$, \footnotesize CV}
\end{axis}
\end{tikzpicture}}
\end{subfigure}
\begin{subfigure}[b]{0.32\textwidth}
\centering
\resizebox{\linewidth}{!}{
\begin{tikzpicture}
\begin{axis}[
    grid = minor,
    xmax=2,xmin=0,
    ymax=10,ymin=0,
    xtick={0,1,2},
    ytick={5,10},
    title={$N=100,T=200$},
    tick label style={/pgf/number format/fixed},
    legend style={draw=none, at={(0.2,0.9)},anchor=north,
    row sep = 3pt}]
\addplot[smooth,tension=0.1,color=black, line width=0.75pt,dashdotted] table[x = c,y=d6] from \dgpbapp;
\addplot[smooth,tension=0.1,no markers, color=blue, line width=0.75pt] table[x = c,y=d60] from \dgpbapp;
\legend{\footnotesize Fixed $c$, \footnotesize CV}
\end{axis}
\end{tikzpicture}}
\end{subfigure}
\begin{subfigure}[b]{0.32\textwidth}
\centering
\resizebox{\linewidth}{!}{
\begin{tikzpicture}
\begin{axis}[
    grid = minor,
    xmax=2,xmin=0,
    ymax=10,ymin=0,
    xtick={0,1,2},
    ytick={5,10},
    title={$N=200,T=200$},
    tick label style={/pgf/number format/fixed},
    legend style={draw=none, at={(0.2,0.9)},anchor=north,
    row sep = 3pt}]
\addplot[smooth,tension=0.1,color=black, line width=0.75pt,dashdotted] table[x = c,y=d9] from \dgpbapp;
\addplot[smooth,tension=0.1,no markers, color=blue, line width=0.75pt] table[x = c,y=d90] from \dgpbapp;
\legend{\footnotesize Fixed $c$, \footnotesize CV}
\end{axis}
\end{tikzpicture}}
\end{subfigure}
\caption{Mean square errors of $(\hat{\Pi}^{\diamond\prime}, \sqrt{N}\hat{\Pi}^{\ast\prime})$ when using fixed $c$ and CV: DGP8}\label{Fig: DGP2app}
\end{figure}

\pgfplotstableread{
c	d1	d2	d3	d4	d5	d6	d7	d8	d9	d10	d20	d30	d40	d50	d60	d70	d80	d90
0	0.3013	0.3066	0.2982	0.1408	0.1433	0.1427	0.0691	0.0706	0.0691	0.2583	0.2600	0.2601	0.1276	0.1283	0.1285	0.0652	0.0645	0.0648
0.05	0.2900	0.2951	0.2876	0.1362	0.1386	0.1383	0.0671	0.0684	0.0671	0.2583	0.2600	0.2601	0.1276	0.1283	0.1285	0.0652	0.0645	0.0648
0.1	0.2805	0.2854	0.2786	0.1326	0.1348	0.1347	0.0656	0.0668	0.0656	0.2583	0.2600	0.2601	0.1276	0.1283	0.1285	0.0652	0.0645	0.0648
0.2	0.2671	0.2715	0.2659	0.1281	0.1299	0.1300	0.0639	0.0649	0.0639	0.2583	0.2600	0.2601	0.1276	0.1283	0.1285	0.0652	0.0645	0.0648
0.3	0.2606	0.2644	0.2598	0.1271	0.1284	0.1287	0.0641	0.0649	0.0641	0.2583	0.2600	0.2601	0.1276	0.1283	0.1285	0.0652	0.0645	0.0648
0.4	0.2608	0.2640	0.2599	0.1296	0.1304	0.1306	0.0662	0.0668	0.0662	0.2583	0.2600	0.2601	0.1276	0.1283	0.1285	0.0652	0.0645	0.0648
0.5	0.2674	0.2698	0.2661	0.1355	0.1357	0.1357	0.0702	0.0705	0.0702	0.2583	0.2600	0.2601	0.1276	0.1283	0.1285	0.0652	0.0645	0.0648
0.6	0.2802	0.2816	0.2780	0.1447	0.1442	0.1440	0.0760	0.0759	0.0760	0.2583	0.2600	0.2601	0.1276	0.1283	0.1285	0.0652	0.0645	0.0648
0.7	0.2990	0.2993	0.2956	0.1571	0.1559	0.1553	0.0836	0.0831	0.0836	0.2583	0.2600	0.2601	0.1276	0.1283	0.1285	0.0652	0.0645	0.0648
0.8	0.3234	0.3225	0.3185	0.1727	0.1707	0.1696	0.0930	0.0921	0.0930	0.2583	0.2600	0.2601	0.1276	0.1283	0.1285	0.0652	0.0645	0.0648
0.9	0.3534	0.3510	0.3467	0.1915	0.1886	0.1869	0.1042	0.1028	0.1042	0.2583	0.2600	0.2601	0.1276	0.1283	0.1285	0.0652	0.0645	0.0648
1	0.3886	0.3847	0.3799	0.2131	0.2094	0.2071	0.1169	0.1151	0.1169	0.2583	0.2600	0.2601	0.1276	0.1283	0.1285	0.0652	0.0645	0.0648
1.5	0.6314	0.6208	0.6129	0.3594	0.3518	0.3457	0.2025	0.1980	0.2025	0.2583	0.2600	0.2601	0.1276	0.1283	0.1285	0.0652	0.0645	0.0648
2	0.9689	0.9504	0.9386	0.5633	0.5509	0.5398	0.3222	0.3142	0.3222	0.2583	0.2600	0.2601	0.1276	0.1283	0.1285	0.0652	0.0645	0.0648
}\dgpcapp

\begin{figure}[htbp]
\centering
\begin{subfigure}[b]{0.32\textwidth}
\centering
\resizebox{\linewidth}{!}{
\begin{tikzpicture}
\begin{axis}[
    legend style={draw=none},
    grid = minor,
    xmax=2,xmin=0,
    ymax=1,ymin=0,
    xtick={0,1,2},
    ytick={0.5,1},
    title={$N=50,T=50$},
    tick label style={/pgf/number format/fixed},
legend style={at={(0.2,0.9)},anchor=north,
    row sep = 3pt}]
\addplot[smooth,tension=0.5,color=black, line width=0.75pt,dashdotted] table[x = c,y=d1] from \dgpcapp;
\addplot[smooth,tension=0.5,no markers, color=blue, line width=0.75pt] table[x = c,y=d10] from \dgpcapp;
\legend{\footnotesize Fixed $c$, \footnotesize CV}
\end{axis}
\end{tikzpicture}}
\end{subfigure}
\begin{subfigure}[b]{0.32\textwidth}
\centering
\resizebox{\linewidth}{!}{
\begin{tikzpicture}
\begin{axis}[
    legend style={draw=none},
    grid = minor,
    xmax=2,xmin=0,
    ymax=0.5,ymin=0,
    xtick={0,1,2},
    ytick={0.25,0.5},
    title={$N=100,T=50$},
    tick label style={/pgf/number format/fixed},
legend style={at={(0.2,0.9)},anchor=north,
    row sep = 3pt}]
\addplot[smooth,tension=0.5,color=black, line width=0.75pt,dashdotted] table[x = c,y=d4] from \dgpcapp;
\addplot[smooth,tension=0.5,no markers, color=blue, line width=0.75pt] table[x = c,y=d40] from \dgpcapp;
\legend{\footnotesize Fixed $c$, \footnotesize CV}
\end{axis}
\end{tikzpicture}}
\end{subfigure}
\begin{subfigure}[b]{0.32\textwidth}
\centering
\resizebox{\linewidth}{!}{
\begin{tikzpicture}
\begin{axis}[
    legend style={draw=none},
    grid = minor,
    xmax=2,xmin=0,
    ymax=0.25,ymin=0,
    xtick={0,1,2},
    ytick={0.12,0.25},
    title={$N=200,T=50$},
    tick label style={/pgf/number format/fixed},
legend style={at={(0.2,0.9)},anchor=north,
    row sep = 3pt}]
\addplot[smooth,tension=0.5,color=black, line width=0.75pt,dashdotted] table[x = c,y=d7] from \dgpcapp;
\addplot[smooth,tension=0.5,no markers, color=blue, line width=0.75pt] table[x = c,y=d70] from \dgpcapp;
\legend{\footnotesize Fixed $c$, \footnotesize CV}
\end{axis}
\end{tikzpicture}}
\end{subfigure}

\begin{subfigure}[b]{0.32\textwidth}
\centering
\resizebox{\linewidth}{!}{
\begin{tikzpicture}
\begin{axis}[
    legend style={draw=none},
    grid = minor,
    xmax=2,xmin=0,
    ymax=1,ymin=0,
    xtick={0,1,2},
    ytick={0.5,1},
    title={$N=50,T=100$},
    tick label style={/pgf/number format/fixed},
legend style={at={(0.2,0.9)},anchor=north,
    row sep = 3pt}]
\addplot[smooth,tension=0.5,color=black, line width=0.75pt,dashdotted] table[x = c,y=d2] from \dgpcapp;
\addplot[smooth,tension=0.5,no markers, color=blue, line width=0.75pt] table[x = c,y=d20] from \dgpcapp;
\legend{\footnotesize Fixed $c$, \footnotesize CV}
\end{axis}
\end{tikzpicture}}
\end{subfigure}
\begin{subfigure}[b]{0.32\textwidth}
\centering
\resizebox{\linewidth}{!}{
\begin{tikzpicture}
\begin{axis}[
    legend style={draw=none},
    grid = minor,
    xmax=2,xmin=0,
    ymax=0.5,ymin=0,
    xtick={0,1,2},
    ytick={0.25,0.5},
    title={$N=100,T=100$},
    tick label style={/pgf/number format/fixed},
legend style={at={(0.2,0.9)},anchor=north,
    row sep = 3pt}]
\addplot[smooth,tension=0.5,color=black, line width=0.75pt,dashdotted] table[x = c,y=d5] from \dgpcapp;
\addplot[smooth,tension=0.5,no markers, color=blue, line width=0.75pt] table[x = c,y=d50] from \dgpcapp;
\legend{\footnotesize Fixed $c$, \footnotesize CV}
\end{axis}
\end{tikzpicture}}
\end{subfigure}
\begin{subfigure}[b]{0.32\textwidth}
\centering
\resizebox{\linewidth}{!}{
\begin{tikzpicture}
\begin{axis}[
    legend style={draw=none},
    grid = minor,
    xmax=2,xmin=0,
    ymax=0.25,ymin=0,
    xtick={0,1,2},
    ytick={0.12,0.25},
    title={$N=200,T=100$},
    tick label style={/pgf/number format/fixed},
legend style={at={(0.2,0.9)},anchor=north,
    row sep = 3pt}]
\addplot[smooth,tension=0.5,color=black, line width=0.75pt,dashdotted] table[x = c,y=d8] from \dgpcapp;
\addplot[smooth,tension=0.5,no markers, color=blue, line width=0.75pt] table[x = c,y=d80] from \dgpcapp;
\legend{\footnotesize Fixed $c$, \footnotesize CV}
\end{axis}
\end{tikzpicture}}
\end{subfigure}

\begin{subfigure}[b]{0.32\textwidth}
\centering
\resizebox{\linewidth}{!}{
\begin{tikzpicture}
\begin{axis}[
    legend style={draw=none},
    grid = minor,
    xmax=2,xmin=0,
    ymax=1,ymin=0,
    xtick={0,1,2},
    ytick={0.5,1},
    title={$N=50,T=200$},
    tick label style={/pgf/number format/fixed},
legend style={at={(0.2,0.9)},anchor=north,
    row sep = 3pt}]
\addplot[smooth,tension=0.5,color=black, line width=0.75pt,dashdotted] table[x = c,y=d3] from \dgpcapp;
\addplot[smooth,tension=0.5,no markers, color=blue, line width=0.75pt] table[x = c,y=d30] from \dgpcapp;
\legend{\footnotesize Fixed $c$, \footnotesize CV}
\end{axis}
\end{tikzpicture}}
\end{subfigure}
\begin{subfigure}[b]{0.32\textwidth}
\centering
\resizebox{\linewidth}{!}{
\begin{tikzpicture}
\begin{axis}[
    legend style={draw=none},
    grid = minor,
    xmax=2,xmin=0,
    ymax=0.5,ymin=0,
    xtick={0,1,2},
    ytick={0.25,0.5},
    title={$N=100,T=200$},
    tick label style={/pgf/number format/fixed},
legend style={at={(0.2,0.9)},anchor=north,
    row sep = 3pt}]
\addplot[smooth,tension=0.5,color=black, line width=0.75pt,dashdotted] table[x = c,y=d6] from \dgpcapp;
\addplot[smooth,tension=0.5,no markers, color=blue, line width=0.75pt] table[x = c,y=d60] from \dgpcapp;
\legend{\footnotesize Fixed $c$, \footnotesize CV}
\end{axis}
\end{tikzpicture}}
\end{subfigure}
\begin{subfigure}[b]{0.32\textwidth}
\centering
\resizebox{\linewidth}{!}{
\begin{tikzpicture}
\begin{axis}[
    legend style={draw=none},
    grid = minor,
    xmax=2,xmin=0,
    ymax=0.25,ymin=0,
    xtick={0,1,2},
    ytick={0.12,0.25},
    title={$N=200,T=200$},
    tick label style={/pgf/number format/fixed},
legend style={at={(0.2,0.9)},anchor=north,
    row sep = 3pt}]
\addplot[smooth,tension=0.5,color=black, line width=0.75pt,dashdotted] table[x = c,y=d9] from \dgpcapp;
\addplot[smooth,tension=0.5,no markers, color=blue, line width=0.75pt] table[x = c,y=d90] from \dgpcapp;
\legend{\footnotesize Fixed $c$, \footnotesize CV}
\end{axis}
\end{tikzpicture}}
\end{subfigure}
\caption{Mean square errors of $\hat{\Pi}_0$ when using fixed $c$ and CV: DGP9}\label{Fig: DGP3app}
\end{figure}

\setlength{\tabcolsep}{18pt}
\begin{table}[htbp]
\centering
\resizebox{0.99\textwidth}{!}{
\begin{threeparttable}
\renewcommand{\arraystretch}{1.35}
\caption{Mean square errors of $\hat{\Pi}$, $\hat{a}$, $\hat{B}$, and $\hat{F}$, and correct rates of $\hat{K}$: DGP7\tnote{\dag}}\label{Tab: MSEDGP1app}
\begin{tabular}{cccccccccc}
\hline\hline
&(N,T)&&$\hat{\Pi}$&$\hat{a}$&$\hat{B}$&$\hat{F}$&&$\hat{K}$&\\
\cline{2-9}
&$(50,50)$    &&2.607&1.127&0.820&0.217&&0.955&\\
&$(100,50)$   &&2.323&1.187&0.667&0.133&&0.990&\\
&$(200,50)$   &&2.111&1.240&0.570&0.095&&1.000&\\
\cline{2-9}
&$(50,100)$   &&1.610&0.962&0.355&0.203&&1.000&\\
&$(100,100)$  &&1.332&1.051&0.306&0.171&&1.000&\\
&$(200,100)$  &&1.155&0.849&0.254&0.114&&1.000&\\
\cline{2-9}
&$(50,200)$   &&1.176&0.720&0.157&0.201&&1.000&\\
&$(100,200)$  &&0.877&0.577&0.124&0.132&&1.000&\\
&$(200,200)$  &&0.707&0.506&0.103&0.091&&1.000\\
\hline\hline
\end{tabular}
\begin{tablenotes}
      \small
      \item[\dag] The mean square errors of $\hat{\Pi}$, $\hat{a}$ , $\hat{B}$, and $\hat{F}$ are given by $\sum_{\ell=1}^{200}\|\hat{\Pi}^{(\ell)}-\Pi\|_{F}^2/200NT$, $\sum_{\ell=1}^{200}\|\hat{a}^{(\ell)}-a\|^2/200N$, $\sum_{\ell=1}^{200}\|\hat{B}^{(\ell)}-BH^{(\ell)}\|_{F}^2/200N$ and $\sum_{\ell=1}^{200}\|\hat{F}^{(\ell)}- F(H^{{(\ell)}\prime})^{-1}\|_{F}^2/{200T}$, where $\hat{\Pi}^{(\ell)}$, $\hat{a}^{(\ell)}$, $\hat{B}^{(\ell)}$, and $\hat{F}^{(\ell)}$ are estimates in the $\ell$th simulation replication, and $H^{(\ell)}\equiv (F^{\prime}M_T\hat{F}^{(\ell)})(\hat{F}^{{(\ell)}\prime}M_T\hat{F}^{(\ell)})^{-1}$ is a rotational transformation matrix. The value of $c$ is chosen from $\{0, 0.05, 0.1, 0.2, \ldots, 0.9,1,1.5,2\}$ by using the 5-fold CV method as outlined in Section \ref{Sec3}.
    \end{tablenotes}
\end{threeparttable}
}
\end{table}

\setlength{\tabcolsep}{12pt}
\begin{table}[htbp]
\centering
\resizebox{0.99\textwidth}{!}{
\begin{threeparttable}
\renewcommand{\arraystretch}{1.35}
\caption{Mean square errors of $\hat{\Pi}^{\diamond}$, $\hat{\Pi}^{\ast}$, $\hat{\mu}$, $\hat\Lambda$, $\hat{\phi}, \hat{\Phi}$, and $\hat{F}$, and correct rates of $\hat{K}$: DGP8\tnote{\dag}}\label{Tab: MSEDGP2app}
\begin{tabular}{ccccccccccccc}
\hline\hline
&(N,T)&&$\hat{\Pi}^{\diamond}$&$\hat{\Pi}^{\ast}$&$\hat{\mu}$&$\hat\Lambda$&$\hat{\phi}$&$\hat{\Phi}$&$\hat{F}$&&$\hat{K}$&\\
\cline{2-12}
&$(50,50)$     &&0.561&0.358&0.208&0.074&0.435&0.078&0.185&&1.000&\\
&$(100,50)$    &&0.455&0.251&0.215&0.069&0.388&0.063&0.114&&1.000&\\
&$(200,50)$    &&0.403&0.193&0.222&0.068&0.338&0.053&0.078&&1.000&\\
\cline{2-12}
&$(50,100)$    &&0.407&0.300&0.108&0.038&0.370&0.045&0.170&&1.000&\\
&$(100,100)$   &&0.311&0.187&0.117&0.035&0.272&0.033&0.107&&1.000&\\
&$(200,100)$   &&0.256&0.130&0.128&0.032&0.209&0.025&0.068&&1.000&\\
\cline{2-12}
&$(50,200)$    &&0.331&0.271&0.054&0.019&0.284&0.030&0.171&&1.000&\\
&$(100,200)$   &&0.219&0.159&0.058&0.016&0.180&0.020&0.098&&1.000&\\
&$(200,200)$   &&0.165&0.100&0.062&0.014&0.123&0.014&0.059&&1.000&\\
\hline\hline
\end{tabular}
\begin{tablenotes}
      \small
      \item[\dag] The mean square errors of $\hat{\Pi}^{\diamond}$, $\hat{\Pi}^{\ast}$, $\hat{\mu}$, $\hat\Lambda$, $\hat{\phi}, \hat{\Phi}$, and $\hat{F}$ are given by $\sum_{\ell=1}^{200}\|\hat{\Pi}^{\diamond(\ell)}-\Pi^{\diamond}\|_{F}^2/200NT$, $\sum_{\ell=1}^{200}\|\hat{\Pi}^{\ast(\ell)}-\Pi^{\ast}\|_{F}^2/200T$,$\sum_{\ell=1}^{200}\|\hat{\mu}^{(\ell)}-\mu\|^2/200N$, $\sum_{\ell=1}^{200}\|\hat{\Lambda}^{(\ell)}-\Lambda H^{(\ell)}\|_{F}^2/200N$, $\sum_{\ell=1}^{200}\|\hat{\phi}^{(\ell)}-\phi\|^2/200$, $\sum_{\ell=1}^{200}\|\hat{\Phi}^{(\ell)}-\Phi H^{(\ell)}\|^2/200$ and $\sum_{\ell=1}^{200}\|\hat{F}^{(\ell)}- F(H^{{(\ell)}\prime})^{-1}\|_{F}^2/{200T}$, where $\hat{\Pi}^{\diamond(\ell)}$, $\hat{\Pi}^{\ast(\ell)}$, $\hat{\mu}^{(\ell)}$, $\hat{\Lambda}^{(\ell)}$, $\hat{\phi}^{(\ell)}$, $\hat{\Phi}^{(\ell)}$, and $\hat{F}^{(\ell)}$ are estimates in the $\ell$th simulation replication, and $H^{(\ell)}\equiv (F^{\prime}M_T\hat{F}^{(\ell)})(\hat{F}^{{(\ell)}\prime}M_T\hat{F}^{(\ell)})^{-1}$ is a rotational transformation matrix. The value of $c$ is chosen from $\{0, 0.05, 0.1, 0.2, \ldots, 0.9,1,1.5,2\}$ by using the 5-fold CV method as outlined in Section \ref{Sec3}.
    \end{tablenotes}
\end{threeparttable}
}
\end{table}

\setlength{\tabcolsep}{18pt}
\begin{table}[htbp]
\centering
\resizebox{0.99\textwidth}{!}{
\begin{threeparttable}
\renewcommand{\arraystretch}{1.35}
\caption{Mean square errors of $\hat{\Pi}_0$, $\hat{\phi}_0$, $\hat{\Phi}_0$, and $\hat{F}$ ($\times 10^{-1}$), and correct rates of $\hat{K}$: DGP9\tnote{\dag}}\label{Tab: MSEDGP3app}
\begin{tabular}{cccccccccc}
\hline\hline
&(N,T)&&$\hat{\Pi}_0$&$\hat{\phi}_0$&$\hat{\Phi}_0$&$\hat{F}$&&$\hat{K}$&\\
\cline{2-9}
&$(50,50)$    &&2.583&0.615&0.081&1.731&&1.000&\\
&$(100,50)$   &&1.276&0.248&0.036&0.862&&1.000&\\
&$(200,50)$   &&0.652&0.131&0.019&0.447&&1.000&\\
\cline{2-9}
&$(50,100)$   &&2.600&0.486&0.050&1.697&&1.000&\\
&$(100,100)$  &&1.283&0.196&0.022&0.832&&1.000&\\
&$(200,100)$  &&0.645&0.083&0.011&0.415&&1.000&\\
\cline{2-9}
&$(50,200)$   &&2.601&0.328&0.030&1.643&&1.000&\\
&$(100,200)$  &&1.285&0.127&0.014&0.804&&1.000&\\

&$(200,200)$  &&0.648&0.056&0.007&0.408&&1.000&\\
\hline\hline
\end{tabular}
\begin{tablenotes}
      \small
      \item[\dag] The mean square errors of $\hat{\Pi}_0$, $\hat{\phi}_0$ , $\hat{\Phi}_0$, and $\hat{F}$ are given by $\sum_{\ell=1}^{200}\|\hat{\Pi}_0^{(\ell)}-\Pi_0\|_{F}^2/200T$, $\sum_{\ell=1}^{200}\|\hat{\phi}_0^{(\ell)}-\phi\|^2/200$, $\sum_{\ell=1}^{200}\|\hat{\Phi}_0^{(\ell)}-\Phi H^{(\ell)}\|_{F}^2/200$ and $\sum_{\ell=1}^{200}\|\hat{F}^{(\ell)}- F(H^{{(\ell)}\prime})^{-1}\|_{F}^2/{200T}$, where $\hat{\Pi}_0^{(\ell)}$, $\hat{\phi}_0^{(\ell)}$, $\hat{\Phi}_0^{(\ell)}$, and $\hat{F}^{(\ell)}$ are estimates in the $\ell$th simulation replication, and $H^{(\ell)}\equiv (F^{\prime}M_T\hat{F}^{(\ell)})(\hat{F}^{{(\ell)}\prime}M_T\hat{F}^{(\ell)})^{-1}$ is a rotational transformation matrix. The value of $c$ is chosen from $\{0, 0.05, 0.1, 0.2, \ldots, 0.9,1,1.5,2\}$ by using the 5-fold CV method as outlined in Section \ref{Sec3}.
    \end{tablenotes}
\end{threeparttable}
}
\end{table}

\subsection{Misspecification and Efficiency}\label{App: G: sub3}
We investigate the performance of the estimators under two scenarios: when homogeneity of $a_i$ and $B_i$ is incorrectly specified, and when it is not effectively used. Specifically, we focus on DGP7 and DGP9. In DGP7, the estimators are implemented with the constraint that $a_i$ and $B_i$ are homogeneous across $i$, corresponding to the formulation in \eqref{Eqn: NuclearNEst}-\eqref{Eqn: aFEst} with $\mathcal{S} = \{1_{N}\otimes \Gamma: \Gamma\in \mathbf{R}^{p\times T}\}$. Since the homogeneity is not true in DGP7, this leads to incorrect specification in the estimation. In DGP9, the estimators are implemented without enforcing the homogeneity constraint, corresponding to the estimators in \eqref{Eqn: NuclearNEst}-\eqref{Eqn: aFEst} with $\mathcal{S} = \mathbf{R}^{Np\times T}$. Although the homogeneity is satisfied in DGP9, the estimation does not leverage this property. The estimators without the homogeneity constraint yield robust results: the mean square errors decrease as $(N,T)$ increases in both DGP7 and DGP9. However, they suffer from efficiency loss in DGP9, where the homogeneity could have been utilized to improve performance. The estimators with the homogeneity constraint exhibit poor performance in DGP7 due to misspecification. The mean square errors fail to decrease with increasing $(N,T)$, highlighting the adverse impact of enforcing an incorrect homogeneity assumption.

\setlength{\tabcolsep}{12pt}
\begin{table}[htbp]
\centering
\resizebox{0.99\textwidth}{!}{
\begin{threeparttable}
\renewcommand{\arraystretch}{1.35}
\caption{Mean square errors of $\hat{\Pi}$, $\hat{a}$, and $\hat{B}$: misspecification and efficiency \tnote{\dag}}\label{Tab: MisEff1}
\begin{tabular}{cccccccccccc}
\hline\hline
&&&&\multicolumn{3}{c}{With Homogeneity}&&\multicolumn{3}{c}{Without Homogeneity}&\\
\cline{5-7}\cline{9-11}
&&(N,T)&&$\hat{\Pi}$&$\hat{a}$&$\hat{B}$&&$\hat{\Pi}$&$\hat{a}$&$\hat{B}$&\\
\cline{2-11}
&\multirow{4}{*}{DGP7}&$(50,50)$    &&2.035&0.756&0.100&&2.607&1.127&0.820&\\
&&$(100,100)$   &&2.400&1.376&0.084&&1.332&1.051&0.306&\\
&&$(200,200)$   &&2.180&0.991&0.085&&0.707&0.506&0.103&\\
&&$(500,500)$   &&2.342&1.192&0.087&&0.303&0.255&0.049&\\
\cline{2-11}
&\multirow{4}{*}{DGP9 ($\times 10^{-1}$)}&$(50,50)$    &&2.583&0.615&0.081&&26.601&13.459&8.781&\\
&&$(100,100)$   &&1.283&0.196&0.022&&13.015&9.392&3.138&\\
&&$(200,200)$   &&0.648&0.056&0.007&&7.013&4.995&1.083&\\
&&$(500,500)$   &&0.257&0.015&0.002&&2.967&2.413&0.396&\\
\hline\hline
\end{tabular}
\begin{tablenotes}
      \small
      \item[\dag] The mean square errors of $\hat{\Pi}$, $\hat{a}$ , and $\hat{B}$ are given by $\sum_{\ell=1}^{200}\|\hat{\Pi}^{(\ell)}-\Pi\|_{F}^2/200NT$, $\sum_{\ell=1}^{200}\|\hat{a}^{(\ell)}-a\|^2/200N$, and $\sum_{\ell=1}^{200}\|\hat{B}^{(\ell)}-BH^{(\ell)}\|_{F}^2/200N$, where $\hat{\Pi}^{(\ell)}$, $\hat{a}^{(\ell)}$, and $\hat{B}^{(\ell)}$ are estimates in the $\ell$th simulation replication, and $H^{(\ell)}\equiv (F^{\prime}M_T\hat{F}^{(\ell)})(\hat{F}^{{(\ell)}\prime}M_T\hat{F}^{(\ell)})^{-1}$ is a rotational transformation matrix where $\hat{F}^{(\ell)}$ the estimate in the $\ell$th simulation replication. The value of $c$ is chosen from $\{0, 0.05, 0.1, 0.2, \ldots, 0.9,1,1.5,2\}$ by using the 5-fold CV method as outlined in Section \ref{Sec3}.
    \end{tablenotes}
\end{threeparttable}
}
\end{table}

\subsection{Comparing Methods}\label{App: G: sub4}
We compare our method with two existing methods: \citet{Fanetal_ProjectedPCA_2016}'s projected-PCA and \citet{Chenetal_SeimiparametricFactor_2021}'s regressed-PCA. To assess the performance of the projected PCA, we consider DGP8 by setting $a_i=0$. The mean square errors of $\hat{F}$ under this method fail to converge and remain significantly large even for large $N$ and $T$ (e.g., close to $1,000$ for $N=800$ and $T=500$), as demonstrated in Figure \ref{Fig: Comparingapp1}. The failure occurs because $x_{it}$ varies over $t$, and $\lambda_i$ does not have zero mean. In contrast, our method is robust to these issues, as demonstrated in Table \ref{Tab: MSEDGP2app}.

To evaluate the performance of the regressed-PCA, we consider two DGPs: DGP10 and DGP11. In both DGPs, $x_{it} = (x_{it,1},x_{it,2}, \ldots,x_{it,p})^{\prime}$, where  $x_{it,1},x_{it,2},$ and $x_{it,3}$ are generated as in Section \ref{Sec6}, and $x_{it,j}$ ($4\leq j \leq p$) are i.i.d. $N(0,1)$ across $i, t,$ and $j$. The settings for $a_i$ and $B_i$ are as follows: in DGP10,
\begin{align}\label{Eqn: phi0Comparing1}
a_i &= \phi_0= \left(
         \begin{array}{ccccc}
          0 & 1 & 1 & 0 & 0_{p-4}^{\prime}\\
         \end{array}
       \right)^{\prime} \text{ and }
\notag\\
B_i &=\Phi_0 = \left(
       \begin{array}{ccccc}
         0 & 0 & 0 & 2 & 0.1\times 1_{p-4}^{\prime} \\
         2 & 0 & 0 & 0 & -0.1\times 1_{p-4}^{\prime} \\
       \end{array}
     \right)^{\prime},
\end{align}
 while in DGP11,
\begin{align}\label{Eqn: phi0Comparing2}
a_i &= \phi_0= \left(
         \begin{array}{ccccc}
          0 & 1 & 1 & 0 & 0_{p-4}^{\prime}\\
         \end{array}
       \right)^{\prime} \text{ and }
\notag\\
B_i &=\Phi_0 = \left(
       \begin{array}{ccccc}
         0 & 0 & 0 & 2 & 0_{p-4}^{\prime} \\
         2 & 0 & 0 & 0 & 0_{p-4}^{\prime} \\
       \end{array}
     \right)^{\prime}.
\end{align}
Here, $0_{p}$ and $1_{p}$ are $p\times 1$ vectors of zeros and ones, respectively. Note that $\Phi_0$ is sparse in DGP11 but not in DGP10. We generate $\varepsilon_{it}$'s and $f_t$'s as in Section \ref{Sec6}. We compare the performance of our method and the regressed-PCA by varying the dimension $p$ while fixing $N=T=50$. Results are shown in Figures \ref{Fig: Comparing2} and \ref{Fig: Comparing2sparse}. In both DGP10 and DGP11, the mean square errors of the regressed-PCA estimators increase rapidly as $p$ grows, often diverging for large $p$. In contrast, our estimators remain stable and exhibit small errors, consistent with the findings in Corollary \ref{Cor: Ex4: 1}. This demonstrates that our method allows $p$ to grow as fast as $N$, whereas the regressed-PCA requires $p$ to grow at a much slower rate to maintain accuracy.

\pgfplotstableread{
T	x1 x2 x3
50	14.942	80.1	65
100	324.83	1724	691
200	1344.7	1224	130
300	404.92	100	1636
400	496.8	191	216
500	316.35	1254	311
600	2166	354	57
700	252	166.6	1559
800	1250	476.7	780
}\comparingappone

\begin{figure}[htbp]
\centering
\resizebox{0.8\linewidth}{!}{
\pgfplotsset{title style={at={(0.5,0.91)}}}
\begin{tikzpicture}
\begin{axis}[
        ybar,
    height=20cm,
    width=20cm,
    enlarge y limits=false,
    axis lines*=left,
    xmax=800,xmin=50,
    ymax=2500,ymin=10,
    xtick={50,100,200,300,400,500,600,700,800},
    ytick={10,500, 1000, 1500, 2000, 2500},
     legend style={at={(0.4,0.9)},
        anchor=north,legend columns=-1,
        /tikz/every even column/.append style={column sep=0.5cm}
        },
        xlabel={$N$},
        ylabel={Mean square errors of $\hat{F}$}
    ]
\addplot[smooth,tension=0.3, mark=square,color=red, line width=1.2pt] table[x = T,y=x1] from \comparingappone;
\addplot[smooth,tension=0.3, mark=star,color=blue, line width=1.2pt] table[x = T,y=x2] from \comparingappone;
\addplot[smooth,tension=0.3, mark=triangle,color=black, line width=1.2pt] table[x = T,y=x3] from \comparingappone;
\legend{$T=50$,$T=200$,$T=500$,}
  \end{axis}
\end{tikzpicture}}
\caption{\citet{Fanetal_ProjectedPCA_2016}'s projected-PCA: DGP8 with $a_i=0$} \label{Fig: Comparingapp1}
\end{figure}

\pgfplotstableread{
p	d1	d2	d3	d4	d5	d6
4	0.083177145	0.008016307	0.228983225	0.061574727	0.006520149	0.304467164
5	0.10711299	0.01148253	0.214454161	0.074261193	0.007903723	0.315967975
6	0.13403181	0.015898057	0.19818683	0.084350005	0.008745007	0.308265715
7	0.150576151	0.018390861	0.200743758	0.1042211	0.010631547	0.32346889
8	0.175633464	0.022366388	0.200266863	0.120703749	0.0124048	0.33422143
10	0.206715705	0.027417704	0.196653376	0.155150488	0.016153966	0.365671615
15	0.262454186	0.037285941	0.19756554	0.271881388	0.026800167	0.438883854
20	0.313423581	0.048705076	0.184508376	0.402475105	0.039952129	0.528912052
30	0.4086562	0.063729277	0.190587707	0.923312901	0.088689784	1.02290514
32	0.426046937	0.06823285	0.185800543	1.1518949	0.106656433	1.217122034
35	0.4378586	0.072316171	0.183401956	1.515478086	0.13989135	1.673967508
38	0.467063985	0.075179351	0.187145371	2.1073151	0.1967811	2.677718792
40	0.480543364	0.079517336	0.1808345	3.106081023	0.276679888	6.744963717
41	0.487435809	0.078883444	0.180951544	3.601622371	0.321614353	6.524892237
42	0.494491214	0.081496126	0.182147399	4.824679931	0.421779399	100
43	0.499098698	0.081454989	0.187254983	6.272282564	0.537362946	101
44	0.511482638	0.085564684	0.186433435	8.881973408	0.738685765	102
45	0.516727337	0.085977093	0.187917729	11.61469557	0.996838865	103
50	0.54068433	0.093318611	0.186520545	107.3941533	1.999799506	104
}\comparingapptwo

\begin{figure}[htbp]
\centering
\begin{subfigure}[b]{0.8\textwidth}
\centering
\resizebox{\linewidth}{!}{
\begin{tikzpicture}
\begin{axis}[
        ybar,
    height=10cm,
    width=20cm,
    enlarge y limits=false,
    axis lines*=left,
    xmax=50,xmin=4,
    ymax=5,ymin=0,
    xtick={4,10,20,30,40,50},
    ytick={0,1,2,3,4,5},
     legend style={at={(0.4,0.9)},
        anchor=north,legend columns=-1,
        /tikz/every even column/.append style={column sep=0.5cm}
        },
        xlabel={$p$},
        ylabel={Mean square errors of $\hat{\phi}_0$}
    ]
\addplot[smooth,tension=0.1,no markers, color=blue, line width=1.2pt] table[x = p,y=d1] from \comparingapptwo;
\addplot[smooth,tension=0.1,color=red, line width=1.2pt,dashdotted] table[x = p,y=d4] from \comparingapptwo;
\legend{\footnotesize Our method, \footnotesize Regressed-PCA}
  \end{axis}
\end{tikzpicture}}
\end{subfigure}

\begin{subfigure}[b]{0.8\textwidth}
\centering
\resizebox{\linewidth}{!}{
\begin{tikzpicture}
\begin{axis}[
        ybar,
    height=10cm,
    width=20cm,
    enlarge y limits=false,
    axis lines*=left,
    xmax=50,xmin=4,
    ymax=1,ymin=0,
    xtick={4,10,20,30,40,50},
    ytick={0,0.2,0.4,0.6,0.8,1},
     legend style={at={(0.4,0.9)},
        anchor=north,legend columns=-1,
        /tikz/every even column/.append style={column sep=0.5cm}
        },
        xlabel={$p$},
        ylabel={Mean square errors of $\hat{\Phi}_0$}
    ]
\addplot[smooth,tension=0.1,no markers, color=blue, line width=1.2pt] table[x = p,y=d2] from \comparingapptwo;
\addplot[smooth,tension=0.1,color=red, line width=1.2pt,dashdotted] table[x = p,y=d5] from \comparingapptwo;
\legend{\footnotesize Our method, \footnotesize Regressed-PCA}
  \end{axis}
\end{tikzpicture}}
\end{subfigure}

\begin{subfigure}[b]{0.8\textwidth}
\centering
\resizebox{\linewidth}{!}{
\begin{tikzpicture}
\begin{axis}[
        ybar,
    height=10cm,
    width=20cm,
    enlarge y limits=false,
    axis lines*=left,
    xmax=50,xmin=4,
    ymax=2,ymin=0,
    xtick={4,10,20,30,40,50},
    ytick={0,0.4,0.8,1.2,1.6,2},
     legend style={at={(0.4,0.9)},
        anchor=north,legend columns=-1,
        /tikz/every even column/.append style={column sep=0.5cm}
        },
        xlabel={$p$},
        ylabel={Mean square errors of $\hat{F}$}
    ]
\addplot[smooth,tension=0.1,no markers, color=blue, line width=1.2pt] table[x = p,y=d3] from \comparingapptwo;
\addplot[smooth,tension=0.1,color=red, line width=1.2pt,dashdotted] table[x = p,y=d6] from \comparingapptwo;
\legend{\footnotesize Our method, \footnotesize Regressed-PCA}
  \end{axis}
\end{tikzpicture}}
\end{subfigure}
\caption{Our method v.s. \citet{Chenetal_SeimiparametricFactor_2021}'s regressed-PCA: DGP10 with $N=T=50$}\label{Fig: Comparing2}
\end{figure}

\pgfplotstableread{
p	d1	d2	d3	d4	d5	d6
4	0.056694961	0.007719957	0.175971707	0.019799592	0.005800371	0.213394465
5	0.096899762	0.013771046	0.173496086	0.032303434	0.007762443	0.222072847
6	0.110592125	0.016612142	0.169115809	0.043950863	0.008704877	0.22461939
7	0.145122598	0.021261218	0.172051838	0.072044234	0.011456238	0.234803109
8	0.163131746	0.023239096	0.16700541	0.084419393	0.012966312	0.244211787
15	0.263274098	0.045149766	0.172223657	0.240091245	0.030595725	0.320133391
20	0.317076996	0.0576389	0.172674863	0.378104945	0.046313272	0.402496511
30	0.419265017	0.084949921	0.179032328	0.941514704	0.109720521	0.812977385
32	0.423771812	0.091815466	0.180946403	1.092432426	0.125639878	0.976689485
35	0.454681214	0.09569021	0.177402091	1.483241419	0.171410772	1.411858915
38	0.475482603	0.108249638	0.178737165	2.170534582	0.251116291	2.313540681
40	0.493187772	0.111249828	0.182845292	2.917486889	0.343720683	5.299735232
41	0.513456224	0.113683441	0.177826403	3.660643061	0.430643673	12.85413099
42	0.52924481	0.120773445	0.180351642	4.480083862	0.519257009	100
43	0.53688432	0.123859068	0.180138072	5.385943458	0.655878435	101
44	0.512447257	0.122964992	0.183130041	8.053279399	0.943069169	102
45	0.541777527	0.126607572	0.183986075	10.71958220	1.237133164	103
50	0.577046566	0.145490594	0.179902573	111.8336545	1.999637977	104
}\comparingapptwosparse

\begin{figure}[htbp]
\centering
\begin{subfigure}[b]{0.8\textwidth}
\centering
\resizebox{\linewidth}{!}{
\begin{tikzpicture}
\begin{axis}[
        ybar,
    height=10cm,
    width=20cm,
    enlarge y limits=false,
    axis lines*=left,
    xmax=50,xmin=4,
    ymax=5,ymin=0,
    xtick={4,10,20,30,40,50},
    ytick={0,1,2,3,4,5},
     legend style={at={(0.4,0.9)},
        anchor=north,legend columns=-1,
        /tikz/every even column/.append style={column sep=0.5cm}
        },
        xlabel={$p$},
        ylabel={Mean square errors of $\hat{\phi}_0$}
    ]
\addplot[smooth,tension=0.1,no markers, color=blue, line width=1.2pt] table[x = p,y=d1] from \comparingapptwosparse;
\addplot[smooth,tension=0.1,color=red, line width=1.2pt,dashdotted] table[x = p,y=d4] from \comparingapptwosparse;
\legend{\footnotesize Our method, \footnotesize Regressed-PCA}
  \end{axis}
\end{tikzpicture}}
\end{subfigure}

\begin{subfigure}[b]{0.8\textwidth}
\centering
\resizebox{\linewidth}{!}{
\begin{tikzpicture}
\begin{axis}[
        ybar,
    height=10cm,
    width=20cm,
    enlarge y limits=false,
    axis lines*=left,
    xmax=50,xmin=4,
    ymax=1,ymin=0,
    xtick={4,10,20,30,40,50},
    ytick={0,0.2,0.4,0.6,0.8,1},
     legend style={at={(0.4,0.9)},
        anchor=north,legend columns=-1,
        /tikz/every even column/.append style={column sep=0.5cm}
        },
        xlabel={$p$},
        ylabel={Mean square errors of $\hat{\Phi}_0$}
    ]
\addplot[smooth,tension=0.1,no markers, color=blue, line width=1.2pt] table[x = p,y=d2] from \comparingapptwosparse;
\addplot[smooth,tension=0.1,color=red, line width=1.2pt,dashdotted] table[x = p,y=d5] from \comparingapptwosparse;
\legend{\footnotesize Our method, \footnotesize Regressed-PCA}
  \end{axis}
\end{tikzpicture}}
\end{subfigure}

\begin{subfigure}[b]{0.8\textwidth}
\centering
\resizebox{\linewidth}{!}{
\begin{tikzpicture}
\begin{axis}[
        ybar,
    height=10cm,
    width=20cm,
    enlarge y limits=false,
    axis lines*=left,
    xmax=50,xmin=4,
    ymax=2,ymin=0,
    xtick={4,10,20,30,40,50},
    ytick={0,0.4,0.8,1.2,1.6,2},
     legend style={at={(0.4,0.9)},
        anchor=north,legend columns=-1,
        /tikz/every even column/.append style={column sep=0.5cm}
        },
        xlabel={$p$},
        ylabel={Mean square errors of $\hat{F}$}
    ]
\addplot[smooth,tension=0.1,no markers, color=blue, line width=1.2pt] table[x = p,y=d3] from \comparingapptwosparse;
\addplot[smooth,tension=0.1,color=red, line width=1.2pt,dashdotted] table[x = p,y=d6] from \comparingapptwosparse;
\legend{\footnotesize Our method, \footnotesize Regressed-PCA}
  \end{axis}
\end{tikzpicture}}
\end{subfigure}
\caption{Our method v.s. \citet{Chenetal_SeimiparametricFactor_2021}'s regressed-PCA: DGP11 with $N=T=50$}\label{Fig: Comparing2sparse}
\end{figure}
\end{appendices}
\spacingset{1.75}
\newpage
\addcontentsline{toc}{section}{References}
\putbib
\end{bibunit}


\end{document}